\documentclass[traditabstract]{aa} % for the abstract without structuration 
\usepackage{graphicx}
\usepackage{txfonts}
\usepackage{natbib}
\bibpunct{(}{)}{;}{a}{}{,}
\newcommand{\ks}{km s$^{-1}$}
\newcommand{\ltsima}{$\; \buildrel < \over \sim \;$}
\newcommand{\simlt}{\lower.5ex\hbox{\ltsima}}
\newcommand{\gtsima}{$\; \buildrel > \over \sim \;$}
\newcommand{\simgt}{\lower.5ex\hbox{\gtsima}}

\begin{document}
   \title{The impact of numerical viscosity in  SPH
simulations of galaxy clusters}

   \subtitle{}

   \author{R. Valdarnini
          \inst{1}
          }

   \institute{SISSA/ISAS, via Bonomea 265, I-34136 Trieste, Italy\\
              \email{valda@sissa.it}
             }

   \date{Received ; accepted }

% \abstract{}{}{}{}{} 
% 5 {} token are mandatory
 
  \abstract{
%   {X-ray observations of galaxy clusters can be profitably used to constrain 
%cosmological parameters provided that in hydrodynamical cluster simulations 
%the thermodynamic of the ICM is correctly reproduced.
%In this context SPH codes are widely used as a tool to investigate the 
%formation and evolution of galaxy clusters in different cosmological 
%scenarios, although the standard artificial viscosity scheme implemented in 
%SPH codes is relatively viscous and affects the level of random gas motions 
%present in the ICM of the simulated clusters.}

 {The goal of this paper is to investigate in N-body/SPH hydrodynamical 
cluster simulations the impact of artificial viscosity on the 
 ICM thermal and  velocity field statistical properties.}
{In order to properly reduce the effects of artificial viscosity a 
time-dependent artificial viscosity scheme is implemented in an SPH code in 
which each particle has its own viscosity parameter, whose time evolution is 
governed by the local shock conditions. 
The new SPH code is verified in a number of test problems with known analytical
or numerical reference solutions and is then used to construct a large set of
N-body/SPH hydrodynamical cluster simulations. These simulations are aimed at
studying  in SPH simulations the impact of artificial viscosity on the 
thermodynamics of the ICM and  its velocity field statistical properties by 
comparing results extracted at the present epoch from runs with different 
artificial viscosity parameters, cluster dynamical states, numerical 
resolution and physical modeling of the gas.}

{Spectral properties of the gas velocity field are investigated by measuring 
for the simulated clusters the velocity power spectrum $E(k)$. The longitudinal 
component $E_c(k)$ exhibits over a limited range a Kolgomorov-like scaling
 $\propto k^{-5/3}$, whilst
the solenoidal power spectrum component $E_s(k)$ is strongly influenced by 
numerical resolution effects.
The dependence of the spectra  $E(k)$ on dissipative effects is found 
to be significant at length scales $\simlt 100-300$~kpc, with viscous damping of the 
velocities being less pronounced in those runs with the lowest artificial 
viscosity.
The turbulent energy density radial profile $E_{turb}(r)$ is strongly affected 
by the numerical viscosity scheme adopted in the simulations, with the 
turbulent-to-total energy density ratios being higher in the runs with the 
lowest artificial viscosity settings and lying in the range between a few 
percent and $\sim10\%$. These values are in accord with the corresponding 
ratios extracted from previous cluster simulations realized using mesh-based 
codes. 

The radial entropy profiles show a
weak dependency on the artificial viscosity parameters of the simulations, with 
a small amount of entropy mixing present in cluster cores.
At large cluster radii, the mass correction terms to the hydrostatic 
equilibrium equation are little affected by the numerical viscosity of 
the simulations, showing that the X-ray mass bias is already estimated 
well in standard SPH simulations.

The results presented here indicate that in individual SPH cluster simulations
at least $N\simgt 256^3$ gas particles are necessary for a correct description
 of turbulent spectral properties over a decade in wavenumbers, whilst radial
profiles of thermodynamic variables can be reliably obtained using 
$N\simgt 64^3$ particles.
Finally, simulations in which the gas can cool radiatively are characterized
by the presence in the cluster inner regions of high levels of turbulence,
generated by the interaction of the compact cool gas core with the 
ambient medium. These findings strongly support the viability  of a turbulent
heating model in which radiative losses in the core are compensated by 
heat diffusion and viscous dissipation of turbulent motion.
}}


   \maketitle
%
%________________________________________________________________

\section{Introduction}

According to the standard hierarchical scenario the formation of larger 
structures is driven by gravity and proceeds hierarchically through merging 
and accretion of smaller size halos. 
Within this scenario galaxy clusters are the most recent and the largest 
virialized objects known in the universe. Their formation and evolution rate 
is then a strong function of the background cosmology, thus making galaxy 
clusters powerful tools for constraining cosmological models
\citep[see][and references therein]{vo05}.
During the gravitational collapse the gaseous component of the cluster is heated
to virial temperatures by processes of adiabatic compression and 
 shock-heating. 
Accordingly, at virial equilibrium most of the baryons in the cluster will
reside in the form of a hot X-ray emitting gas, which is commonly referred to as
the intracluster medium (ICM).

A basic feature of cluster formation occurring in this scenario is the 
large bulk flow motions 
($\sim 1\,000$ \ks) induced in the ICM by major 
merging and gas accretion.
The relative motion between the flow and the ambient gas generates,
at the interface, hydrodynamical instabilities which lead to the development 
of large eddies and to the injection of turbulence in the ICM.
These eddies will in turn form smaller eddies, thereby transferring some of 
the merger energy to smaller scales and generating a turbulent velocity field
with a spectrum expected to be close to a Kolgomorov spectrum.
Additional processes which can stir the ICM are galactic motions and AGN 
outflows, although the contribution to random gas motions from the latter is 
expected to be
relevant only in inner regions of the cluster.

Theoretical estimates for the amount of turbulence present in the ICM are 
dominated by uncertainties in the determination of the kinematic viscosity
$\nu$ of the medium. In the absence of magnetic fields, classical values lie in the
range $Re=U L/\nu\sim 100$ \citep{su03}, where $U$ is the characteristic 
injection scale and $V$ is the characteristic  velocity.
In a magnetized plasma Reynolds numbers are expected to be much higher because
of the reduction in the transport coefficients and the subsequent suppression
of viscosity due to the presence of magnetic fields \citep{ia08}.
These uncertainties in the estimate of the Reynolds numbers of the ICM indicate 
that the classical value can be considered as a lower limit to the level of 
turbulence present in the ICM. A conservative assumption is therefore to 
consider the ICM as moderately turbulent.

On the observational side, turbulence in the ICM could be directly detected using 
high resolution X-ray spectroscopy to measure turbulent velocities through 
emission line broadening. Unfortunately this approach is still below the current 
limit of detectability.
Nevertheless, indirect evidence for the presence of turbulence in the ICM has been 
provided by a number of authors. 
From spatially resolved gas pressure maps of the Coma cluster \cite{su04} measured  
a fluctuation spectrum consistent with the presence of turbulence. Other
observations which suggest the presence of turbulent motion are the lack of 
resonant scattering from the He-like iron $K_{\alpha}$ line at $6.7$ kev
\citep{ch04} and the spreading of metals through the ICM \citep{re06}.

ICM properties are expected to be significantly affected by turbulence in 
a variety of ways. Primarily, the energy of the mergers will be redistributed 
through the cluster volume by the decay of large scale eddies with a turnover
time of the order of a few Gigayears.
Turbulence generated by substructure motion in the ICM  
has been proposed as a heating source to solve the `cooling flow'
present in cluster cores  \citep{fu04a}, the heating mechanism being
the  shock dissipation of the acoustic waves generated by turbulence.
%Heating of the cluster cores through shock dissipation of acoustic waves generated 
%by turbulence has been proposed as a heating source to solve the 'cooling flow'
%problem \citep{fu04}.
Moreover, the usefulness of clusters for precision cosmology relies on accurate
 measurements  of their gravitating mass and can be significantly affected by the 
presence of turbulence in the ICM. X-ray mass estimates are based on the 
assumption that the gas and the total mass distribution are spherically symmetric
and in hydrostatic equilibrium \citep{ras06,nag07a,je07,pi08,lau09}. 
However, the presence of random gas motions implies
additional pressure support which is not accounted for by the hydrostatic 
equilibrium equation.

Turbulence in the ICM may also play an important role in non-thermal phenomena, 
such as the amplification of seed magnetic fields via dynamo processes 
\citep{do02,su06}
and the acceleration of relativistic particles by magnetohydrodynamic waves 
\citep{br07}.
 The transport of metals in the ICM is also likely to be driven by turbulence 
\citep{re06}.

Numerical simulations provide a valuable tool with which to follow in a 
self-consistent manner the complex hydrodynamical flows which take place during
the evolution of the ICM.
In particular, hydrodynamical simulations of merging clusters showed that moving
substructures can generate turbulence in the ICM through shearing instabilities
\citep{ro97,no99a,ta00,ri01,fu04a,ta05,do05,ia08,va09a,pl09}.

The solution of the hydrodynamic equations in a simulation depends on the
adopted numerical method and in cosmology hydrodynamic codes which are used 
to perform simulations of structure formation can be classified into two main
categories: Eulerian, or grid based codes, in which the fluid is evolved on
a discretized mesh \citep{st92,ry93,no99b,fr00,te02}, and Lagrangian methods 
in which the fluid is tracked following the evolution of particles of fixed 
mass \citep{mo05}.
Both of these methods have been widely applied to investigate the formation
and evolution of galaxy clusters 
\citep[cf.][and references  therein]{bk09}.

The main advantage of Lagrangian or smoothed particle hydrodynamics (SPH) codes 
\citep{hk89,sp01,sp05,we09} is that they can naturally follow the development of
matter concentration, but they have the significant drawback that in order to properly
model shock structure they require in the hydrodynamic equations the presence 
of an artificial viscosity term  \citep{mo83}.

 Eulerian schemes such as the Parabolic Piecewise Method \citep{co84} 
implemented in ENZO \citep{no99b,os05a} and FLASH codes \citep{fr00}, 
 are instead characterized by the lack of artificial viscosity and by a better 
shock resolution when compared against SPH codes.
The development of adaptative mesh refinement (AMR) methods, 
in which the spatial resolution of the Eulerian grid is locally 
refined according to some selection criterion \citep{be89,kr97,no05},
led to a substantial increase in 
the dynamic range of cosmological simulations of galaxy clusters and  a better
 capability in following the production of turbulence in the ICM induced by 
merger events \citep{ia08,ma09,va09a,va10}.

 In principle, application of different codes to the same test problem with
identical initial setups should lead to similar predictions.
However, a comparison between the results produced by AMR and SPH codes in
a number of test cases reveals several differences \citep{fr99,os05b,ag07,wa08,
ta08,mi09}.
\cite{ag07} showed that the formation of fluid instabilities is artificially
suppressed in SPH codes with respect AMR codes because of the difficulties 
of SPH codes in properly modeling the large density gradients which develop 
at the fluid interfaces.
The problem has been  re-analyzed by \cite{wa08}, who concluded that the
origin of the discrepancies is due partly to the artificial viscosity 
formulation implemented in SPH and mainly to the Lagrangian nature of SPH, 
which inhibits the mixing of thermal energy \citep{pr08}.

Moreover, a long standing problem between the two numerical approaches occurs
 in non-radiative simulations of a galaxy cluster, where a discrepancy is 
produced at the cluster core by the two codes 
in the radial entropy profile of the gas.  
\citep{fr99,os05b,wa08,mi09}. 
 To investigate the origin of this discrepancy, \cite{mi09} compared the final entropy
profiles extracted from simulation runs of idealized binary merger cluster 
simulations.
They found that in the cluster central regions ( $\sim$ few per cent of the 
virial radius) the entropy profile of Eulerian 
 simulations is a factor $\sim2$ higher than in the SPH runs.
The authors argue that the main source of this difference in the amplitude of 
central entropy is strictly related to the amount of mixing present in the two 
codes. 

In the Eulerian codes it is the numerical scheme which forces the fluids to be
mixed below the minimum cell size.
This is in contrast with SPH simulations, in which some degree of fluid 
undermixing is present owing to the Lagrangian nature of the numerical method.
When compared with the results of cluster simulations performed with AMR codes,
 entropy generation through fluid mixing is therefore inhibited in SPH 
simulations. 
Although a certain degree of overmixing could be present in mesh-based codes
\citep{sp09}, it appears then worth pursuing any improvement in the SPH 
method which leads to an increase in the amount of mixing present in SPH 
cluster simulations.
This is motivated by the strong flexibility that Lagrangian methods posses 
in tracking large
variation in the spatial extent and in the density of the simulated fluid.
 Fluid mixing is expected to increase if viscous damping of
random gas motions if effectively reduced.
This effect occurs in SPH simulations because of the presence of an artificial
viscosity term in the  hydrodynamic equations, which is necessary to properly 
model shocks but introduces a numerical viscosity. 

In the standard formulation of SPH, the strength of artificial viscosity for
approaching particles is controlled by a pair of parameters which are fixed 
throughout the simulation domain. This renders the numerical scheme relatively
viscous in regions far away from the shocks, with subsequent pre-shock 
entropy generation and the damping of turbulent motions as a side effect.
Given these difficulties \cite{mm97} proposed a modification  to the original scheme 
in which each particle has its own viscous coefficient, whose time evolution
is governed by certain local conditions which depends on the shock strength.
The benefit of this method is that the artificial viscosity is high in supersonic
flows where it is effectively needed but quickly decays to a minimum value in 
the absence of shocks. 
As a consequence, the numerical viscosity is strongly reduced in regions away 
from shocks and the modeling of turbulence generated by shearing motions
is greatly improved. 
 
Given the shortcomings of SPH codes which affect the development of turbulence
in  hydrodynamic cluster simulations, it is therefore interesting to conduct a 
study to analyze the effect that the adoption of a time-dependent artificial
viscosity has on ICM properties of SPH cluster simulations.
This is the goal of the present work, in which a time-dependent artificial viscosity
 formulation is implemented in a SPH code with the purpose of studying the
differences induced in the ICM random gas motion of simulated clusters 
with respect the standard artificial viscosity scheme.

 To this end, a test suite of simulations is presented in which the ICM final
profiles and turbulent statistical properties provide the quantitative 
measures used to compare runs with different  artificial viscosity parameters,
cluster dynamical states, numerical resolution and physical modeling of the gas.
A similar study has already been undertaken in pioneering work by \cite{do05}, 
who analyzed the role of numerical viscosity and the level of random gas
motions in a set of cosmological SPH cluster simulations.
Here several aspects of a time-dependent  artificial viscosity implementation
in SPH are discussed in a more systematic way.

Specifically, we study the effects on the development of turbulence in the ICM
which come from varying the simulation parameters which govern the time 
evolution of the artificial viscosity strength in the new formulation.
Moreover, in addition to the adiabatic gas dynamical simulations 
a set of cooling runs was constructed in which the physical modeling of the 
gas includes radiative cooling, star formation and energy feedback.

The paper is organized as follows.  In Sect. \ref{sph.sec} we present the 
hydrodynamical method  and the implementation of the artificial viscosity 
scheme.
The construction of the set of simulated cluster samples used to perform 
comparisons
between results extracted from SPH runs with different artificial viscosity 
parameters, is described in Sect. \ref{sample.sec}.
Sect. \ref{stat.sec} provides an introduction to several statistical methods 
used in homogeneous isotropic turbulence
to characterize statistical properties of the velocity field of a medium. 
Some numerical tests are discussed in Sect. \ref{test.sec}, in order  
to assess the validity of the code and its shock
resolution capabilities.
The results of the cluster simulations are presented in Sect. \ref{cluster.sec}, 
while Sect. \ref{concl.sec} summarizes the main conclusions.

%__________________________________________________________________

\section{Description of the code}
\label{sph.sec}
Here we provide the basic features of the numerical scheme used to 
follow the hydrodynamics of the fluid; for a comprehensive
review see \cite{mo05}.
%\subsection{Numerical Method}
\subsection{The hydrodynamical method}
\label{hydro.sec}
The hydrodynamic equations of fluid motion are solved according to the 
SPH method, in which the fluid is described within the domain by a 
collection of $N$ particles with mass $m_i$, velocity $\vec v_i$, density
$\rho_i$ and a thermodynamic variable such as the specific thermal energy 
$u_i$ or the entropy $A_i$. The latter is related to the particle pressure
$P_i$ by $P_i=A_i\rho_i^{\gamma}$, where $\gamma=5/3$ for a monoatomic gas.
The density estimate $\rho(\vec r)$ at the particle position $\vec r_i$ 
is given by 

   \begin{equation}
 \rho_i=\sum_j m_j W(r_{ij},h_i)~,
    \label{rho.eq}
   \end{equation}
where $W(|\vec r_i-\vec r_j|,h_i)$ is the $B2$ or cubic spline kernel
which has compact support and is zero for $|\vec r_i-\vec r_j|\geq2h_i$
\citep{mo05}. The sum (\ref{rho.eq}) is over a finite number of 
particles and the smoothing length $h_i$  is a varying 
quantity which is implicitly defined through the equation \citep{sh02}
   \begin{equation}
  \frac{4 \pi (2h_i)^3 \rho_i}{3}=N_{sph} m_i~,
    \label{hrho.eq}
   \end{equation}
with $N_{sph}$ being the number of neighboring particles within a radius 
$2h_i$. Typical choices lie in the range $N_{sph}\sim 33-50$, here $N_{sph}=33$
is used.
\begin{table}
\caption{Summary of the AV parameters used in the simulations. First column
is the index label which identifies the runs with different AV parameters. 
The index $0$ is for the standard AV scheme, other indices refer to different 
values of the $(\alpha_{min},l_d)$ pair used in the new time-varying AV scheme.
The last column gives the value of decay constant 
$\delta_{\alpha}\sim 0.447 /l_d$ defined by \cite{mm97}.}  
\label{visc.tab}      % is used to refer this table in the text
\centering                          % used for centering table
\begin{tabular}{c c c c c}        % centered columns (4 columns)
\hline\hline                 % inserts double horizontal lines
AV & $\alpha_{min}$ & $\alpha_{max}$ & $l_d$ & $\delta_a \sim$\\    
\hline                        % inserts single horizontal line
   0 & 1. & 1. & -- & --\\      % inserting body of the table
   1 & 0.1 & 1.5 & 0.1 & 4.5\\      % inserting body of the table
   2 & 0.1 & 1.5 & 0.2 & 2.25 \\
   3 & 0.1& 1.5 & 0.5 & 1 \\
   4 & 0.1 & 1.5 & 1 & 0.5\\
   5 & 0.01 & 1.5 & 1 & 0.5\\ 
\hline                                   %inserts single line
\end{tabular}
\end{table}

The equation of motion for the SPH particles is given by 

   \begin{equation}
     \frac {d \vec v_i}{dt}=-\sum_j m_j \left[
     \frac{P_i}{\Omega_i \rho_i^2} 
  \vec \nabla_i W_{ij}(h_i) +f_j \frac{P_j}{\Omega_j \rho_j^2}
   \vec \nabla_i W_{ij}(h_j)
   \right]~,
  \label{fsph.eq}
   \end{equation}
where the coefficients $\Omega_i$ are defined as 
   \begin{equation}
   \Omega_i=\left[1-\frac{\partial h_i}{\partial \rho_i} 
   \sum_k m_k \frac{\partial W_{ik}(h_i)}{\partial h_i}\right]~,
    \label{fh.eq}
   \end{equation}
 and in the momentum equation (\ref{fsph.eq}) account for the effects due to 
the gradients of the smoothing length $h_i$ \citep{mo05}.
The momentum equation must be 
generalized by including an additional viscous pressure term which in SPH 
is needed to represent the effects of shocks. This is achieved in SPH by introducing 
an artificial viscosity (herefater AV) term with the purpose of converting
kinetic energy into heat and preventing particle interpenetration during
shocks. The new term is given by

%%%%%%%%%%%%%%%%%%%%%%%%%%%%%%
\begin{table*}[ht]   
\caption{
\label{clu.tab} 
Main cluster properties and simulation parameters of the baseline sample.
From left to the right : the dynamical subset class , 
the cosmological sample membership,
the corresponding sample index of the cluster, 
the present cluster mass $M_{200}$  within $r_{200}$ in units of 
$h^{-1} M_{\odot}$, the radius $r_{200}$ is units of Mpc,
the number of gas particles $N_{gas}$ inside the $L_c/2$ sphere at $z=z_{in}$, 
 the same but for the dark matter particles,
  the mass of the gas particles $m_{gas}$ in $M_{\odot}$,
 the same but for the dark matter particles,
 the gravitational softening parameter for the gas in kpc and the
 root mean square plane average cluster power ratio $\bar {\Pi}_3(r_{200})$}
\centering                            
\begin{tabular}{ccccccccccc}        
\hline\hline                 
 dynamical state & sample & index &$M_{200}[h^{-1} M_{\odot}]$ & 
$r_{200}$[Mpc]  & $N_{gas}$ & $N_{dm}$ & $m_{gas}[M_{\odot}]$ &
 $m_{dm}[M_{\odot}]$ & $\varepsilon_{gas}$[kpc] & $\bar {\Pi}_{3}(r_{200})$ \\ 
\hline        

{\it perturbed(P)} &  ${\mathrm S_8}$ & $1$ & $6\cdot10^{14}$ &  $1.96$ &
$220175 $ &$220144 $ & $3\cdot10^9$& $1.5\cdot10^{10}$ & $25.4$ & $-3.99$\\  

%{\it perturbed(P)} &  ${\mathrm S_8}$ & $1$ & $5.5\cdot10^{14}$ &  $1.01$ &
%$ 72311$ &$ 72264$ & $9.17\cdot10^9$& $4.74\cdot10^{10}$ & $36.8$ & $-3.99$\\  

&  ${\mathrm S_4}$ & $5$ &  $2.4\cdot10^{14}$ &  $1.45$ &
$ 221007$ &$ 220976$ & $2\cdot10^9$& $1.1\cdot10^{10}$ & $22.2$ & $-4.6$\\  
%&  ${\mathrm S_4}$ & $5$ &  $2.26\cdot10^{14}$ &  $1.42$ &
%$ 72375$ &$ 72688$ & $6.2\cdot10^9$& $3.2\cdot10^{10}$ & $32.3$ & $-4.6$\\  

&  ${\mathrm S_4}$  &$16$ &  $2\cdot10^{14}$ &  $1.37$ &
$ 221167$ &$ 221136$ & $1.7\cdot10^9$& $8.7\cdot10^{9}$ & $21$ & $-6.07$\\  
%&  ${\mathrm S_4}$  &$16$ &  $2.2\cdot10^{14}$ &  $1.4$ &
%$ 72703$ &$ 72656$ & $5.1\cdot10^9$& $2.6\cdot10^{10}$ & $30.3$ & $-6.07$\\  

&  ${\mathrm S_2}$ &$105$ & $6\cdot10^{13}$ &  $0.91$ &
$ 221391$ &$221360 $ & $2.5\cdot10^8$& $1.3\cdot10^{9}$ & $11.1$ & $-4.65$\\  
%&  ${\mathrm S_2}$ &$105$ & $6.45\cdot10^{13}$ &  $0.935$ &
%$ 72855$ &$ 72808$ & $7.7\cdot10^8$& $4\cdot10^{9}$ & $16.1$ & $-4.65$\\  

\hline       
                       
{\it quiescent (Q)} &  ${\mathrm S_4}$ &$11$ & $5.7\cdot10^{14}$ &  $1.94$ &
$ 220759$ &$220728 $ & $1.5\cdot10^9$& $7.6\cdot10^{9}$ & $20$ & $-7.46$\\  
%{\it quiescent (Q)} &  ${\mathrm S_4}$ &$11$ & $6.2\cdot10^{14}$ &  $1.98$ &
%$ 72599$ &$ 72552$ & $4.5\cdot10^9$& $2.3\cdot10^{10}$ & $29$ & $-7.46$\\  

&  ${\mathrm S_4}$ &$19$ & $7\cdot10^{14}$ &  $2.07$ &
$ 219631$ &$219600 $ & $1.7\cdot10^9$& $8.7\cdot10^{9}$ & $21$ & $-8.5$\\  
%&  ${\mathrm S_4}$ &$19$ & $6.4\cdot10^{14}$ &  $2.01$ &
%$ 72175$ &$ 72128$ & $5.1\cdot10^9$& $2.6\cdot10^{10}$ & $30$ & $-8.5$\\  

&  ${\mathrm S_2}$ &$13$ & $2.4\cdot10^{14}$ &  $1.44$ &
$ 221823$ &$221792 $ & $6.4\cdot10^8$& $3.3\cdot10^{9}$ & $15.1$ & $-8.67$\\  
%&  ${\mathrm S_2}$ &$13$ & $2.58\cdot10^{14}$ &  $1.48$ &
%$ 73135$ &$ 73088$ & $1.94\cdot10^9$& $1.0\cdot10^{10}$ & $22$ & $-8.67$\\  

&  ${\mathrm S_2}$ &$110$ & $9.4\cdot10^{13}$ &  $1.05$ &
$ 222111$ &$222080 $ & $2.5\cdot10^8$& $1.3\cdot10^{9}$ & $11.1$ & $-8.27$\\  
%&  ${\mathrm S_2}$ &$110$ & $9.74\cdot10^{13}$ &  $1.07$ &
%$ 73159$ &$ 73112$ & $7.7\cdot10^8$& $4\cdot10^{9}$ & $16.1$ & $-8.27$\\  

\hline  

\hline                           
\end{tabular}
\end{table*}

   \begin{equation}
   \left (\frac {d \vec v_i}{dt}\right )_{visc}=-\sum_i m_j \Pi_{ij} \vec \nabla_i \bar W_{ij}~,
    \label{fvis.eq}
   \end{equation}
where the term  $\bar W_{ij}= \frac{1}{2}(W(r_{ij},h_i)+W(r_{ij},h_j))$ is the
symmetrized kernel and $\Pi_{ij}$ is the AV tensor.
To follow the thermal evolution of the gas, an entropy-conserving
approach \citep{sh02} is used here and entropy is generated at a rate 
 
   \begin{equation}
  \frac {d A_i}{dt} =\frac{1}{2}\frac{\gamma-1}{\rho_i^{\gamma-1}}
  \sum_j m_j \Pi_{ij} \vec v_{ij}\cdot \nabla_i \bar W_{ij}
    -\frac{\gamma-1}{\rho_i^{\gamma}} \Lambda(\rho_i,T_i)~,
    \label{avis.eq}
   \end{equation}

where  $\vec v_{ij}= \vec v_i - \vec v_j$, $T_i$ is the particle temperature  and 
the additional term $\Lambda(\rho_i,T_i)$
accounts for the radiative losses of the gas, if present. 
 In the following, simulations in which the cooling term $\Lambda$ is absent 
from Eq. (\ref{avis.eq}) will be referred to as adiabatic.
The expression for the AV tensor $\Pi_{ij}$ is

   \begin{equation}
\Pi_{ij} = \left\{ 
\begin{array} {ll}
   \frac{-\alpha_{ij} c_{ij} \mu_{ij}+\beta_{ij} \mu_{ij}^2}
   {\rho_{ij}} f_{ij}  & \mbox{if $\vec v_{ij} \cdot \vec r_{ij} > 0$} \\
   0 & \mbox{otherwise~,}  
   \end{array}
\right. 
   \label{pvis.eq}
   \end{equation}
and so  $\Pi_{ij}$ is non-zero only for approaching particles. Here 
 scalar quantities with the subscripts $i$ and $j$ denote arithmetic averages,
$c_i$ is the sound speed of particle $i$, the parameters $\alpha_i$ and $\beta_i$ 
regulate the amount of AV and $f_i$ is a controlling factor which reduces 
the strength of AV in presence of shear flows. The $\mu_{ij}$ term is
given by 
   \begin{equation}
  \mu_{ij}=\frac{h_{ij} \vec v_{ij} \cdot \vec r_{ij}}{ r_{ij}^2+\eta h_{ij}^2}~,
   \label{muvis.eq}
   \end{equation}
where the factor $\eta=10^{-2}$ is included to prevent numerical divergences.
In order to limit the amount of AV generated in shear flows, \cite{ba95} 
proposed the following expression for $f_i$ :

   \begin{equation}
  f_i=\frac {|\vec \nabla \cdot \vec v|_i}
  {|\vec \nabla \cdot \vec v|_i+|\vec \nabla \times \vec v|_i+\eta^{\prime}_i}~,
   \label{fdamp.eq}
   \end{equation}
where $(\vec \nabla \cdot \vec v)_i$ and $(\vec \nabla \times \vec v)_i$
are the standard SPH estimates for divergence and curl \citep{mo05}, and
the factor $\eta^{\prime}_i=10^{-4} c_i/h_i$ is inserted to prevent numerical 
divergences.
This expression for $f_i$ is effective in suppressing AV in pure shear flows,
for which the condition
 $|\vec \nabla \times \vec v|_i>> |\vec \nabla \cdot \vec v|_i$ holds.

The strength of the AV in the standard SPH formulation  
 is given by $\beta_i=2 \alpha_i$ and $\alpha_i=\mathrm{const}\equiv \alpha_0$,
with $\alpha_0=1$ being a common choice for the viscosity coefficient 
\citep{mo05}.
 In the following this parametrization will be referred to as the 
`standard' AV.

\cite{mo97} proposed a modification to this parametrization for AV based on an 
analogy with the Riemann problem. In a number of test problems, results obtained
using his `signal velocity' formulation are found to be equivalent or 
slightly improved
with respect to the standard AV scheme. However, this new formulation is not
introduced here in order to avoid any further source of difference in the results
produced by the SPH code, additional to those due to the choice of the 
time-dependent AV scheme parameters.

In simulations in which the gas is allowed to cool radiatively, cold gas 
in high density regions is subject to star formation and gas particles 
are eligible to form star particles. Energy and metal feedback is returned 
from these star particles to their gas neighbors by supernova explosions, 
according to the stellar lifetime and initial mass function. 
For a detailed description see also \cite{va06}.
%{\bf HERE }

% \begin{equation}
%(\vec \nabla \cdot \vec v)_i=-\frac{1}{\rho_i}\sum_j m_j \vec v_{ij} 
% \cdot \nabla_i W _{ij}
% \label{div.eq}
% \end{equation}

% \begin{equation}
%(\vec \nabla \times \vec v)_i=\frac{1}{\rho_i}\sum_j m_j \vec v_{ij} 
% \times \nabla_i W _{ij}
% \label{rot.eq}
% \end{equation}

\subsection{The new artificial viscosity formulation}
\label{alfa.sec}
The standard AV formulation is successfull in properly resolving shocks
but at the same time generates unwanted viscous dissipation in 
regions of the flow which are not undergoing shocks. \cite{mm97} proposed that 
the viscous coefficient should be different for each particle and 
left free to evolve in time under the local conditions. Following 
\cite{mm97}
   \begin{equation}
  \frac {d \alpha_i}{dt} =-\frac{\alpha_i-\alpha_{min}}{\tau_i} +{\tilde S}_i~,
    \label{alfa.eq}
   \end{equation}
where
   \begin{equation}
  \tau_i=\frac{h_i}{c_i ~l_d}
    \label{tau.eq}
   \end{equation}
 is a decay time-scale which regulates, through the dimensionless parameters 
$l_d$, the time evolution of $\alpha_i(t)$ away from shocks. The parameter 
$\alpha_{min}$ is the minimum value to which $\alpha_i(t)$  is allowed to 
decay. The source term 
${\tilde S}_i$ is given by
   \begin{equation}
 {\tilde S}_i=f_i S_0 {max}(-(\vec \nabla \cdot \vec v)_i,0)
(\alpha_{max}-\alpha_i)\equiv S_i (\alpha_{max}-\alpha_i)~,
    \label{salfa.eq}
   \end{equation}
 and is constructed in such a way that it increases in the presence of 
 shocks. Following a suggestion of \cite{mm97}, the damping factor $f_i$ is 
inserted to account for the presence of vorticity. The scale factor 

   \begin{equation}
 S_0=\ln( \frac{5/3+1}{5/3-1}) / \ln( \frac{\gamma+1}{\gamma-1})~
   \end{equation}
 is chosen such that the peak value of $\alpha_i$ 
is independent of the equation of state. 
The source term equation (\ref{salfa.eq})
 is of the form proposed by \cite{ro00}, which allows a better response in 
the presence of shocks with respect to the original formulation.
In a number of test simulations \cite{ro00} found that appropriate values 
for the parameters $\alpha_{max},\alpha_{min},l_d$ are $1.5,0.05$ and $0.2$,
 respectively.

From the viewpoint of the description of the fluid flow velocities, the most 
significant parameters are $\alpha_{min}$ and $l_d$. Since the goal of the SPH 
simulations presented here is to investigate the effect on  ICM fluid flows 
of the numerical viscosity, the set of runs will be performed with the 
parameter $\alpha_{max}$ set to the value $\alpha_{max}=1.5$ (see Sect. 
\ref{tubea.sec}), whilst a range of values will be considered
for the AV parameters  $\alpha_{min}$ and $l_d$. 
However, a lower limit to the timescale $\tau_i$ is set by the minimum time 
taken to propagate through the resolution length $h_i$, so that the value 
$l_d=1$ sets an upper limit to the parameter $l_d$.

The different implementations of AV used in the simulations are summarized in 
Table \ref{visc.tab}. In the simulations which incorporate the new 
time-dependent AV scheme, five different pairs of values have been chosen 
for the parameters $(\alpha_{min},l_d)$, while simulations in which the AV is 
modeled according to the standard formulation are used for reference purposes.
 Simulations with different AV schemes or parameters will then be labeled by 
the corresponding index of Table \ref{visc.tab}.

However, if the decay parameter $l_d$ approaches unity, the time-dependent 
AV scheme discussed here may fail to properly evolve 
the viscosity parameters $\alpha_i$ when a shock is present.
To avoid these difficulties an AV scheme which generalizes the \cite{ro00}
source term expression is proposed here. Since the implementation  of
the new source term equation is strictly related to the shock tube problem
discussed in Sect. \ref{testa.sec}, the modifications to Eq. (\ref{salfa.eq}) 
will be presented in Sect. \ref{testa.sec} together with the shock tube tests.

\section{Sample construction of simulated clusters}
\label{sample.sec}
 In order to investigate the effects of numerical viscosity on 
 ICM random gas motions a large ensemble of hydrodynamical cluster simulations 
was created  by performing, for a chosen AV test case, SPH hydrodynamical 
simulations  using a baseline sample consisting of eight different initial 
conditions for the simulated clusters. 
These initial conditions are those of simulated clusters which are part of a 
large set produced in an ensemble of cosmological simulations (see later), 
and  these clusters are chosen at the present epoch with the selection criterion
of covering a wide range in virial masses and dynamical properties.
This choice was made in order to study, using the hydrodynamical simulations, 
the 
dependencies of the ICM gas velocities on these cluster properties.
The number of eight clusters was a compromise between the need  for
obtaining results of sufficient statistical significance and of keeping the
computational cost at a minimum level, given the range of parameters explored.

The initial conditions of the  baseline sample were constructed according to
the following procedures \citep{va06}, a similar approach has been followed
by \citet{do05} to construct their set of high resolution hydrodynamical 
cluster simulations. An N-body cosmological simulation is first run with a 
comoving box of size 
$L_2=200h^{-1}$Mpc. 
The cosmological model assumes a flat geometry with the present
matter density $\Omega_\mathrm{m}=0.3$, vacuum energy density 
$\Omega_\mathrm{\Lambda}=0.7$, $\Omega_\mathrm{b}=0.0486$, and with $h=0.7$  
being the value of the Hubble constant $H$ in units of $100$ \ks
{Mpc$^{-1}$. The scale 
invariant power spectrum is normalized to $\sigma_\mathrm{8}=0.9$ on an 
$8 \, h^{-1}$ Mpc scale at the present epoch.

A cluster catalog was generated by identifying clusters in the simulation 
at $z=0$ using a  friends-of-friends (FoF) algorithm to detect overdensities in 
excess of $\sim200 \Omega_\mathrm{m}^{-0.6}$.
The sample comprises $N_2=120$ clusters ordered according to the value of 
their mass $M_{200}$, where 

   \begin{equation}
   M_{\Delta}= (4 \pi/3) \, \Delta\, \rho_\mathrm{c} \, r_{\Delta}^3
   \end{equation}

 denotes the mass contained in a sphere of radius
$ r_{\Delta}$ with mean density $\Delta$ times the critical density 
$\rho_\mathrm{c}$.

Each of these clusters was then resimulated individually using an N-body+SPH 
simulation performed in 
physical coordinates starting from the initial redshift $z_{in}=49$. 
The integration was made by first locating the cluster
center at $z=0$  and identifying all of the simulation particles  lying
within $r_{200}$. These particles are then located at $z_{in}$ and 
a cube of size $L_c\propto M_{200}^{1/3}$ , enclosing all of them, is found.
 A lattice of $N_L=74^3$ is set inside the cube and to each node is
associated a gas particle with its mass and position, a similar lattice is set for 
dark matter particles.
% N_L=35 , 51 , 74
The particle positions are then perturbed, using the same random realization
as for the cosmological simulations. 
 The particles kept for the hydrodynamic simulation 
 are those whose perturbed positions lie within a sphere
of radius $L_c/2$ from the cluster center.
To model tidal forces, the sphere is surrounded out to a radius 
$L_c$ by an external shell of dark matter particles extracted from a cube of 
 size $2L_c$  centered in the same way as the original cube
and consisting of $N_L=74^3$ grid points.
 The particles are evolved up to the present time using an entropy-conserving 
SPH code, described in Sect. \ref{hydro.sec}, combined with a treecode 
gravity solver. Particles are allowed
to have individual timesteps and their gravitational softening parameter is set
 according to the scaling $\varepsilon_i \propto m_i^{1/3}$, where $m_i$ is 
the mass of particle $i$. The relation is  normalized  by
  $\varepsilon_i =15~( m_i/6.2\cdot10^8 M_{\odot})^{1/3}$~kpc. 
The set of these individual cluster hydrodynamical simulations is then 
denoted as sample $S_2$.

The whole procedure is then repeated twice more in order to generate the cluster samples 
$S_4$ and $S_8$ from cosmological simulations with box size $L_4=400h^{-1}$Mpc 
and 
$L_8=800h^{-1}$Mpc. The number of clusters $N_m$ in these samples 
is chosen such that the mass $M_{200}$ of the $N_m-th$ cluster of sample
$S_m$ is greater that the mass $M_{200}$ of the most massive cluster of sample 
$S_{m/2}$, and samples $S_8$ and $S_4$ consist of  $N_8=10$  and $N_4=33$ 
clusters, respectively.
The final cluster sample $S_{all}$ is constructed by combining all of the 
samples $S_m$ generated from the three cosmological runs. 

In order to provide the baseline sample for the hydrodynamical simulations,
eight clusters were chosen from those of sample $S_{all}$, with the 
selection criterion being the construction of a representative sample of the 
cluster dynamical states and masses.
\begin{table}
\caption{The main parameters used in simulations of different 
resolution. The first column gives the label which denotes the runs with 
different numbers of  particles, $N_L$ is the number of grid points
set inside the $L_c$ cube, and the last two columns indicate the approximate 
number of gas particles and total particles used in the cluster simulations.}  
\label{resol.tab}      % is used to refer this table in the text
\centering                          % used for centering table
\begin{tabular}{c c c c }        % centered columns (4 columns)
\hline\hline                 % inserts double horizontal lines
{\it sim resolution} & $N_L$ & $\sim N_{gas}$ & $\sim N_{tot}$ \\    
\hline                        % inserts single horizontal line
   $\mathrm {LR}$ & $35^3$ & 24\,000 & 72\,000  \\      % inserting body of the table
   $\mathrm  {MR}$& $51^3$ & 73\,000 & 216\,000 \\      % inserting body of the table
   $\mathrm {HR}$ & $74^3$ & 220\,000 & 660\,000  \\
\hline                                   %inserts single line
\end{tabular}
\end{table}

 As an indicator of the cluster dynamical state, we adopt the power ratios
$P_r/P_0$. The quantity $P_r$ is proportional to the square of the 
$r$-th moments of the X-ray surface brightness $S_X(x,y)$,
 as measured within a circular aperture of radius $R_{ap}$ 
in the plane orthogonal to the line of sight \citep{bu95}.
 Here the power ratio $P_3/P_0$, or equivalently 
$\Pi_3(R_\mathrm{ap})=\log_{10} (P_3/P_0)$, is used as an estimator of the 
 cluster dynamical state since it gives an unambiguous detection of 
asymmetric structure. 
For a fully relaxed configuration $\Pi_\mathrm{r}\rightarrow - \infty$.

For the simulated clusters of sample $S_{all}$, the power ratios are evaluated 
at $z=0$ in correspondence of the aperture radius $R_{ap}=r_{200}$. 
To minimize projection effects, the average quantity
 $\bar {\Pi}_3(r_{200})=\log_{10} (\bar {P}_3/\bar {P}_0)$ is used to 
estimate the  cluster dynamical state, where $\bar {P}_r$ it is 
the {\it rms} plane average of the moments $P_r$ evaluated along the three 
orthogonal lines of sight. 
Clusters of sample $S_{all}$ are then sorted according to their values of 
 $\bar {\Pi}_3(r_{200})$  and eight clusters are extracted from the sample.
The four of them with the lowest values of  $\bar {\Pi}_3(r_{200})$ 
among the cluster power ratio distribution are chosen and these relaxed 
clusters will be denoted as quiescent ($Q$) or relaxed clusters.
The remaining four are chosen with the opposite criterion of having the
power ratios with the highest values among those of the 
$\bar {\Pi}_3(r_{200})$ distribution.
These clusters are then labeled as perturbed or $P$ clusters.
Within each subset, clusters are chosen with the additional 
criterion of having their virial mass distribution as wide as possible.
 The cluster mass at $r_{200}$  is used in place of the virial mass and 
Table \ref{clu.tab} lists the main cluster properties and simulation
parameters of the  baseline sample constructed according to the above criteria.

The numerical convergence of the results is assessed by studying the dependence 
of the final profiles on the numerical resolution adopted in the simulations.
In addition to the hydrodynamical simulations realized from the 
baseline sample using various AV prescriptions, 
 a set of mirror runs with a different number of simulation particles
was then performed with the aim of studying the stability of the final 
results against the numerical resolution of the simulations. 
 Because of the large number of AV test simulations performed here, 
the stability against resolution of the simulation results 
for the baseline sample is 
investigated by comparing the chosen profiles with the 
corresponding ones obtained from parent simulations with lower resolution.
A sufficient condition for the numerical convergence of the profiles is given
by their stability against simulations performed with higher resolution.
However, the approach undertaken here provides significant indications 
about the stability of the baseline sample and at the same time allows 
a substantial reduction in the amount of computational resources needed
to assess numerical convergence.

The initial conditions of the mirror runs are therefore constructed using the
same procedures as previously described for the baseline sample, the
only difference being the number $N_L$ of grid points set inside 
the $L_c$ cubes. 
With respect to the baseline sample, which will be denoted
as high resolution (HR), low (LR) and medium resolution (MR) runs 
are obtained by setting $N_L=35^3$ and $N_L=51^3$, respectively. 
Table \ref{resol.tab} lists some basic parameters.
Individual runs have their label obtained by combining the different 
labels specified in the Tables.

%                                     Two column figure (place early!)
%______________________________________________ Gamma_1 (lg rho, lg e)
   \begin{figure*}
   \centering
   \includegraphics{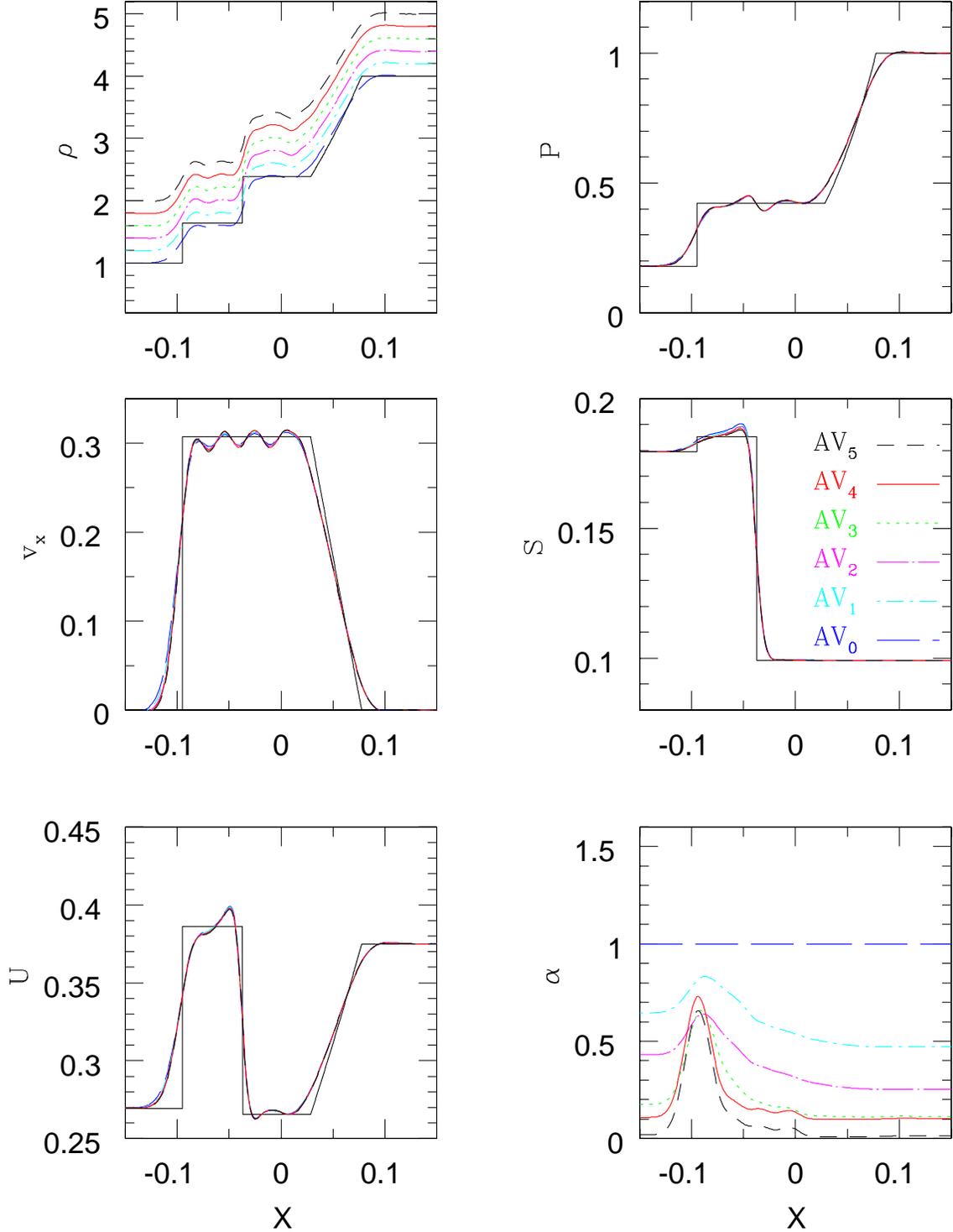}
   \caption{Results at $t=0.12$ from the shock tube test for 3D SPH
runs with different AV parameters. The profiles are projected along the 
shock front. From top to bottom are plotted: density, velocity and thermal
energy ( left column); pressure, entropy and viscosity parameter (right
column). The solid black line represents the analytical solution, while lines
with different styles and colors are the profiles of the SPH runs 
with different AV parameters. Different runs are labeled according 
to Table \ref{visc.tab} and the relationship with the corresponding profiles 
is illustrated in the entropy panel.  In the density panel profiles
of different runs have been shifted vertically to better illustrate their
relative differences.}
   \label{tubea.fig}%
    \end{figure*}
%

%
%_____________________________________________________________

\section{Statistical measures}
\label{stat.sec}
Here we present several statistical tools with which will be studied the 
impact that the strength of AV has on the statistical properties of
the turbulent velocity field $\vec u(\vec x)$ of the simulated clusters. 
 Homogeneous isotropic turbulence is commonly studied using 
the spectral properties of the velocity field $\vec u(\vec x)$. Its Fourier 
transform is defined as 
   \begin{equation}
 \vec {\tilde u}(\vec k)= \frac{1}{(2\pi)^3} \int \vec u (\vec x) 
 e^{-\imath 2 \pi \vec k \cdot \vec x}d^3x~,
    \label{vkft.eq}
   \end{equation}
 and the ensemble average velocity power spectrum  $\mathcal{P}(k)$  is 
given by

   \begin{equation}
< \vec {\tilde u}^{\dagger}(\vec k^{\prime}) \vec {\tilde u}(\vec k)>=
\delta_D(\vec k^{\prime}-\vec k) \mathcal{P}(k).
    \label{pwft.eq}
   \end{equation}
The  energy spectrum function $E(k)$ is then defined as 

   \begin{equation}
E(k)=2 \pi k^2 \mathcal{P}(k), 
%\vec {\tilde u}(\vec k)^2|
    \label{eka.eq}
   \end{equation}
 where $k\equiv|\vec k|$, and the integral of $E(k)$ is the mean kinetic energy per 
unit mass
   \begin{equation}
\int_0^{\infty} E(k) d~k= \frac{1}{2} <u^2>.
%\vec {\tilde u}(\vec k)^2|
    \label{ekb.eq}
   \end{equation}

In the case of incompressible turbulence the energy spectrum follows the 
Kolgomorov scaling $E(k) \propto k^{-5/3}$. 
In contrast, the regime of compressible turbulence  exhibits 
large variation in the gas density and a generalized energy spectrum 
can be considered by introducing  in Eq. (\ref{vkft.eq}) a weighting function  
 $\vec u(\vec x) \rightarrow  \vec u_{w}(\vec x)\equiv w(\vec x) \vec u(\vec x)$, 
where
 $w(\vec x)$  is proportional to some power of the density.
For compressible turbulence  a  natural choice  is $w(\vec x)\propto 
\sqrt {\rho(x)}$,
 in this case the integral of $E(k)$ is just the kinetic energy density :
$ \int E(k) d~k= \frac{1}{2} <\rho u^2>$.
As noticed by \cite{ki09} the spectral properties of compressible turbulence 
are better described using this weighting scheme, which 
allows a more accurate estimate of the power at small scales
 where mass accumulation occurs  because of shocks.  
Alternatively, \cite{kr07} demonstrated that for supersonic isothermal 
turbulence, a Kolgomorov scaling for the spectrum function $E(k)$
is recovered  using $w(\vec x)\propto {\rho(x)}^{1/3}$. 
Here the spectral properties of compressible turbulence will be studied 
using a generalized energy spectrum with density weighting given by
 $w(\vec x)\propto {\rho(x)}^{1/2}$.
This is the same choice adopted by \cite{ki09} and
provides for the energy spectrum $E(k)$ a physical  reference to the kinetic
energy density.

To study the energy spectrum in cluster simulations, spectral quantities  were
constructed as follows. A cube of side $L_{sp}$ with $N_g^3$ grid points is 
first placed at the cluster center and 
 hydrodynamic variables $A(\vec x)$ are estimated at the grid points 
$\vec x_p$ according to the SPH prescription

   \begin{equation}
 A(\vec x_p)= \sum_i A_i\frac{m_i}{\rho_i} W(|\vec x_p-\vec x_i|,h_i),
    \label{sphvar.eq}
   \end{equation}

where $A(\vec x)$ denotes either the components of the 
velocity field $\vec u(\vec x)$ or
the  weighting function $w(\vec x)$.
% and only gas particles which lie within 
%a distance $L_{sp}/2$ from the cluster center are considered for the summation. 
% The weighting function  is set
%to $w(\vec x) =(\rho(\vec x)/\rho_0)^{1/2}$, where
% $\rho_0=\sum \rho(\vec x_p)/N_g^3$ is the mean density within the cube.

The weighting function is set
to $w(\vec x) =\rho(\vec x)^{1/2}/S_w$, where  
 $S_w=\sum \rho(\vec x_p)^{1/2}/N_g^3$. This normalization choice guarantees 
$\sum w(\vec x_p)=N_g^3$, regardless of the power density exponent used in 
the weighting.
The discrete transforms of 
$\vec u_{w}(\vec x_p)$ are then computed using fast Fourier transforms and an 
angle-averaged density-weighted power spectrum 
$\mathcal{P}^d(k)=< |\vec {\tilde u_{w}}^d(\vec k)|^2>$
is generated as a function of $k=|\vec k|$, binning the quantity 
$|\vec {\tilde u_{w}}^d(\vec k)|^2$ in spherical shells of radius 
$k$ and averaging in the bins.

As an estimator of the the energy density (\ref{eka.eq}) one can therefore use
$E(k)=2 \pi k^2 \mathcal{P}^d(k) (\frac{L_{sp}}{2\pi})^3$. However, 
 in order to consistently compare  density-weighted power spectra extracted 
from different clusters and boxes  it is useful to define  a rescaled 
turbulent power spectrum as  

   \begin{equation}
E(k)=\frac{1}{L_{sp} \sigma^2_v}\left [ 2 \pi k^2 \mathcal{P}^d(k) 
\left (\frac{L_{sp}}{2\pi} \right )^3 \right ],
    \label{pow.eq}
   \end{equation}
where $\sigma_v=\sqrt {G M_{200}/r_{200}}$. 
The dimensionless form of this power spectrum allows one to compare 
  curves of $E(k)$ extracted from different clusters
as a function of ${\tilde k}\equiv k_rL_{sp}/2\pi$, where $k_r=|\vec k|$.
The spectral properties of the gas velocity fields will be studied
using the density-weighted power spectrum defined according to Eq.
 (\ref{pow.eq}), whereas volume-weighted  ($w=1$) spectra will be considered
for comparative purposes.
 
Moreover, the longitudinal and solenoidal components of the power spectrum $E(k)$ 
will be studied separately. For doing this, the shear and compressive parts of 
the velocity $\vec u(\vec x)$ are defined respectively in the $\vec k-$space 
as

\begin{eqnarray} 
\vec {\tilde u}(\vec k)_{shear}&=& \frac{\vec k \times \vec {\tilde u}(\vec k)}  {|\vec k|}~,\\
 \vec {\tilde u}(\vec k)_{comp}&=& \frac{\vec k \cdot \vec {\tilde u}(\vec k)}
  {|\vec k|} \,.
   \label{pvisc.eq}
   \end{eqnarray}

The corresponding power spectra are used to define the power spectrum 
decomposition, $E(k)=E_s(k)+E_c(k)$ \citep{ki09,zh10}.

In addition to the spectral properties of the velocity $\vec u(\vec x)$, 
it is useful to investigate the scaling behavior of the velocity structure
functions. These are defined by

   \begin{equation}
  \mathcal{S}_p(\vec r)\equiv <|u(\vec x+\vec r)-\vec u(\vec x)|^p>,
    \label{spu.eq}
   \end{equation}
where $p$ is  the order of the function.
% and the averaging is taken over all the 
%particle pairs in the simulation separated by the distance $\vec r\leq r_{200}$.
For computing the structure functions, a random subsample of $N_s$ 
particles is constructed; these particles are extracted from the set of 
gas particles which satisfy the constraint of being located within a distance 
 $r_{200}$ from the cluster center. For each of these sample particles $s$, the
 relative velocity difference $\Delta \vec u=\vec u(\vec x_i+\vec r_{si})-
\vec u(\vec x_s)$
is computed for  all of the gas particles $i$  separated by a distance 
$r_{si}\leq r_{200}$, and the quantity $|\Delta \vec u|^p$  is binned 
in the corresponding radial bin. 
Final averages are obtained by dividing  the binned quantities by the corresponding
number of pairs belonging to the radial bin. 

As for the power spectrum, structure functions have been computed separately 
for both transverse and longitudinal components. These are respectively 
defined as $\Delta u_{\perp}=\Delta \vec u  \times \vec r_{si}/ |\vec r_{si}|$ and
$ \Delta u_{\parallel}=\Delta \vec u  \cdot \vec r_{si}/ |\vec r_{si}|$.
 Here the study will be restricted to second-order ($p=2$) structure functions.
Moreover, in order to compare
structure functions of different clusters, these will be rescaled 
according to $S_p(\vec r)=\mathcal{S}_p(\vec r)/\sigma_v^p$ and radial
distances will be expressed in units of $r_{200}$.
Density-weighted  velocity structure functions are defined using 
 the weighted field $\vec u_{w}(\vec x)$ in Eq. (\ref{spu.eq}).

%As outlined in the Introduction,  turbulent velocity fields in the ICM 
%are characterized by the presence of eddies over a wide range of 
%spatial scales. 
Finally, the energy content of the turbulent velocity field is another
quantity used to investigate the dependence of the velocity field 
properties on the amount of AV present in the simulations.
In order to properly consider the contribution of random gas motion to the
turbulent energy budget, it is however necessary to separate from the 
velocity field $\vec u(\vec x)$ the part due to the streaming motion.
A useful approach is then to define a spatial decomposition which
separates the velocity into a small-scale and a large-scale component \citep{ad00}.
More specifically, 

   \begin{equation}
 <u(\vec x,t)>=\int_D f(\vec x -\vec x^{\prime}, H) \vec u (\vec x^{\prime}, t)
d^3 \vec x^{\prime}~,
    \label{filt.eq}
   \end{equation}

where $f(\vec x -\vec x^{\prime},H)$ is a shift-invariant kernel which 
acts as a low-pass filter, $H$ being a filtering scale. 
Velocity fields are always considered here at a given 
time slice, so that 
the dependence on $t$ will be removed  from Eq. (\ref{filt.eq}). 
The small-scale turbulent velocity field 
 $\tilde{\vec u}(\vec x)$ is then defined as
   \begin{equation}
   \tilde{\vec u}(\vec x)=\vec u(\vec x)- <u(\vec x)>.
    \label{filtu.eq}
   \end{equation}
 
 This method of decomposing a turbulent velocity field is also known
as Reynolds decomposition and is commonly employed 
in Large Eddy Simulations \citep[LES,][]{me00}.
The filtering kernel chosen here to define the large-scale velocity field
is a triangular-shaped cloud function (TSC) \citep{ho88}.
This choice is similar to that adopted by \cite{do05} when studying
the ICM velocity fields, and is motivated by the compact support properties
of the kernel, the TSC being a second-order scheme, which from a computational
viewpoint greatly reduces the complexity of estimating the mean field 
velocity (\ref{filt.eq}).
The choice of the filtering scale $H$ for the TSC spline will be discussed in 
Sect. \ref{cluster.sec}; here is worth noting that after application of 
Eq. (\ref{filt.eq}) to $\vec u(\vec x)$, the harmonic content of the 
small-scale 
turbulent velocity field $\tilde {\vec u}(\vec x)$ is defined by the 
condition $kH>>1$.

   \begin{figure}
%   \centering
   \includegraphics[width=10.2cm]{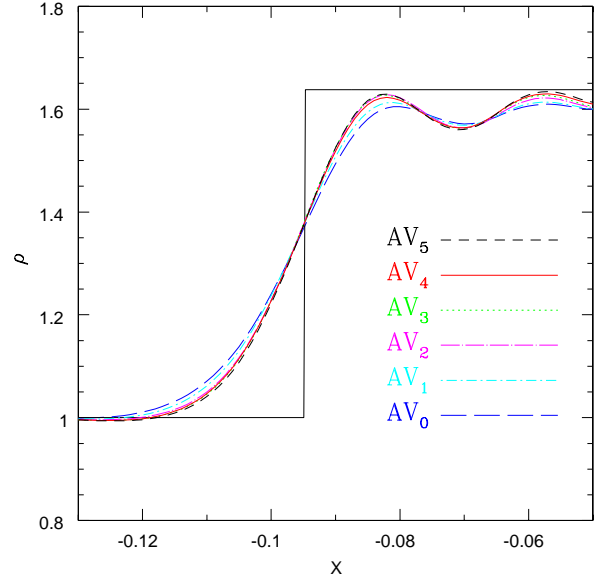}
   \caption{A closer view at the shock front of the density profiles
displayed in the top left panel of Fig. \ref{tubea.fig}.}
   \label{tubeb.fig}%
    \end{figure}

\section{Test simulations}
\label{test.sec}
In this section simulation results are presented obtained by applying the
SPH implementation described in the previous sections to several test problems.
These tests have known solutions provided by analytical models or 
independent numerical methods, and against these solutions are compared 
results from the SPH runs with the purpose of validating the code and 
assessing its performance. Section \ref{testa.sec} is dedicated to the shock 
tube problem and Sect. \ref{testb.sec} to the 3D collapse of a cold gas 
sphere. Both of these tests have been chosen because they have been widely
used in the literature and moreover they allow comparisons with previous 
SPH runs in which a time-dependent  AV scheme was implemented.

\subsection{The shock tube problem}
\label{testa.sec}
\subsubsection{Initial condition set-up and parameter calibration of the 
time-dependent AV scheme}
\label{tubea.sec}
The Riemann shock tube problem is a test commonly used for SPH codes. Its main
advantage is that it admits an analytical solution following the propagation
of a shock wave in a medium initially at rest.
Here the initial condition setup consists of an ideal fluid with $\gamma=5/3$ 
initially at rest at $t=0$. An interface at $x=0$ separates the  fluid
on the right with density and pressure $(\rho_1,P_1)= (4, 1)$ from the
fluid on the left with $(\rho_2,P_2)=(1, 0.1795)$. 
The fluid is then allowed to evolve freely and at later times there is
 a shock wave propagating toward the left, followed by a rarefaction
wave and a contact discontinuity. The resulting profiles are given by 
the analytic solution \citep{ra91}.

The one-dimensional Riemann shock tube problem  is often used to 
test hydrodynamic codes, but the numerical results are often of
limited validity because numerical effects which may arise in 3D
calculations, such as particle streaming, are absent or reduced when there is only
 one degree of freedom.
For this reason the shock test is carried out here in three dimensions.
To construct the initial conditions for the SPH runs, a cubic box 
of side-length unity was filled with $10^6$ equal mass particles.
Of these one million particles, $800\,000$ were placed in the right-half
 of the cube, while $200\,000$ were placed in the left-half. 
The particles in the two halves of the cube were extracted from two independent
uniform glass-like distributions contained in a unit box. The two
distributions had  $1.6 \cdot 10^6$ and $4\cdot 10^5$ particles, respectively.

   \begin{figure*}
   \centering
   \includegraphics[width=15.2cm,height=14.2cm]{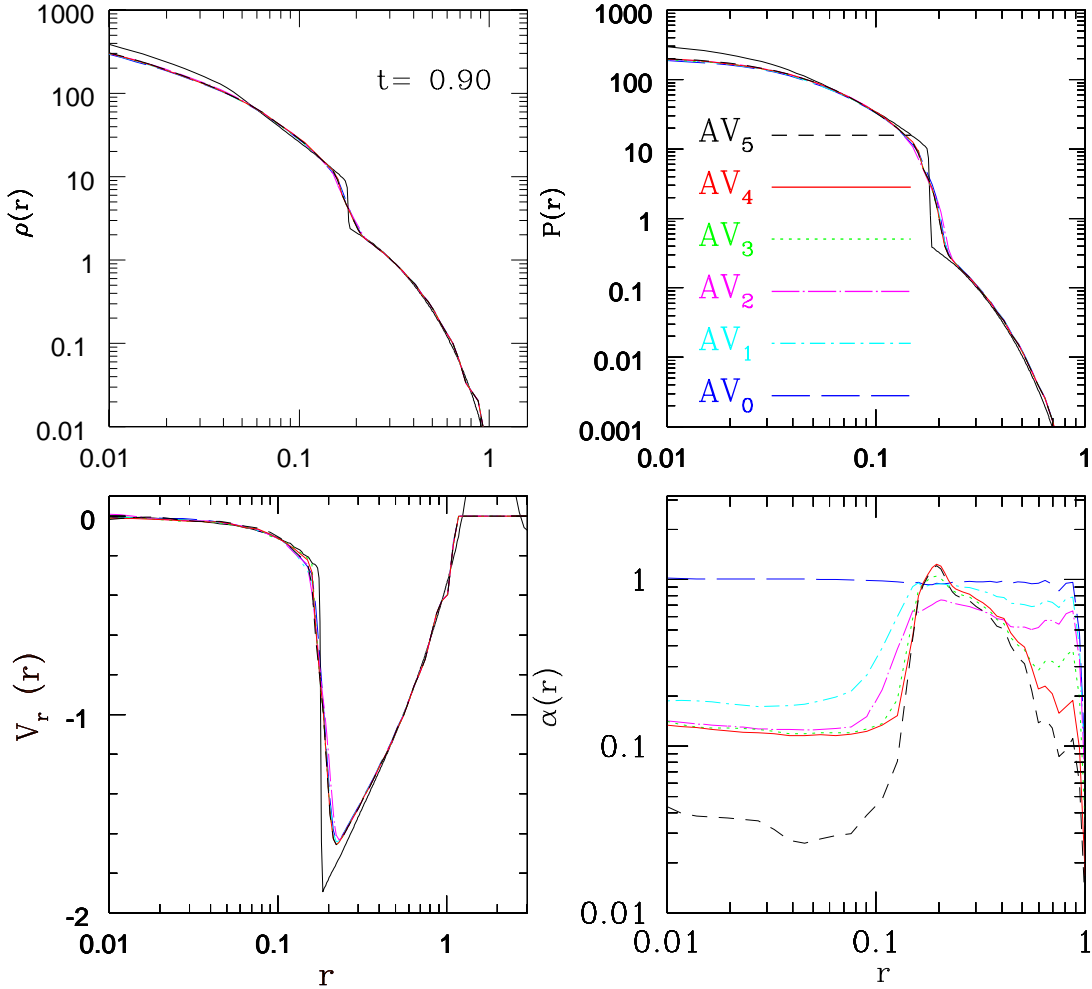}
   \caption{Results at $t=0.9$ for the adiabatic collapse of a cold
gas sphere. Clockwise from the top left panel: radial density profiles of
density, pressure, viscosity parameter $\alpha$ and radial velocity.
The black solid lines are the results of the 1D PPM calculation of \cite{st93}.
SPH runs with different AV parameters are labeled in the same way as in the
 shock tube test panels of Fig .\ref{tubea.fig}.}
   \label{coldb.fig}%
    \end{figure*}
These initial conditions have been chosen because they are the same 
as those adopted by Tasker et al. (2008, hereafter T08), in their Sect. 3.1, 
to study the shock tube problem. 
Those authors used a variety of numerical problems with known analytical 
solutions to compare the behavior of different astrophysical codes and
their ability to resolve shocks. It is therefore of particular interest 
to compare the results of T08 with those obtained from the
shock tube SPH simulations performed here using different AV strengths. 
 The SPH runs were realized imposing periodic boundary conditions along 
the $y$ and $z$ axes and the results were examined at $t=0.12$.
Runs with the time-dependent AV scheme have their viscosity parameters
initialized to one. 

Before proceeding to discuss the behavior of the shock tube tests, it is 
necessary to assess the validity, under different conditions, of the AV 
scheme introduced in Sect. \ref{alfa.sec}.
As outlined in this section, the time evolution of the viscosity parameter 
can be affected if  very short damping time scales are imposed.
To better illustrate this point it is useful to rewrite Eq. (\ref{alfa.eq})
using the second of the equalities of Eq. (\ref{salfa.eq}).  The new 
Eq. (\ref{alfa.eq}) is then

   \begin{equation}
  \frac {d \alpha_i}{dt} =-\frac{\alpha_i}{\tau^{\prime}_i} + \frac{q_i}
{\tau^{\prime}_i}~,
    \label{alfanew.eq}
   \end{equation}
where
   \begin{equation}
  \tau^{\prime}_i=\frac{\tau_i}{1+S_i \tau_i}~,
    \label{taunew.eq}
   \end{equation}
and  $q_i$ is a modified source term
   \begin{equation}
 q_i=\frac{\alpha_{min}+S_i\tau_i \alpha_{max}}{1+S_i\tau_i}.
    \label{qsou.eq}
   \end{equation}

Neglecting variations in the coefficients, the solution to Eq. 
(\ref{alfanew.eq}) at times $t> t_{in}$ is given by
   \begin{equation}
  \alpha_i(t)=q_i+(\alpha_i(t_{in}) -q_i)\exp ^{-(t-t_{in})/\tau^{\prime}_i}~,
    \label{alfatn.eq}
   \end{equation}
which shows that $\alpha_i(t)$ approaches the asymptotic value $\alpha_{min}$ 
in the absence of shocks, and $\alpha_{max}$ if a shock is present.
However, for the source term $q_i$, the condition $q_i\simeq \alpha_{max}$ holds
only in the shock regime $S_i \tau_i \gg1$.
If this condition is not satisfied and $S_i \tau_i \simlt 1$, the peak value
$\alpha_{peak}$ of $\alpha_i(t)$ will be smaller than $\alpha_{max}$.
In particular $\alpha_{peak}$ will depend on the chosen value of 
the decay parameter $l_d$. This dependence of $\alpha_{peak}$ on $l_d$ 
is absent for strong shocks, but for mild shocks introduces the unwanted 
feature  that for short decay timescales ($l_d\rightarrow1$)
the peak value of $\alpha_i(t)$ 
at the shock front might be below the AV strength necessary to properly treat 
shocks.

In fact, application of the SPH code implemented according to the AV scheme
described in Sect. \ref{alfa.sec} to the shock tube problem with the initial 
condition setup  studied here, 
shows that at the shock front location the peak in the AV parameter  
ranges from unity down to $\sim0.2$ for $l_d=1$.
Previous numerical tests \citep{mm97,ro00} showed that for this 
 shock tube problem there is good agreement with the exact results using 
$l_d=0.2$, and at the shock front the peak in the viscosity parameter is then
approximately $\sim0.6-0.7$.

In order to mantain the same shock resolution capabilities also in those 
cases  studied using short time scales with $l_d=1$, a correction factor 
$\zeta$ is introduced so as to compensate in the source term $q_i$ the 
smaller values of $S_i\tau_i$ with
respect to the small $l_d$ regimes. This is equivalent to considering a
higher value for $\alpha_{max}$, so that in Eq. (\ref{qsou.eq}) 
$\alpha_{max}$ is substituted by $\alpha_{max}\rightarrow \zeta \alpha_{max}$.
The calibration of the correction factor $\zeta$ is achieved by requiring
that for the current shock tube problem, the SPH runs with AV decay
parameters $l_d>0.2$ should have at the shock front the same value 
 $\alpha_{peak}$ as that for the $l_d=0.2$ case, i.e.
$\alpha_{peak}\sim0.6-0.7$ for $l_d>0.2$.
This choice guarantees that the viscosity parameter reaches the correct 
size  at the shock front, but away from it quickly decays according
to the chosen value of $l_d$.
After several tests it has been found that the choice 
$\zeta=MAX((l_d/0.2)^{0.8},1)$ yields satisfactory results and all of the 
numerical tests shown in this paper incorporate 
in the source term (\ref{salfa.eq}) the modification

   \begin{equation}
\left\{ 
\begin{array} {l}
   \alpha_{max} \rightarrow \zeta \alpha_{max}~, \\
   \zeta=MAX((l_d/0.2)^{0.8},1).
   \end{array}
\right .
   \label{zeta.eq}
   \end{equation}
 Note that the validity of this setting has been established 
for the shock strength of the shock tube problem considered here, 
nonetheless the results of the other tests indicate that this 
parametrization is appropriate to make the peak value 
of the viscosity parameter  at the shock location independent of the chosen value
for the decay parameter $l_d$.

\subsubsection{Results}
\label{tubeb.sec}
The results of the shock tube tests are shown in Figures \ref{tubea.fig} 
and \ref{tubeb.fig}. In each panel, different profiles refer to runs with 
different AV parameters. The solid line is the analytical solution, which
exhibits a shock front at $x=-0.095$ and a contact discontinuity at $x=-0.033$.
The profiles shown in Figures  \ref{tubea.fig}  and  \ref{tubeb.fig} can be 
compared with the corresponding Figures 2 and 3 of T08. 

The fixed AV simulation is in good agreement with the GADGET2 run of 
T08, although  a closer view of the density profile in the proximity 
of the shock front in Fig. \ref{tubeb.fig} shows post-shock ringing features
 which are absent in T08. 
This happens because close to the shock the amount of AV generated by the 
code allows particle interpenetration.
Note that in order to avoid post-shock  oscillations in the density, T08 
explicitly made the choice of setting  the viscosity parameter in the GADGET2 run
 to ArtBulkVisc=2, whereas here the fixed AV run was performed with
the viscosity parameter $\alpha$ set to unity.
The post-shock oscillations in velocity are very similar in amplitude and 
behavior to the ones produced by the GADGET2 run of T08, and similarly for 
the glitch in pressure at the contact discontinuity.
The spike in thermal energy and entropy at the contact discontinuity 
is originated by the initial discontinuity in the density profile, which has 
been left unsmoothed at the beginning of the simulation. These features are 
present also in T08, although here the height of the spikes is a bit smaller.

The profiles of the simulations in which a time-dependent AV scheme has been 
implemented show a behavior very similar to those of the fixed AV run.
The post-shock oscillations in the density have a tendency to be amplified 
as the AV scheme uses shorter decay time scales, but the effect is minimal.
This can be seen in the top left panel of Fig. \ref{tubea.fig},  in which 
the density profiles have been shifted vertically to illustrate the
effect better.
The differences in the other profiles as a function of the AV coefficients 
are very small, with the post-shock entropy being the quantity with the
strongest dependence and the smallest values being those of the AV runs 
with the lowest viscosity parameter ($l_d=1$).
Finally, the radial profiles of the time-dependent artificial viscosity
parameters $\alpha$ are consistent with the discussion of Sect. \ref{alfa.sec}:
their peaks are located approximately at the shock front and their post-shock
radial decay is faster for those AV runs with the shortest decaying time scales.
   \begin{figure*}
   \centering
   \includegraphics[width=16.2cm,height=8.0cm]{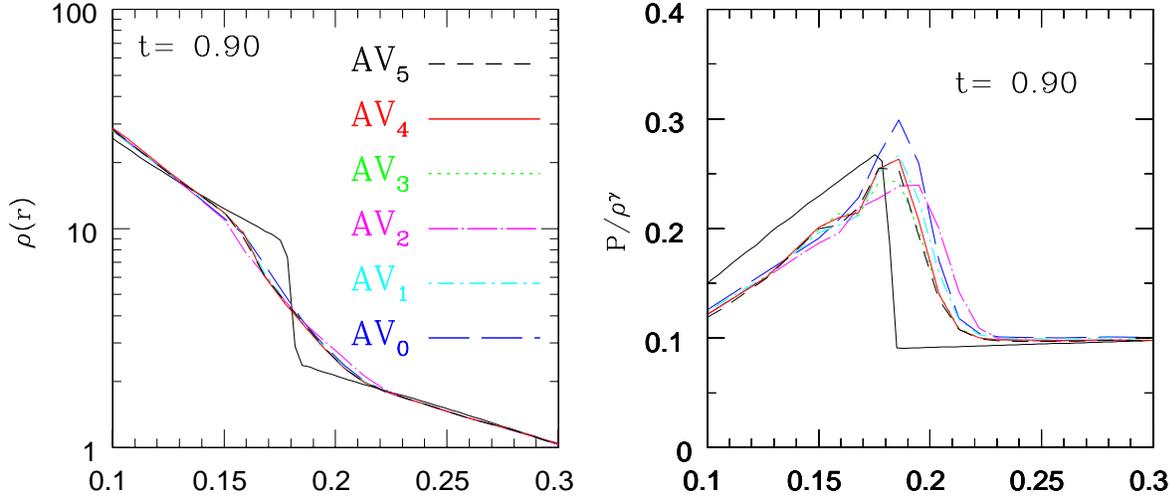}
   \caption{The same as in Fig. \ref{coldb.fig}, radial profiles of
density and entropy are shown in the proximity of the shock front.}
   \label{coldc.fig}%
    \end{figure*}

To summarize, the simulations of the shock tube problem performed with the 
SPH code presented here using the standard AV scheme show results in good
agreement with the analytic solution and with the ones obtained by other 
authors using similar initial condition setup and parameters.
The profiles of the runs with a time-dependent AV scheme agree well with
those of the fixed AV run, thus indicating shock resolution properties 
comparable to those of the standard AV scheme, but with a reduced AV
strength elsewhere.

\subsection{Collapse of a cold gas sphere}  
\label{testb.sec}
The Riemann shock tube test of the previous section illustrates the ability 
of the code to resolve discontinuities, but the shock strengths produced
by the test are much smaller than those which develop during the gravitational
collapse of astrophysical objects. A more stringent test for SPH codes in which 
gasdynamics is modeled including self-gravity is therefore to study a 3D collapse
problem. The test used by \cite{ev88} follows in time the adiabatic collapse
of a gas sphere. This test has been widely used by many authors 
\citep{hk89,st93,da97,wa04,sp05,va09,we09} as one of the standard tests for SPH 
codes.
The shock strength which develops during the collapse is comparable to that 
of the \cite{na93} self-similar collapse test.

The gas cloud is spherically symmetric and initially at rest, with mass $M$, 
radius $R$ and density profile
   \begin{equation}
   \rho(r)=\frac{M}{2\pi R^2}\frac {1}{r}~.
  \label{rhocl.eq}
   \end{equation}

The gas obeys the ideal gas equation of state with $\gamma=5/3$ and the
thermal energy per unit mass is initially set to $u=0.05 GM/R$. 
 The chosen time unit is the cloud free-fall time scale 
$t_{ff}=(\pi^2/8)\sqrt{R^3/GM}$
\footnote{ Note that this time normalization is the same as that of \cite{hk89} 
and differs by a factor $\sqrt{\pi^2/8}$ from that of \cite{ev88} and 
by $\pi^2/8$ with respect to that adopted by \cite{st93}.}
and the SPH simulations are performed using units for which $G=M=R=1$.  
The cloud begins to collapse under its own gravity and the gas in the core
is heated until the growth of pressure is sufficient for the 
infalling material to bounce back and a shock wave then propagates outwards.
Most of the kinetic energy is converted into heat at the epoch
of maximum compression of the gas, which occurs at $t\sim1.1$.

The initial conditions for the simulations are constructed by stretching the radial
coordinates of a glass-like uniform distribution of $N=88\,000$ equal mass particles
contained within a sphere of unit radius.
Radial coordinates are transformed according to the rule 
$r\rightarrow r^{\prime}= r^{3/2}$, so as to generate the density profile 
(\ref{rhocl.eq}). The particle smoothing lengths are adjusted  according to
Eq. (\ref{hrho.eq}) with $N_{sph}=50$ neighbors and the gravitational softening 
length is taken as $\varepsilon_g=0.02$. This choice for the simulation parameters
 allows one to compare the test calculations performed here with the 
results presented for the same collapse problem by \cite{we09} in their Sect. 7.1.

Fig. \ref{coldb.fig} shows the radial dependence at $t=0.9$ of the spherically 
averaged profiles for density, pressure, radial velocity and time-dependent 
viscosity.
The solid line is for the high-resolution 1D PPM simulation of \cite{st93}, which for
practical purposes can be considered as being an exact solution.
The differences between the profiles with different AV parameters are minimal and
the post-shock radial decay of the viscosity parameters is in accord  with 
the expectation of the model.
The profiles of the SPH runs are consistent with previous findings \citep{sp05,we09}
 and reproduce the overall features of the PPM solution, 
with some differences at the shock front, which at this epoch is 
located at $r\sim0.18$. To illustrate these differences, Fig. \ref{coldc.fig} 
shows a 
closer view of the density and entropy profiles in the proximity of the shock front. 

A feature common to all of the runs is a significant amount of pre-shock 
entropy generation. This is inherent in the AV implementation of the SPH code, 
which near to the bounce is switched on by the strong convergence of the flow,
 therefore generating dissipation \citep{wa04}. 
Clearly, different settings of the AV parameters are reflected 
 in the code's ability to model shocks and compressions which in turn
 generates the corresponding differences in the pre-shock entropy profiles.
All of the simulations share the basic attribute of having a shock front 
located outside the position indicated by the reference PPM 
entropy profile. The location of the shock front has a weak dependence on the
 implemented  AV settings, the time-dependent AV runs with the shortest 
decay time scales (runs AV$_4$ and AV$_5$) being the closest to the 
PPM reference position.
Moreover, the fixed AV run overpredicts the entropy peak whilst the other 
time-dependent AV simulations are in rough agreement with the PPM one. 
Finally, the post-shock entropy profile of all the runs is below the corresponding
reference PPM profile.
   \begin{figure*}
   \centering
 \includegraphics[width=17.2cm,height=13.2cm]{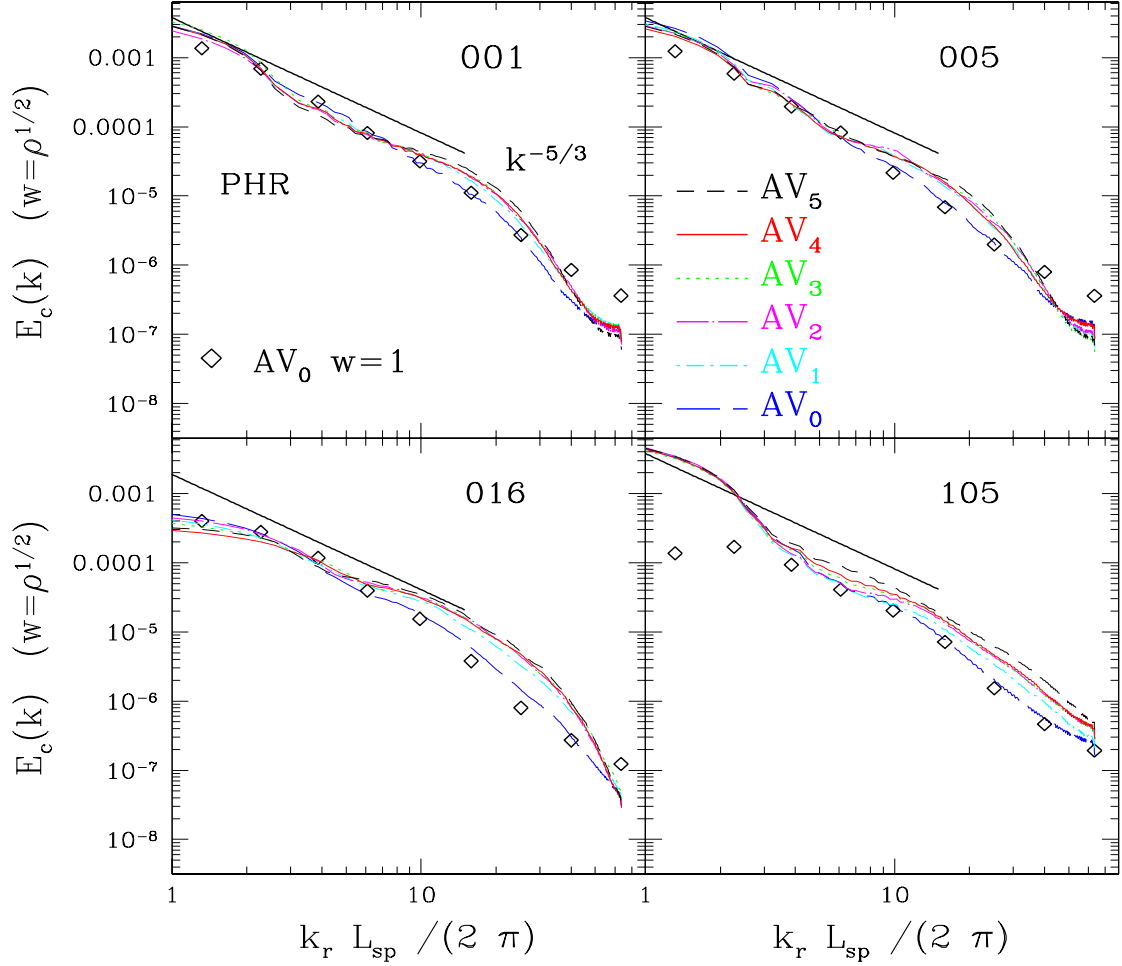}
   \caption{Compressive components of the density-weighted velocity power 
spectra (\ref{pow.eq}) 
are shown at $z=0$ as a function of the dimensionless wavenumber 
${\tilde k}\equiv k_rL_{sp}/2\pi$, where $k_r=|\vec k|$.
The spectra are extracted from the high resolution (HR) runs of the four 
dynamically perturbed test clusters using a cube of size $L_{sp}=r_{200}$ with
$N_g^3=128^3$ grid points and are shown up to the wavenumber 
${\tilde k}=N_g/2$. In each panel, different lines refer to runs 
with different AV viscosity parameters, the line style and color coding  is the
same as in the previous figures. The solid black line indicates the Kolgomorov
scaling, while open diamonds correspond to the volume-weighted velocity power 
spectrum of the standard AV runs. 
}
   \label{PHRc.fig}%
    \end{figure*}

The density and entropy panels of Fig. \ref{coldc.fig}  can be compared with 
Figures 7 and 8 of \cite{we09}.
 For the chosen parameter settings, the AV$_1$ run corresponds to the \cite{we09}
model with $\delta_a=5$ and $\alpha_{\star}=0.1$, whilst AV$_2$ corresponds to the
 $\delta_a=2$, $\alpha_{\star}=0.1$ model.
A comparison shows that the results presented here 
share some common properties with those of \cite{we09}, but at the same time there 
are also several differences. For instance, with respect to the reference PPM
entropy profile, pre-shock entropy production and a lower entropy level behind
the shock is common to both of the tests. 
However, in \cite{we09},  pre-shock heating for the time-dependent AV models 
 occurs at radii larger than for the fixed AV run  and the differences in the 
entropy profiles (as well as in the density) for different AV runs 
are here reduced  more.

It is difficult to ascertain the origin of the different behavior of the two codes
in the proximity of the shock front, given also the fact that the differences in the
corresponding profiles are not dramatic.
However, the SPH formulation implemented in the two codes differs in several
aspects, so that even results produced for the same test problem and using 
the same initial conditions might be different.
The thermal evolution of the gas is followed here according to an entropy-conserving
scheme, while in the \cite{we09} code it is the specific internal energy which is 
integrated.
Moreover, unlike \cite{we09}, the equation of motion (\ref{fsph.eq}) incorporates 
the terms due to the presence of smoothing length gradients. 
The reader is referred to \cite{sh02} for a thorough discussion of the 
relative differences between simulation results produced using the two SPH 
formulations.

To summarize, for the test problem considered the results presented in this 
section agree with previous findings and validate the code, as well as 
showing its 
capability to properly model shocks which develop during the collapse of
self-gravitating objects. In accordance with the results of Sect. \ref{tubeb.sec},
 profiles extracted from simulations
with a time-dependent AV  scheme exhibit a behavior which is very similar to the
corresponding ones for the fixed AV model, thus showing shock 
resolution capabilities analogous to those of the standard AV scheme for these
models.
Finally, the peaks of the viscosity parameters at the shock front appear almost
independent of the chosen value of the AV decay parameter $l_d$, thereby 
supporting the proposed parametrization (\ref{zeta.eq}).
   \begin{figure*}
   \centering
   \includegraphics[width=17.2cm,height=13.2cm]{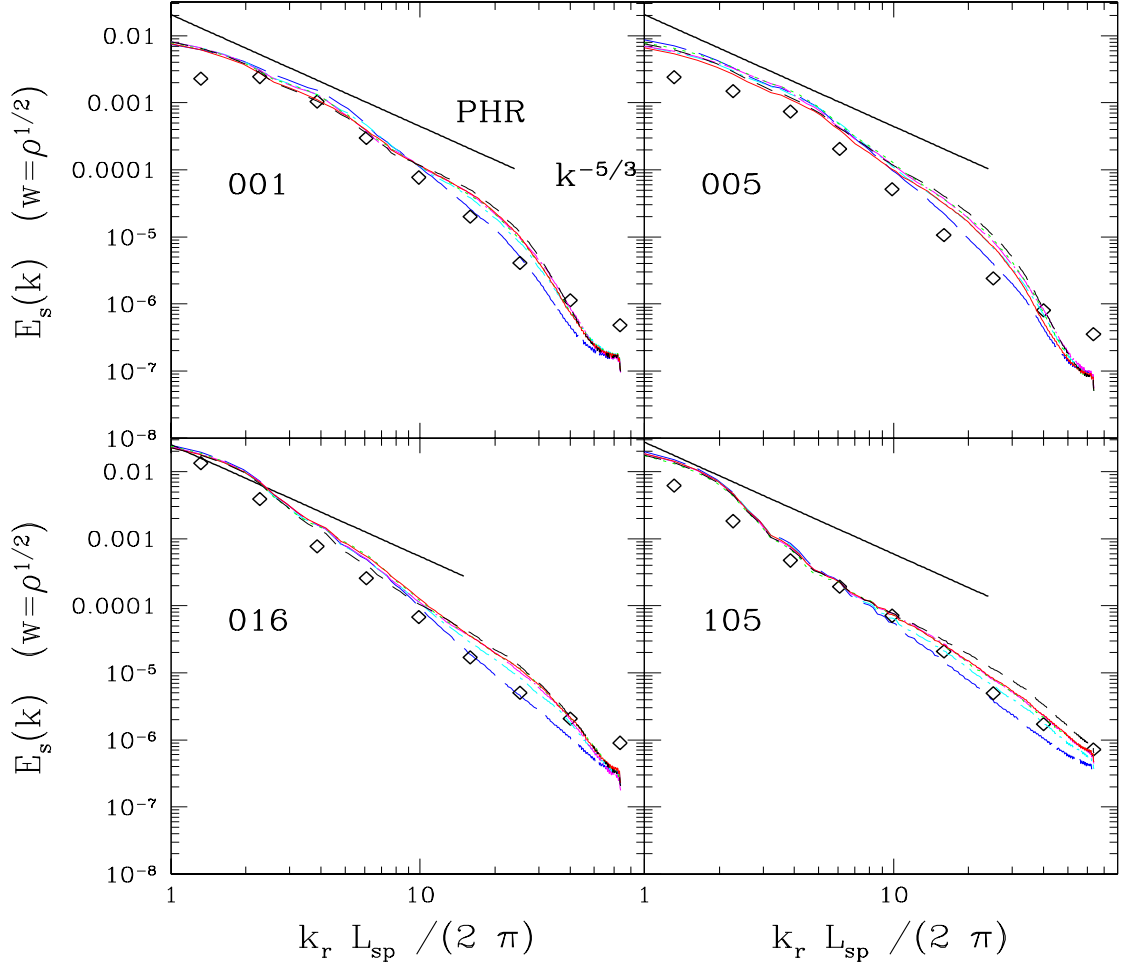}
   \caption{The same as in Fig. \ref{PHRc.fig}, but for the shearing components
of the velocity power spectra.}
   \label{PHRs.fig}%
    \end{figure*}

\section{Cluster simulations}
\label{cluster.sec}
In this section the impact of numerical viscosity on the ICM velocity field
of simulated clusters is studied using the statistical tools presented in
Sect. \ref{stat.sec}. The purpose of the analysis is also to obtain 
useful indications about the numerical constraints which are necessary 
in order to adequately describe turbulence in simulations of the ICM using SPH 
codes.
 
\subsection{Velocity power spectra}  
\label{clustera.sec}
Fourier spectra extracted from hydrodynamical simulations are expected to 
exhibit a dependence on both the simulation resolution and the 
hydrodynamical method used in the simulations.
Usually, resolution issues are debated in a dedicated section with the
purpose of assessing the stability of the simulation results. Here, however, 
the interpretation of the results is strictly related to the resolution
employed in the simulations, so that the two arguments will be discussed
together.

In order to measure the velocity Fourier spectra of the simulated clusters,
a cube of side length $L_{sp}$ is placed at the cluster center, 
 the latter is defined as the location where the gas density reaches its
maximum value.
As described in Sect. \ref{stat.sec}, Fourier transforms 
of the density weighted velocity field are evaluated by first estimating 
interpolated quantities at the cube grid points and then performing a 3D FFT
of the sampled values.
 The choice of the cube side length $L_{sp}$ and of the number of grid points 
$N_g^3$ is however limited by several arguments which strongly constrain the 
possible choices. 

At variance with studies of supersonic turbulence 
using hydrodynamical simulations (Kitsionas et al. 2009, herefater K09; 
Price \& Federrath 2010)
 here the gas distribution is driven by gravity and because of the 
 Lagrangian nature of the SPH code,
the bulk of the particle distribution is located in the central cluster regions.
For the cluster simulations presented here, at the present epoch
about half of the cluster mass is contained within a radius of $\sim r_{200}/3$.
Therefore, in order to minimize this aspect of the resolution effects, the size 
of the cube should be kept as small as possible, but in this case 
most of the large-scale modes will be missed in the spectral analysis,
in particular, those modes corresponding to the merging and accretion 
processes of the cluster substructure.
%which are the main sources of energy injection into the medium.
As a compromise between these two opposite needs, the side-length of the 
cube is set to $L_{sp}=r_{200}$ (the scaling $L_{sp}\propto r_{200}$ is chosen 
in order to consistently compare velocity spectra extracted from different 
clusters). 

The choice of the number of grid points $N_g^3$ in the cube or , equivalently, of the
grid spacing $L_{sp}/N_g$, depends on the effective SPH resolution of the 
simulations.
In SPH the smallest spatially resolvable scale is set by the values of the 
gas smoothing lengths $h_i$. As already outlined, because of the  Lagrangian 
nature of SPH simulations, the gas particle number density increases in 
high-density regions and this in turn implies a subsequent decrease in the 
 gas smoothing lengths $h_i$. If in these regions the SPH spatial resolution 
becomes higher than the grid resolution of the cube, SPH kernel interpolation 
at the cube grid points will appear in the $k-$space as a extra power at 
the highest wave number 
permitted by the grid. This effect has been noticed by K09 and can be avoided
provided that the cube grid spacing is chosen suitably small.
For the HR cluster simulations, values of the gas  smoothing lengths $h_i$ 
in the cluster cores range from $h_i\sim 20 $~kpc for the most massive
clusters down to $h_i\sim 5$~kpc for the least massive ones.
Similar values for the grid spacing are obtained setting $N_g=128$ grid points
along each of the spatial axis of the cube. 
However, the number of cube grid points cannot be made arbitrarily high 
because of the need for avoiding undersampling effects in the estimate of 
SPH variables at the grid points. As a rule of thumb, 
the number $N_g$ should not exceed  $\sim 2 N_p^{1/3}$, with 
$N_p$ being the number of SPH particles.

Finally, when calculating the velocity power spectra
non-periodicity effects should be taken into account by adopting a
zero-padding technique \citep{va09a}. 
However, application of this method to the clusters presented here requires 
some care since the choice of the cube side length $L_{sp}$ is dictated
by the previous constraints so that a cube of volume $L_{sp}^3$ occupies at 
the cluster center $\sim 1/4$ of the volume of a sphere of radius $r_{200}$ 
with its origin centered as the cube. 
Hence, a power estimate based on the non-periodicity assumption would likely 
underestimate the velocity power spectrum at  wavenumbers
$\sim \pi/L_{sp}$, in particular
for those clusters which are undergoing a major merging event. 
 The procedure adopted here  should then provide  
 a better indication of the spectral behavior of the ICM velocity 
field at spatial scales of the order of the cube size.

Having chosen the cube parameters, velocity power spectra have been evaluated
at the present epoch for the ensemble of simulated clusters constructed by
collecting simulations of the the cluster HR baseline sample 
performed with different AV prescriptions.
The resulting spectra are shown in Figures \ref{PHRc.fig} to \ref{QHRs.fig} 
and provide a 
quantitative comparison between statistical properties of the longitudinal
and solenoidal velocity field spectral components for clusters simulated with 
different AV settings and with different dynamical states.
Each panel in the figures corresponds to an individual cluster and the curves
shown in the panels  are the velocity power spectra extracted from simulations 
of the considered cluster realized using different AV parameters.

A first conclusion to be drawn from the spectra shown in the figures is that,
for a given cluster and power spectrum component, at large scales
the velocity power spectra do not show systematic differences 
between different AV runs and therefore the effects of numerical viscosity 
can be considered negligible on these scales.
Moreover, dissipative effects in the kinetic energy are relatively less 
important 
for those runs in
which the parameter settings of the time-dependent AV scheme correspond 
to the shortest decay time scales for the viscosity parameter $\alpha(t)$. 
In fact, at high wavenumbers the power spectra of the AV$_5$ runs are higher than
those of the standard AV runs (AV$_0$) by a factor of $\sim2$ and in certain cases 
even by a factor of $\sim10$ ( see, for example, the longitudinal  power spectrum
of cluster $110$ of the relaxed subsample). 
These behaviors are shared by the velocity power spectrum components of all of 
the simulated clusters and because of the sample size can be considered as
systematic.
   \begin{figure*}
   \centering
   \includegraphics[width=17.2cm,height=13.2cm]{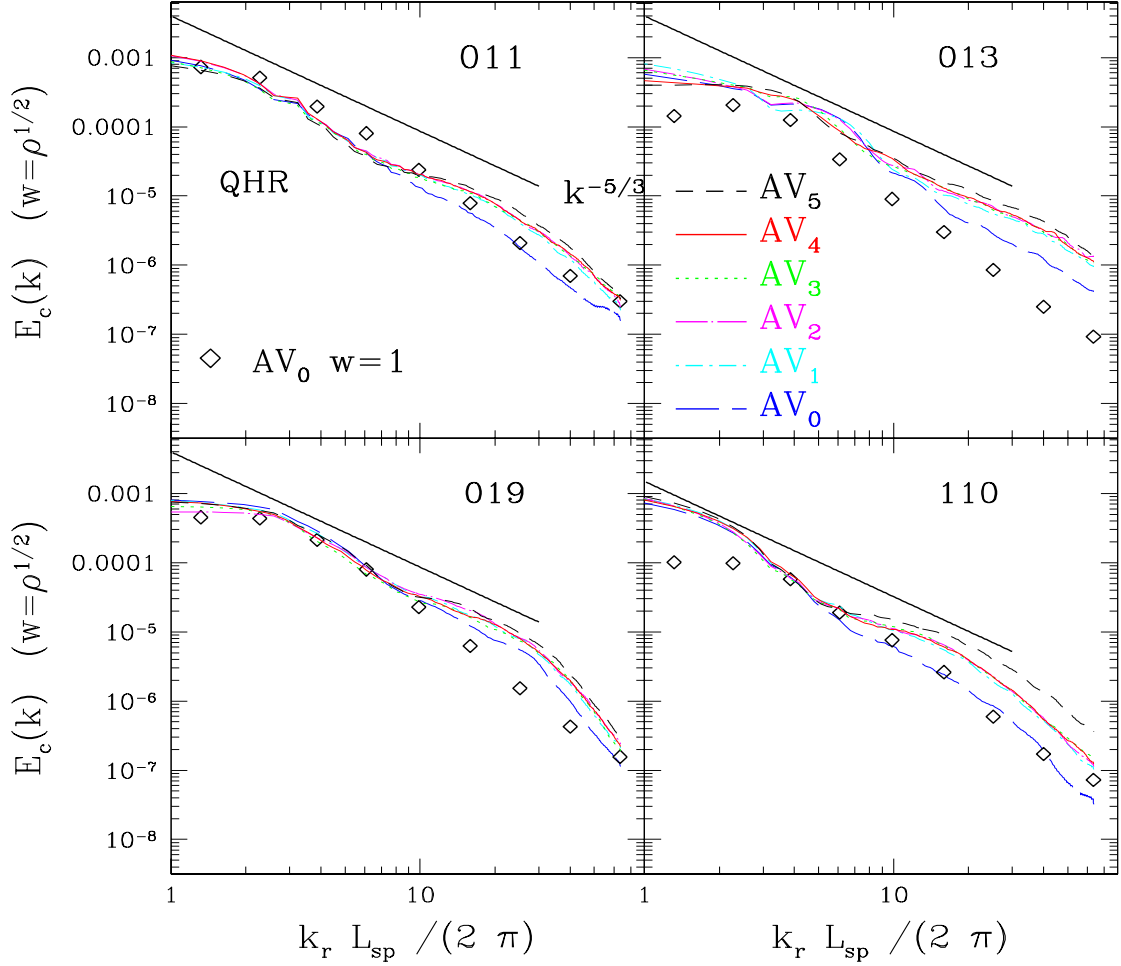}
 \caption{The same as in Fig. \ref{PHRc.fig}, but the spectra here are computed 
for the quiescent test clusters.}
   \label{QHRc.fig}%
    \end{figure*}

The use of Fourier spectra allows one to study in a quantitative way  
the scale dependency of dissipative effects due to numerical viscosity.
All of the spectra exhibit a maximum at ${\tilde k}\simlt 2$, at spatial 
scales of $\sim r_{200}/2$.  This is a key signature in the power spectra of the
ICM velocity field, in which merging and accretion processes driven by 
gravity at cluster scales are the primary sources of energy injection into the 
ICM. Similar findings have been obtained by \cite{va09a}, who measured 
the velocity power spectrum of simulated clusters in cosmological simulations 
using the AMR Eulerian code ENZO.
   \begin{figure*}
   \centering
   \includegraphics[width=17.2cm,height=13.2cm]{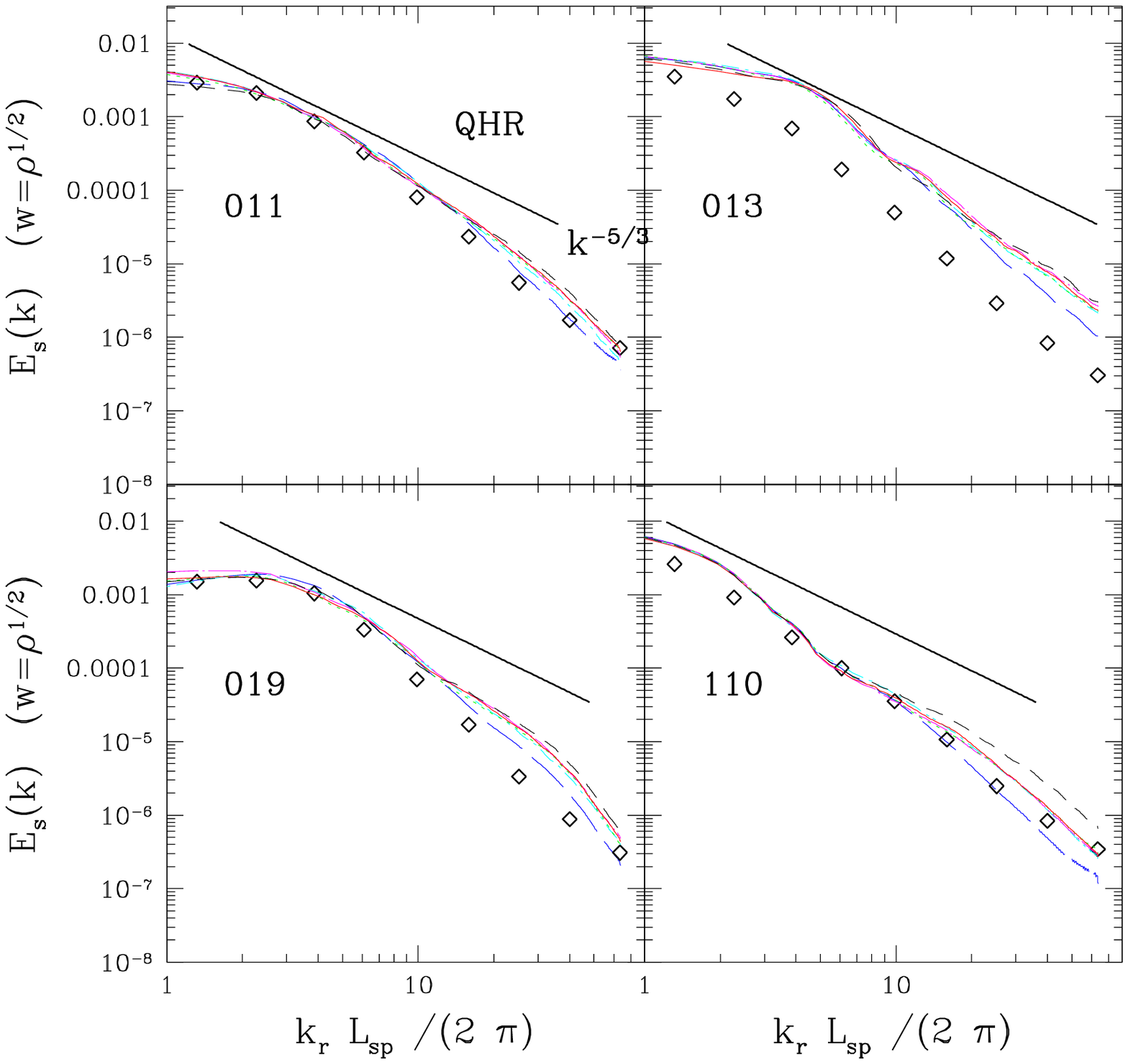}
   \caption{As in Fig. \ref{PHRs.fig}, but for the quiescent test clusters.}
   \label{QHRs.fig}%
    \end{figure*}

Spectra of runs with different AV settings begin to deviate at wavenumbers 
 ${\tilde k}\simgt 6={\tilde k_d}$. 
This indicates that for the present simulation resolution (HR) 
the effects of numerical viscosity on the formation of eddies 
due to  Kelvin-Helmholtz instabilities generated by substructure motion  
\citep{ta05,su06}  become significant at spatial scales 
$l_{diss}=\pi/ k_d\sim r_{200}/10\sim100-200$~kpc, in accordance with
the behavior of the velocity structure functions (see later).
 The generation of these instabilities leads to the development 
of turbulent motions which, as expected, are stronger in those runs with the 
lowest numerical viscosity.
There are no large differences seen between the shape of the spectra extracted
from clusters in different dynamical states. However, the power spectra of 
relaxed clusters, in comparison with the spectra of the dynamically 
perturbed clusters,
show a certain tendency to flattening at large scales and 
smaller amplitudes at the smallest cube wavenumber.

For comparative purposes, each panel shows for the considered case the 
volume-weighted velocity power spectrum of the standard AV run. The behavior of 
the other volume-weighted spectra reflects the similarities seen in the 
corresponding density-weighted
spectra of simulations with different AV settings and for the sake of clarity
these spectra are not shown in the panels.

At large scales, volume-weighted and density-weighted spectra do not show 
significant differences, with few expections such as the longitudinal 
power spectra of cluster $105$ of the perturbed subsample. 
%or cluster $11$ of the relaxed subsample. 
Variations in the behavior of density-weighted spectra 
and volume-weighted spectra are expected to appear at smaller scales,
when gas density gradients become significant. Note, however, that there 
are no systematic differences between density and volume-weighted spectra
as a function of wavenumber versus the cluster dynamical state and/or 
power spectrum decomposition.

   \begin{figure*}
   \centering
   \includegraphics[width=17.2cm,height=13.2cm]{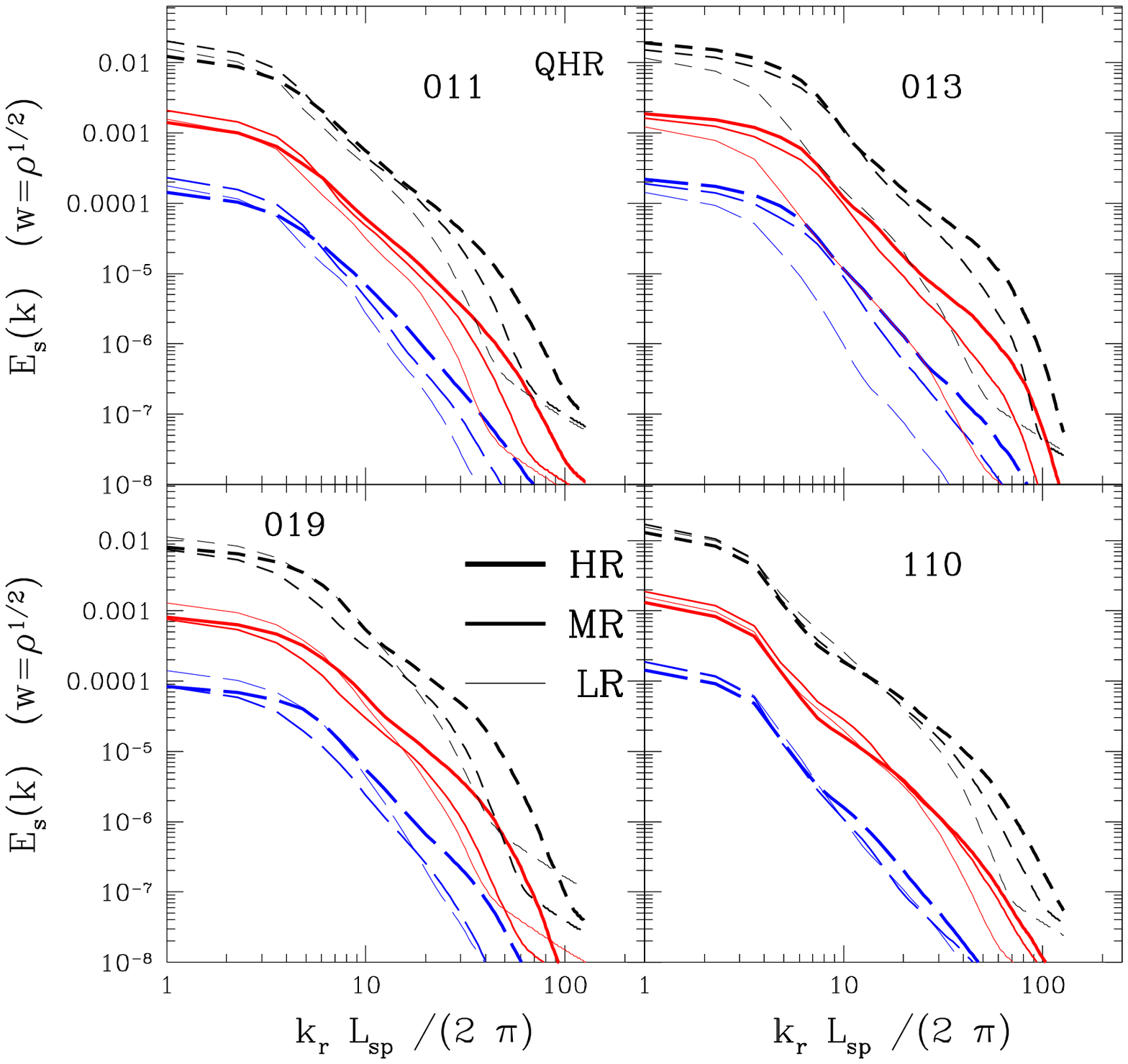}
   \caption{Shearing components of the density-weighted velocity power 
spectra  are shown at $z=0$ for the runs of the relaxed test clusters as a 
function of the wavenumber. As in the previous figures, different line styles 
are for  different AV runs.
In each panel  runs with different resolution are displayed with different line 
thickness. For the sake of clarity, not all of the AV runs are plotted and
runs with different AV parameters have been  equispaced from top to bottom 
by a factor of $10$. To better illustrate the wavenumber dependence, spectra 
have been computed here using $N_g^3=256^3$ grid points.} 
   \label{RESQus.fig}%
    \end{figure*}

At high wavenumbers, in certain cases density-weighted spectra 
can have values below those of volume-weighted spectra.
This is interpreted as a resolution effect, which is stronger as runs
with lower resolution (MR or LR) are considered. As already outlined,  
the gas distribution is driven by gravity and because of the 
  Lagrangian nature of the SPH code, for wavelength modes approaching 
the grid spacing the contribution of low density regions which are 
undersampled becomes dominant. This effect is not present in all of the
clusters because it is strongly dependent on the degree of gas  inhomogeneity
which in turn depends on the individual cluster dynamical history.

Although the aim of the paper is to assess the effectiveness of the 
time-dependent AV scheme in reducing the numerical viscosity in SPH 
cluster simulations, nevertheless the velocity power spectra presented here 
can be used to obtain useful indications about the description of turbulence 
in the ICM when SPH simulations are used.

Below the driving scales ${\tilde k}\sim 1-2$, the longitudinal component of 
the energy spectra exhibit an approximate power-law behavior close to the 
Kolgomorov scaling that extends down to dissipative scales. These features
are not shared by the solenoidal component of the spectra. To study the 
behavior of the shearing part of the spectra, it is useful to introduce 
the ratio $\Psi(k)\equiv E_c(k)/(E_c(k)+E_s(k))$ which measures as a function 
of the wavenumber the fractional contribution of the longitudinal velocity
power spectrum component. This ratio lies in the range $\Psi(k)\sim 0.3-0.5$ 
at large scales and rises above ${\tilde k}\sim 5=k^{res}_s$, the latter is 
defined as approximately the wavenumber at which occurs the bending in $\Psi(k)$.
While the ratio at large scales can be understood in terms of fully developed 
turbulence, the rise in $\Psi(k)$ at small wavenumbers is a consequence 
of the steep decrease in the shearing part of the spectra in this range 
of scales, which is interpreted here as a resolution effect.
This behavior shows that a fundamental issue to be addressed is the resolution
dependence of the velocity power spectra extracted from the simulations.

K09 and \cite{pr10} studied supersonic isothermal turbulence using both SPH
and Eulerian based codes. Both groups reach the conclusion that 
differences in the simulation results are mainly due to the resolution used, 
 with dissipative effects specific to the codes confined at high wavenumbers. 
In fact, equivalent results in the measured  velocity power spectra are 
obtained using an equivalent number of resolution elements $N$ for each direction 
in space in simulations with different codes. For example, SPH simulations
with $N^3=512^3$ particles would give comparable results to those obtained using
 the same number of grid cells in Eulerian simulations.
   \begin{figure*}
   \centering
   \includegraphics[width=17.2cm,height=13.2cm]{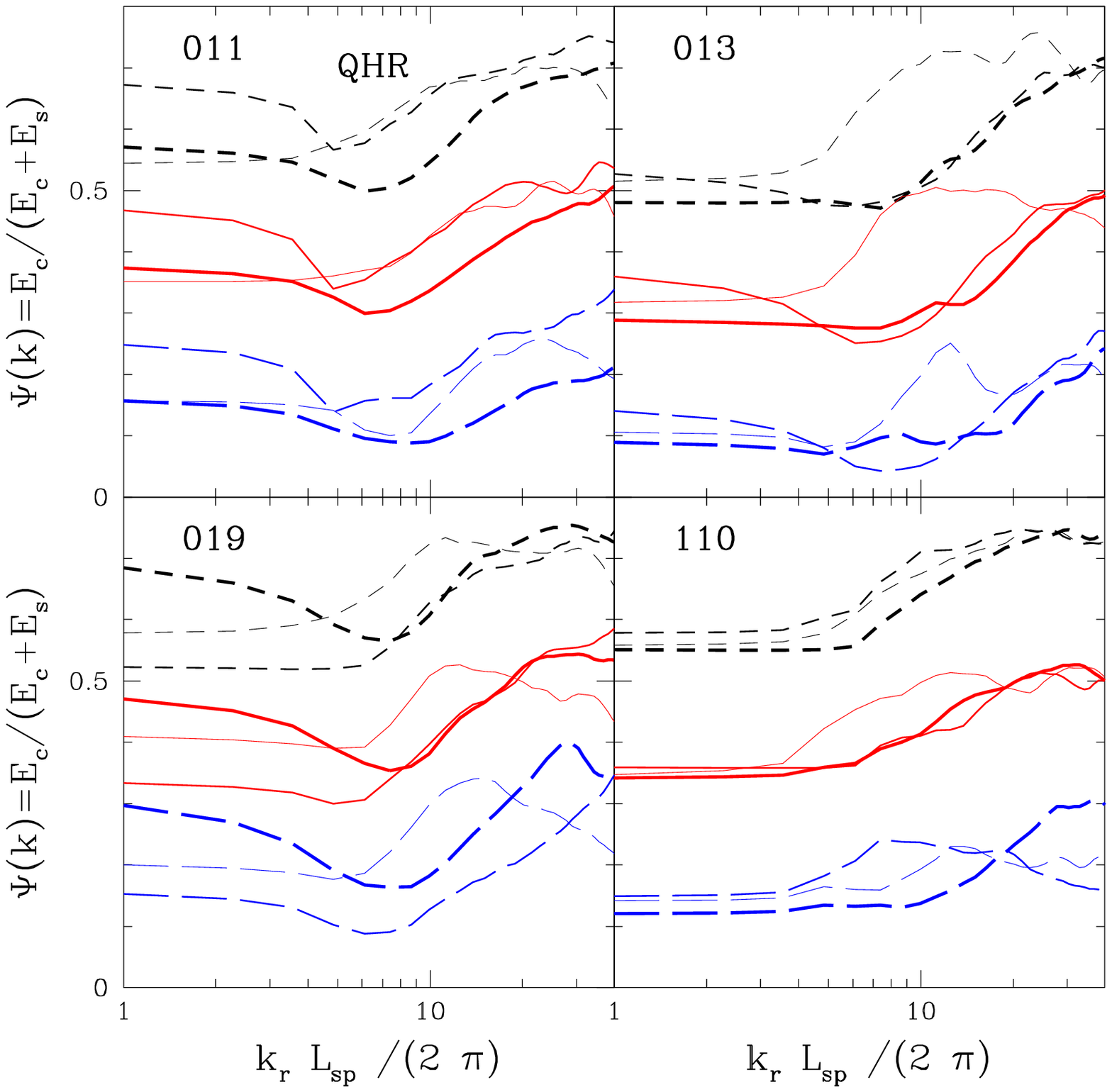}
   \caption{The ratio of the longitudinal to total velocity power 
spectra  are shown at $z=0$ for runs of the relaxed test clusters as a 
function of the wavenumber. The line coding is the same as in the previous 
figure; runs with different AV parameters have been  equispaced from top to 
bottom by a factor $0.2$.} 
   \label{RESQpsi.fig}%
    \end{figure*}

Exploiting this correspondence one can then interpret the dependence of
the velocity power spectra on resolution effects using the results 
of \cite{fe10}, who studied  interstellar turbulence using a FLASH code with up
to $N^3=1024^3$ cell elements.
From the simulations the authors (see their sect. 6) measure 
in the extracted Fourier spectra $E(k)$ a rise in $\Psi(k)$ 
starting at $k\sim 40\sim N/26$ 
which they explain as a resolution effect.
Basically, this happens because the number of computational 
elements needed to accurately resolve rotational modes is larger than that 
necessary to describe longitudinal modes, for which only one degree of 
freedom is present. In fact, longitudinal modes appear to be well described 
down to $k\sim N/10$, suggesting a difference in the resolution scale by 
about a factor of $\sim$ three. 

Application of these results to the present simulations  suggests that
the solenoidal part of the spectra is barely resolved in the HR runs, 
with a resolution scale 
 $k^{res}_s\sim N/26\sim 64/26\sim3$, whilst the longitudinal component 
is resolved down to $k^{res}_c\sim N/10\sim6$. 
To corroborate these findings, Fig. \ref{RESQus.fig} shows  
 the solenoidal component of the density-weighted 
velocity power spectra   versus the wavenumber for simulations of different
resolution for the subsample of relaxed clusters.
All of the spectra show a well-defined tendency towards shallower slopes 
as the simulation resolution is increased, with a reduction of the dissipative
 range towards higher wavenumbers. 
This spectral behavior is also partly explained by the larger number of 
matter clumps which are now resolved at small scales as the mass resolution is
increased, thus adding power to the spectra at high wavenumbers.
At large scales a slight decrease in the amplitude of the spectra can
occur because of resolution effects, as the resolution is increased 
clumps of matter accreting at $r\sim r_{200}$ survive more easily 
therefore reducing the amount of turbulence injected into the medium
via ram pressure stripping mechanisms.
This effect can be seen in  Fig. \ref{RESQus.fig} for clusters $011$ and $019$,
for which the large scale part of the spectra decreases  as the resolution
is increased. 
Moreover, note that in certain
cases there is a plateau in the spectra at high wavenumbers which is 
removed as the resolution is increased. 

Fig. \ref{RESQpsi.fig} shows 
the ratio $\Psi(k)$, for the same test runs. Although there are large 
cluster-to-cluster variations, nevertheless cluster $110$ offers a 
particularly neat example of the shift of $k^{res}_s$ towards higher
wavenumbers as the resolution is increased.
This dependence on the numerical resolution also confirms that the
predominance of the compressive components in the velocity power spectra 
in not due to the presence of shocks at small scales, with the primary 
injection mechanisms occurring at cluster scales.
However, the estimates of $k^{res}_s$ inferred from the application of other
simulation results might be too pessimistic. 
From Fig.  \ref{RESQpsi.fig}  it can be seen that for clusters $013$ and $110$, 
 $\Psi(k)$ is stable down to ${\tilde k}\sim 6-7$. These values are a factor 
of $\sim$two higher than those estimated from previous scalings.
   \begin{figure*}
   \centering
   \includegraphics[width=17.2cm,height=13.2cm]{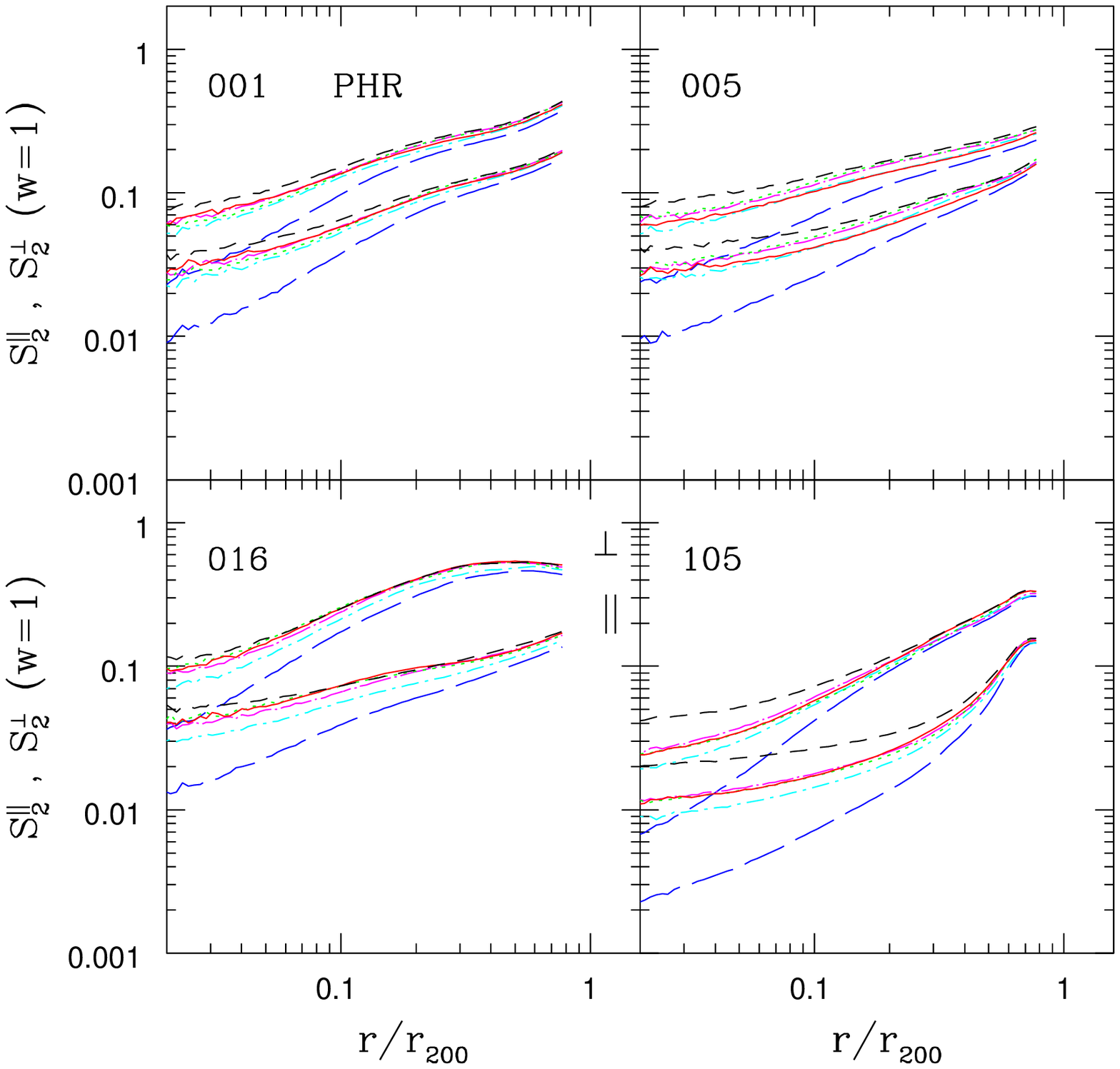}
 \caption{Transverse ($S_{\parallel}$) and longitudinal ($S_{\perp}$) 
second-order normalized velocity structure functions at $z=0$ for 
 HR runs of the  perturbed test clusters as a function of $r/r_{200}$.}
   \label{PHRsvel.fig}%
    \end{figure*}

Similar underestimates are found for  the dependence of the dissipative scale set by 
numerical viscosity versus  the numerical resolution.
According to K09, length-scales defined by the condition $k>N/32$ are affected by
dissipative mechanisms specific to the code. This implies that dissipative
effects here should start acting from ${\tilde k}\sim 2$, whereas the 
 spectral behavior indicates that at least down to ${\tilde k}\sim 6={\tilde k_d}$ 
there are no significant differences between spectra extracted from 
runs with different AV settings.

An important issue which needs to be addressed is the dependence on the numerical
resolution of the dissipative scales identified from the spectral behavior of 
the velocity power spectra. For doing this the use of the longitudinal power 
spectrum 
components is more suitable because for these spectra, numerical resolution effects 
are less severe than for the solenoidal components.
In fact, a comparison as in Fig. \ref{RESQus.fig}  of the behavior of the 
longitudinal part of the density-weighted velocity power spectra for 
simulations of different resolutions, shows that the dissipative
wavenumber ${\tilde k_d}$ scales approximately as $N$, suggesting that 
in order to resolve eddies down to length-scales of the order of galaxy-sized
objects ($\sim 50$~kpc) simulations would be required with at least $N^3=256^3$ gas 
particles. However, the spatial range of $l_{diss}$ derived from the spectra 
of Figures \ref{PHRc.fig}-\ref{QHRs.fig} is in rough agreement with that 
expected using analytical models \citep{su06} to estimate the  coherence 
velocity field length-scale associated with minor mergers in clusters.
Moreover, \cite{ma09} have recently presented an Eulerian-based AMR code
with subgrid modeling for turbulence at unresolved scales. The authors
studied the evolution of turbulence in cluster simulations and argue that 
most of the turbulent energy is injected into the medium at length scales of
$\sim125-250h^{-1}$~kpc (cf. their Fig. 6). This consistency between different
 independent estimates of the characteristic turbulent length scales therefore
suggests that for the HR runs, spectral features due to dissipative effects might 
be of physical origin rather than due to numerical resolution effects.

To summarize, application of previous findings \citep{ki09,fe10,pr10} to the 
simulations presented 
here produces  several discrepancies in the resolution dependence of 
characteristic scales, which appear less severe here than those found by 
these authors. 
A clarification of these issues clearly has to await simulations with much 
higher resolution; nevertheless it is suggested that physical differences 
in the problems considered might play a role. For instance, gravity is here the 
driving force which dominates the dynamics of the collapsed object, thereby 
creating, owing to the adaptative nature of the SPH codes in space,
a `scale dependence' in resolution as one moves outwards from the cluster 
center.
As a consequence, the effective resolution of the simulations performed here might
be higher than that estimated using simple scaling arguments.
  
However, these results indicate that the present simulation resolution is clearly 
inadequate to fully describe the spectral behavior of the ICM velocity field, 
therefore raising the question
of what is the minimum number of particles needed in order to adequately resolve
turbulence in SPH simulations. This is strictly related to the minimum
spatial scale which must be resolved in the velocity Fourier spectra.
Requiring for rotational motion at least one order of magnitude of
spectral resolution, implies that $N^3=512^3$ particles
are needed for a correct description of turbulence in SPH simulations of the 
ICM. Taking into account the previous caveats, the same results could be 
obtained using $N^3=256^3$ particles.

%These findings are in accord with the 
%behavior of the spectra extracted here and  confirm the connection between 
%SPH and grid codes in terms of resolution elements.

In addition to spectral analysis, the scale dependence of dissipative effects
can be studied using the velocity structure functions which provide information
in physical space about the velocity field self-correlation. 
In particular, the second order structure function $\mathcal{S}_2(\vec r)$
is complementary to Figures \ref{PHRc.fig}-\ref{QHRs.fig} for obtaining 
informations about the effects of numerical viscosity 
on the velocity field statistics.
To evaluate the structure functions is a computationally demanding task since
the number of pairs scales approximately as $\propto N_{gas}^2$.  
The choice of the subsample size $N_s$ is a compromise between the needs  
of adequately sample the structure functions and of reducing the computational cost. 
The parallel and transverse second order structure functions have been 
calculated here setting  $N_s=N_{gas}/8$, corresponding approximately to 
$\sim5\cdot10^9$ pairs 
for the HR runs. This choice for $N_s$ is found to give convergent 
results when tested against higher values. 
Note that a uniform random number extraction of $N_s$ particles from the 
simulation particles does not correspond to an approximately uniform 
sampling in space, as the bulk of the gas particles ($\sim 1/2)$ are
concentrated within distances $\simlt r_{200}/3$ from the cluster center.

The parallel and transverse second order velocity structure functions are
shown in Fig. \ref{PHRsvel.fig} for the HR runs using volume-weighted 
velocities. All of the functions exhibit a growing behavior with increasing
 pair separation $\tilde r= r/r_{200}$, thereby showing how in the ICM
turbulence is substained by motions on cluster scales. 
For two of the clusters ($1$ and $5$), the radial dependence is approximately
a power-law, a result which can be taken as indicative of a scaling range, although 
this issue needs to be clarified with simulations of much higher resolution.
   \begin{figure*}
   \centering
   \includegraphics[width=17.2cm,height=13.2cm]{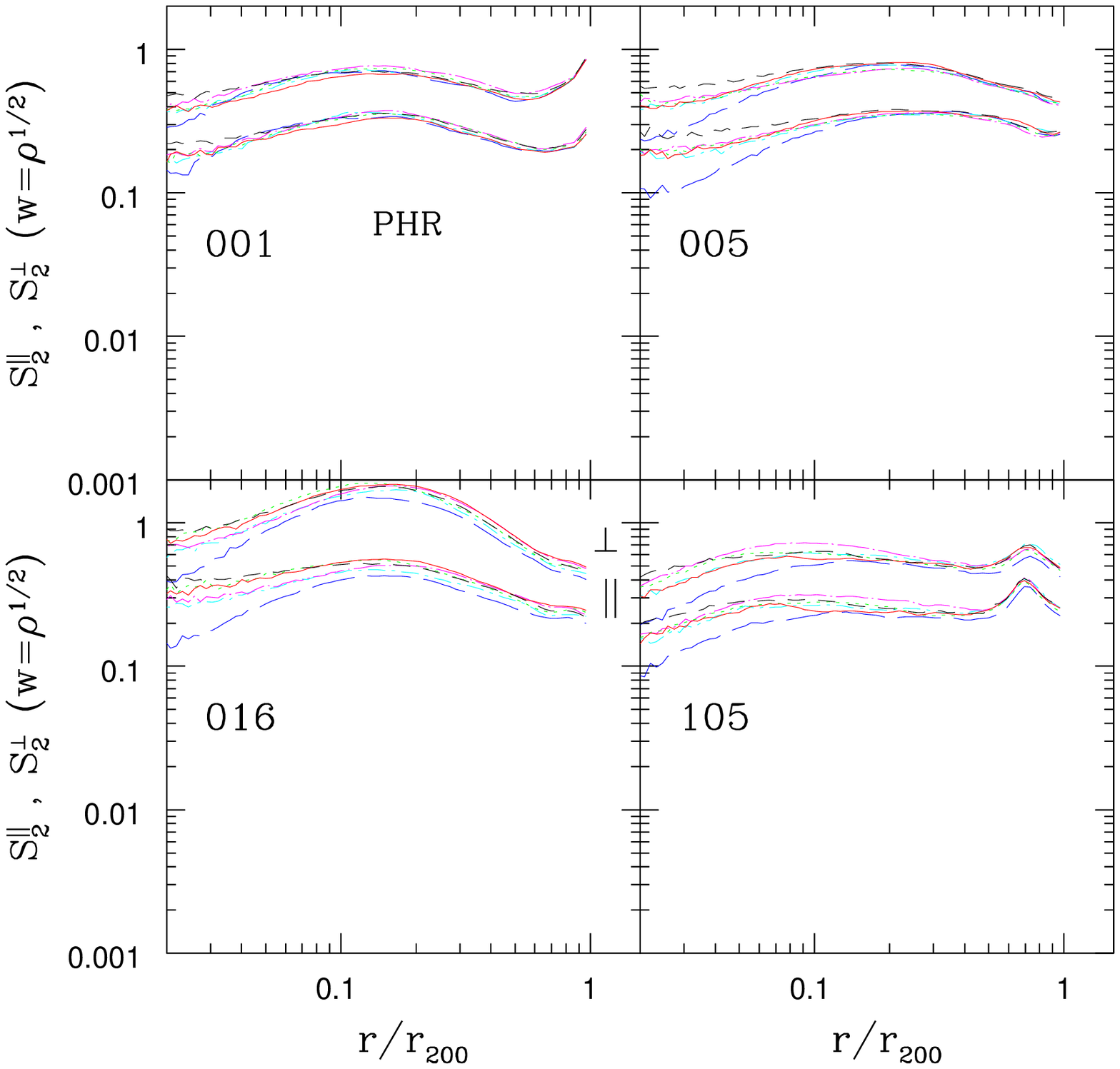}
   \caption{As in Fig. \ref{PHRsvel.fig}, but for the density-weighted
   velocity structure functions.}
   \label{PHRsvelu.fig}%
    \end{figure*}

Assuming as reference profile that of the structure functions produced by the
standard AV runs, structure function profiles extracted from the runs with a 
time-dependent AV scheme at small scales deviate significantly from the 
reference profile. The deviation increases as $\tilde r$ decreases and is
wider as the decaying viscosity parameter $l_d$ approaches unity. This 
behavior clearly shows the effectiveness of the new numerical viscosity 
scheme in reducing the viscous damping  of velocity away from shocks, and 
therefore of being the code better able to follow the development of 
turbulent motion.  Profiles of the structure functions extracted from 
simulations
with lower resolution are noisier than the profiles presented here but 
preserve the
basic features shown in Fig. \ref{PHRsvel.fig}, therefore  supporting the 
validity of these arguments.

Structure functions derived using a density-weighted velocity are 
shown in Fig. \ref{PHRsvelu.fig}. They exhibit 
the same behavior as for  the volume-weighted functions but with a 
much shallower radial dependence. This is due to the adopted weighting scheme, 
for which the contribution of those pairs  located in high density regions 
is enhanced with respect to the volume-weighted functions. This preferentially
happens at small scales because in SPH codes most of the particles are
located in high density regions.
Finally, a comparison with the spectra shown in  
Figures \ref{PHRc.fig}-\ref{QHRs.fig}, shows an approximate correspondence 
between the length scales 
($\sim r_{200}/10)$ at which dissipative effects become significant.

   \begin{figure*}
   \centering
   \includegraphics[width=17.2cm,height=13.2cm]{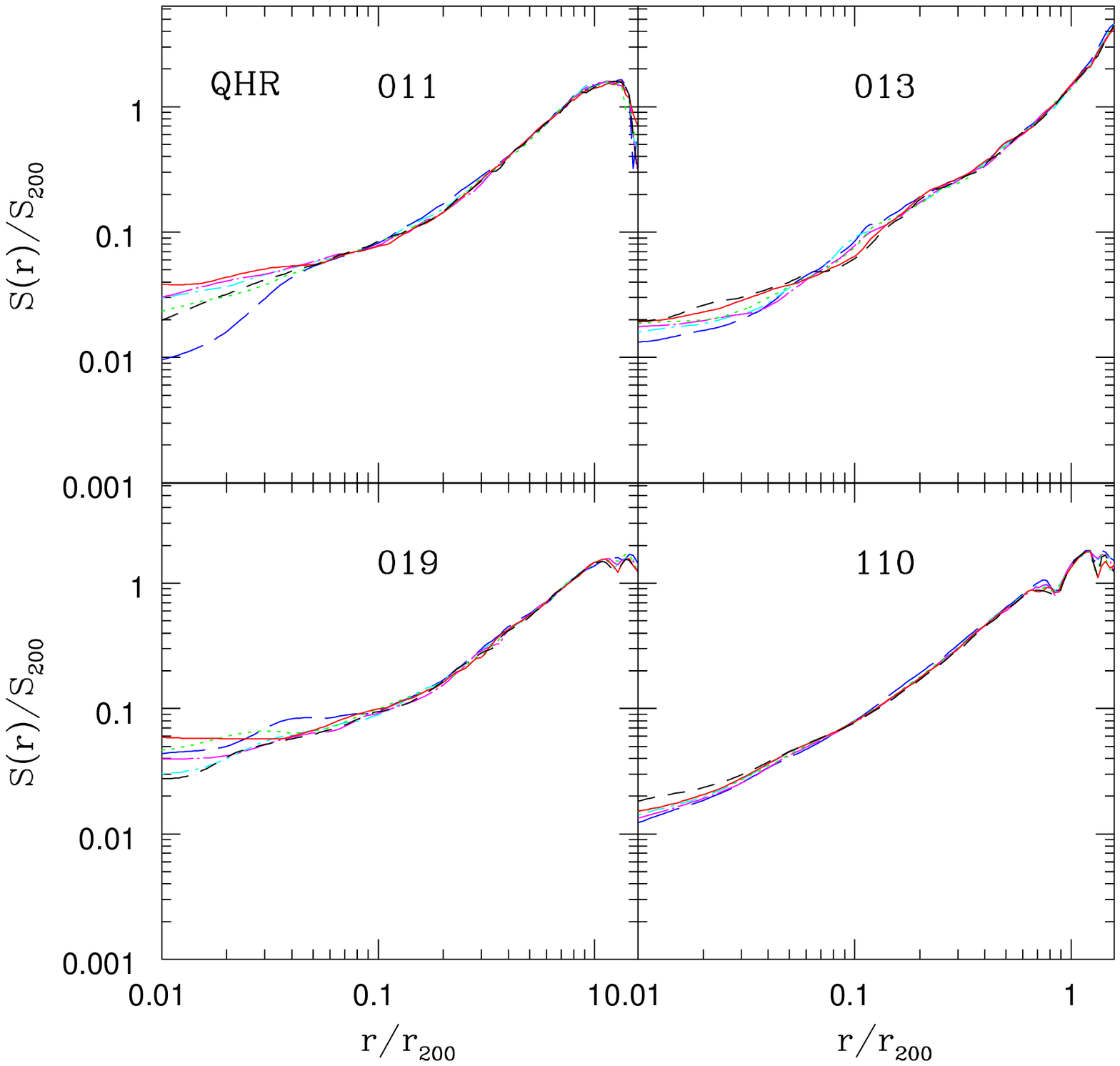}
   \caption{Final radial entropy profiles  for the HR runs of the
relaxed test clusters. 
 Gas entropy  is defined as $S(r)=k_B T(r)/\mu m_p \rho_g^{2/3}$ and  is
plotted in units of $S_{200}=\frac{1}{2}\left [\frac{2 \pi}{15} \frac 
{G^2 M_{200}}{f_bH(z)}\right]^{2/3}$.
As in Fig. \ref{PHRc.fig}, in each panel different lines
refer to runs with different AV parameters, the line coding being the same.}
   \label{QHRentr.fig}%
    \end{figure*}
\subsection{Average radial profiles}  
\label{clusterb.sec}
Measurements of the average radial profiles of thermodynamic variables, 
such as density or temperature, extracted from spatially resolved 
X-ray observations allow one to probe ICM properties and test the 
predictions of cosmological cluster simulations. This is achieved
by comparing the observed ICM profiles with those extracted from mock X-ray 
data. It is therefore important to examine in SPH cluster simulations 
the impact  of numerical viscosity on the radial profiles of the 
variables considered.

For doing this, the final radial profiles  
 of the gas entropy $S(r)=k_B T(r)/\mu m_p \rho_g^{2/3}$
are shown in Fig.  \ref{QHRentr.fig} for the simulated clusters of the relaxed 
subsample. As in Figures \ref{PHRc.fig}-\ref{QHRs.fig}, each panel corresponds
to an individual cluster and within each panel different 
curves are for runs with different AV settings. In order to compare the 
entropy profiles of different clusters, the entropy has been normalized to

   \begin{equation}
S_{200}=\frac{1}{2}\left [\frac{2 \pi}{15} \frac {G^2 M_{200}}{f_bH}\right]^{2/3}~,
  \label{entr.eq}
   \end{equation}

where $f_b=\Omega_\mathrm{b}/\Omega_\mathrm{m}$ 
is the global baryon fraction.

Discussion of the dependence of the final profiles on the AV scheme 
used in the simulations is restricted here to the entropy only, since both
density and temperature profiles exhibit quite similar behavior.
A feature common to all of the entropy profiles  is  
the relatively small scatter between those profiles extracted from runs
for a given cluster with different strengths of the numerical viscosity.
The variations in the entropy profiles due to the AV implementation appear 
to be confined to the inner cluster regions ($\sim10\%$ of $r_{200}$).
As already outlined in the Introduction,  a fundamental issue to be 
investigated is the amount of entropy present in cluster cores 
\citep{ag07,wa08,mi09}.
Viscous damping of random gas motion is expected to be reduced in low-viscosity
runs, thereby increasing fluid mixing and possibly the accretion at the cluster 
center of gas with higher entropy, with a subsequent rise of 
the entropy floor in the core. 

The profiles of Fig. \ref{QHRentr.fig} illustrate that the amount of gas 
mixing due to numerical viscosity is not significant, although there is 
a certain tendency for the core entropy to be influenced by the AV strength 
of the simulation.
For instance, cluster $11$ exhibits a large difference of $\sim5$ between the 
core entropy $S(r_c=0.01r_{200})$ of run AV$_0$ and that of run AV$_4$. 
For two other clusters ($12$ and $110$), the entropy profiles of the low-viscosity
runs have a moderate increase with respect to that of the standard AV run
as the radial coordinate decreases, with central differences being smaller than 
a factor of $\sim$ two. Finally, there is one cluster ($19$) for which $S(r_c)$ of run AV$_4$ is below that of run AV$_0$.
This behavior is shared by the entropy profiles of the perturbed cluster 
subsample and it indicates that the amount of entropy mixing in cluster cores 
induced by
numerical viscosity effects is modest or quite negligible.

The stability of the results is assessed  in Fig. \ref{RESQentr.fig} 
by contrasting  for the same test clusters the entropy profiles 
 of simulations with different resolution. In each panel the 
profiles of only three AV runs are plotted (AV$_2$, AV$_4$ and AV$_5$) 
to avoid overcrowding.
From Fig. \ref{RESQentr.fig} it can be seen that there is little variation
between the entropy profiles of runs performed with the same AV settings
but with different resolution.
 In the AV$_5$ test case, for several clusters the profiles of the HR runs 
show  some deviations at small distances from those with lower resolution. 
Interpretation of this difference as a systematic effect due to  
inadequate resolution requires some care however, since it is absent in
the profiles of the other AV runs. Moreover, even for the test case AV$_5$,  
in one cluster ($110$) the core entropy $S(r_c)$ does not vary with the 
simulation resolution.

To summarize, the reliability of the entropy profiles of simulated clusters 
with respect to variations in the simulation resolution is supported by the 
profiles of Fig.  \ref{RESQentr.fig}, for which one sees the lack of any systematic 
dependence of $S(r_c)$ with varying numerical resolution.
This indicates that the profiles of Fig. \ref{QHRentr.fig} can be reliably  
used to draw the conclusion that in SPH simulations the amount of entropy
present in cluster cores due to numerical viscosity effects can be considered 
negligible.
The smallness of AV effects in determining the core entropy of simulated 
clusters therefore indicates that the discrepancy, which originates at the 
cluster centers between SPH and grid-based Eulerian codes in the produced 
entropy profiles of adiabatic simulations, is inherent to the method 
itself rather than being a problem of the AV scheme implemented in the SPH codes.
In particular, the lack of significant AV effects indirectly confirms that
the primary source of the entropy discrepancy between the two hydrodynamical 
 approaches resides in the treatment of flow discontinuities in SPH 
simulations \citep{ag07}.
   \begin{figure*}
   \centering
   \includegraphics[width=17.2cm,height=13.2cm]{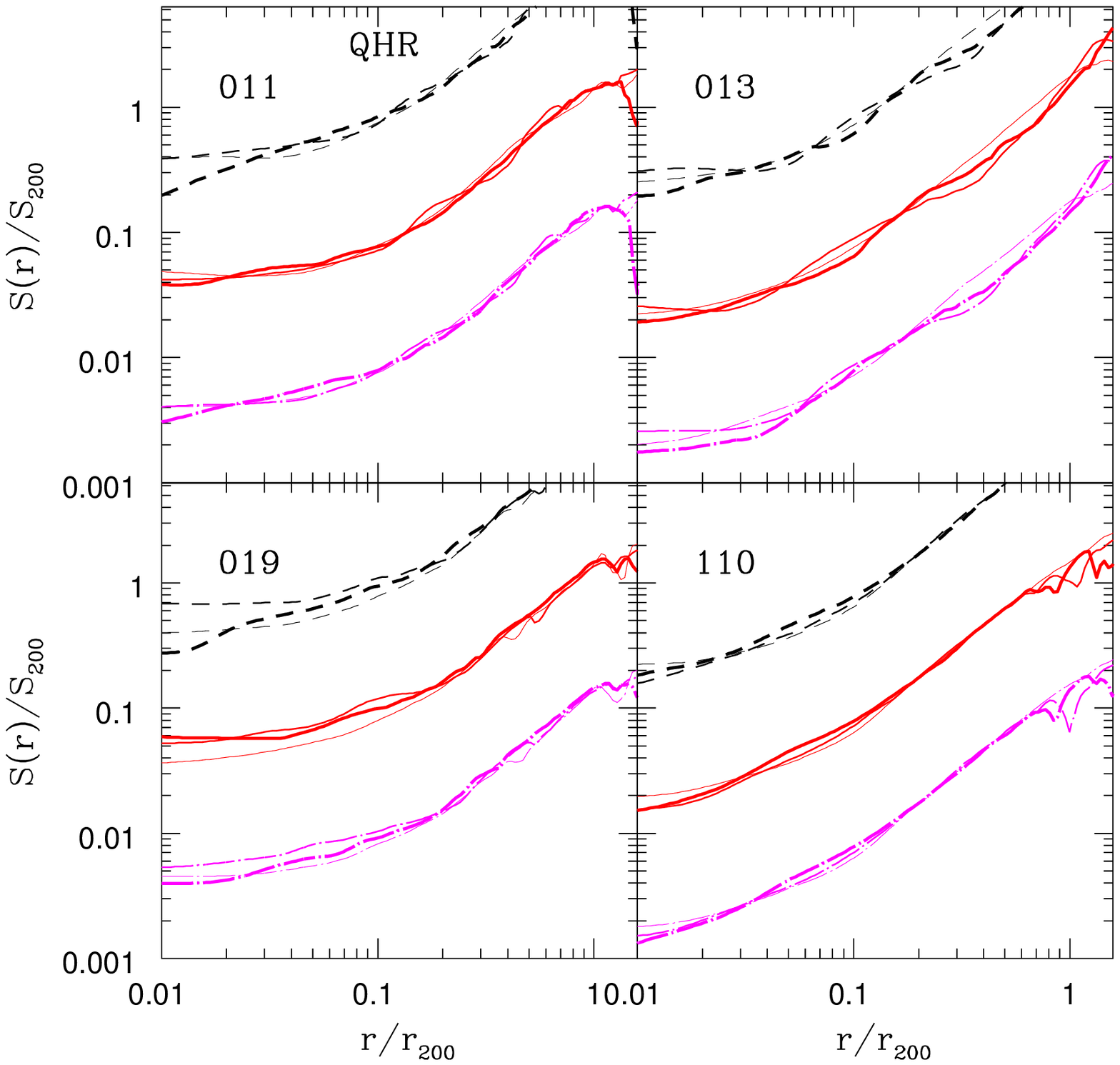}
   \caption{As in Fig. \ref{RESQus.fig}, but for the entropy profiles.}
   \label{RESQentr.fig}%
    \end{figure*}

To overcome these difficulties \cite{pr08} and \cite{wa08} proposed to add
an explicit diffusion term in the SPH thermal energy equation, with the 
purpose of smoothing the discontinuities present in the thermal energy at
the fluid interfaces. 
The introduction of this diffusion term leads to a certain amount of gas mixing
and in turn to an establishment of an entropy core in simulated clusters 
\citep{wa08}, which appears to qualitatively agree with that produced 
by Eulerian codes.
However, generation of vorticity associated with fluid instabilities is
severely affected by viscous damping due to numerical viscosity.
Using a suite of 2-D numerical tests \cite{pr08} showed that 
the treatment of Kelvin-Helmholtz instabilities can be greatly improved in SPH 
 codes provided that the hydrodynamic equations are reformulated so as to include 
 thermal diffusion as well as variable AV coefficients.
 It would then be interesting to reinvestigate for the same test sample
presented here the growth of central entropies using SPH simulations
in which the hydrodynamic equations have been modified by adding an artificial
thermal energy dissipation term.

The radial profiles $\alpha(r)$ of the viscosity parameters are shown in 
Fig. \ref{RESQalfa.fig} for the relaxed subsample. For each cluster, only 
the profiles of runs AV$_2$ and AV$_5$ are shown, the radial behavior 
of the profiles for the other AV runs being intermediate between the two.
Within each panel, the profiles of the AV test cases are accompanied by the
viscosity profiles extracted from the corresponding low and medium resolution
runs. This is in order to estimate the stability of the profiles with respect 
to varying the numerical resolution of the simulations.
From the figure it can be seen that for a given AV setting there is little
scatter between the profiles extracted from runs with different resolutions,
 with some differences near to the shock front which are 
 not significant however. The profiles of the perturbed cluster subsample
are similar to those shown here, but with a wider scatter because of the 
different dynamical history of these clusters.
The radial behavior of the profiles is in agreement with what was expected:
note that for two clusters ($19$ and $110$), the peak values $\alpha_{peak}$ 
of $\alpha(r)$  at the shock front for the run AV$_5$ are significant smaller 
than those for run AV$_2$. This is not the case for the peak values of 
run AV$_4$ (not shown here), which  are similar to those 
 of run AV$_2$, indicating therefore that for 
weak shocks (say $\alpha_{peak}\simlt0.3$) the peak value of the 
viscosity parameter is influenced by the floor value $\alpha_{min}$ 
of the gas particles ahead of the shock.
   \begin{figure*}
   \centering
   \includegraphics[width=17.2cm,height=13.2cm]{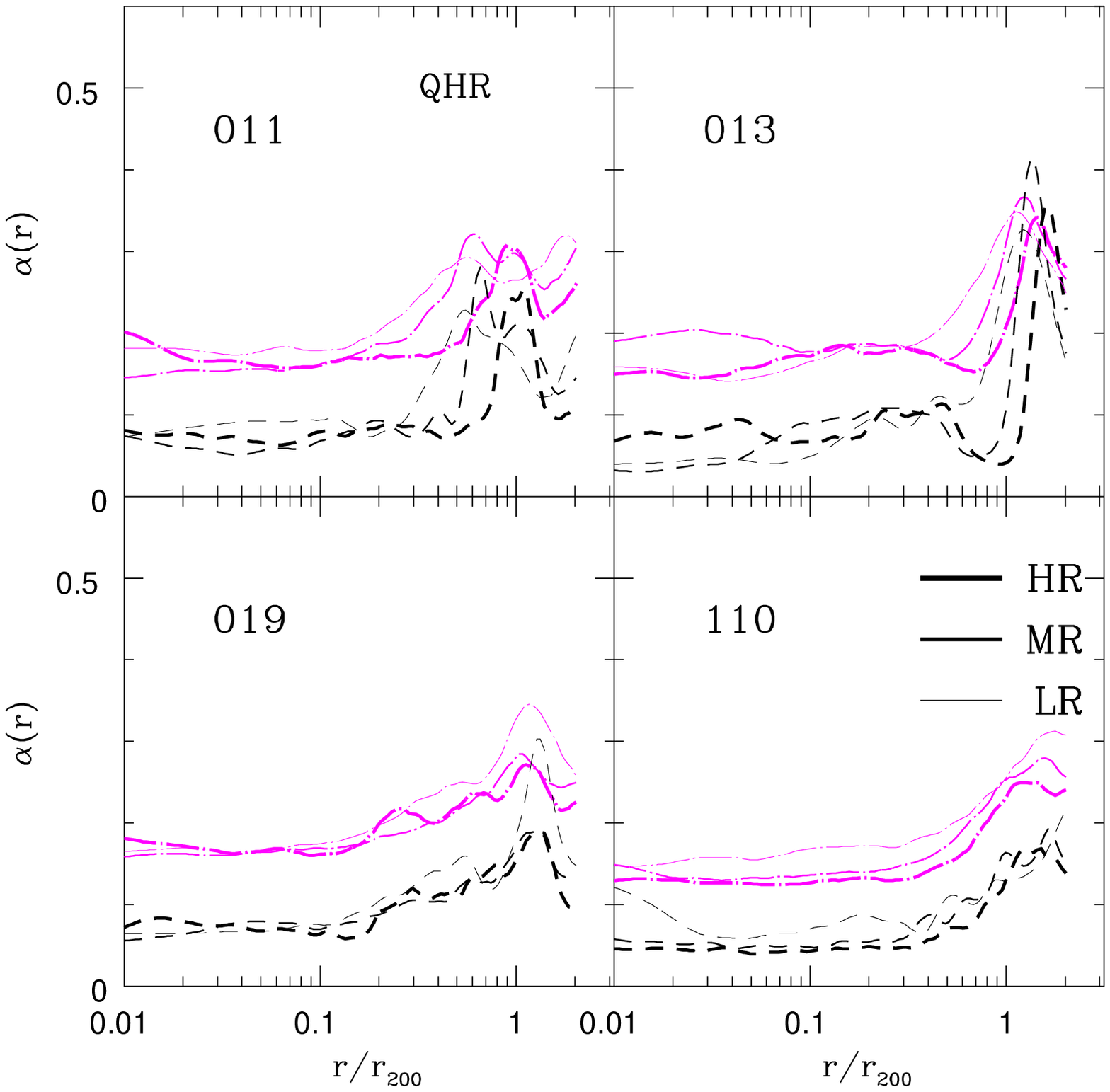}
   \caption{As in Fig. \ref{RESQus.fig}, final profiles of the viscosity 
parameter $\alpha$ are shown for the runs of the relaxed test clusters 
with different AV parameters and resolution. In each cluster panel the 
profiles have been plotted without any shifting between the various cases.}
   \label{RESQalfa.fig}%
    \end{figure*}

One of the most important aspects in which viscous damping of random gas 
motion is expected to influence ICM properties, is in the ICM energy 
budget.
In order to investigate the impact of numerical viscosity on the energy content
of the ICM, it is necessary however to properly disentangle from the velocity
field ${\vec u}(\vec x)$ the contribution of bulk motions from the small-scale 
chaotic parts. Following \cite{do05}, a turbulent velocity field 
 $\tilde{\vec u}(\vec x)$ can then be defined by subtracting from 
 ${\vec u}(\vec x)$ a local mean velocity $<u(\vec x)>$. According to 
Eq. \ref{filtu.eq}, the latter is defined by using a TSC window function
to interpolate gas density and velocities at the grid points of a cubic mesh
with grid spacing $H$. The mean velocity $<u(\vec x)>$ is defined by 
averaging over the gas particles which belong to the grid cell; subtraction of
 the filtered velocity from the velocity
field $u(\vec x)$ is expected to remove the contribution of large scale 
motions and leave a velocity $\tilde{\vec u}(\vec x)$ with spectral content 
dominated by those wavenumbers for which $kH>>1$. 
The choice of the filtering scale $H$ is somewhat arbitrary: \cite{do05} 
found that a mesh spacing of the order of $\sim 30$~kpc yields consistent 
results; in agreement with the choices of Sect. \ref{clustera.sec}  
the scaling $H\propto r_{200}$ is adopted here, allowing  
the velocity field $\tilde{\vec u}(\vec x)$ of clusters of 
different masses to be compared in a self-similar way. 
The choice of $H$ is a compromise between the needs of removing the contribution 
of bulk flow motions to $u(\vec x)$ as much as possible and the constraints 
imposed by the numerical resolution of the simulation for which  values 
of $H$ which are too small leads to an undersampling of local estimates.
These effects are expected to be significantly reduced by requiring at 
least $N_c\sim10^2$ gas particles in the cube cells.
 From this constraint, the allowed range for the grid spacing $H$ can 
be estimated using the average gas density radial profiles.
For the clusters presented here, $\rho/\rho_c \sim5$ at the cluster 
radius $r_{200}$ and for the HR runs this translates into a range of values
for $H$ between $\sim 740$~kpc for the most massive clusters down to 
$\sim 350$~kpc for the least massive ones.

Such filtering scales are too large to effectively remove bulk motion
components from the velocity field $u(\vec x)$. As a compromise, the 
contribution of turbulent energy can be investigated by restricting the
study to the inner parts of the cluster. At the radius $r\sim r_{200}/2$ the 
cluster gas densities lie around $\rho/\rho_c \sim30$ and the range for $H$ 
is now between $\sim 400$~kpc and $\sim 180$~kpc, which corresponds to
  $H\sim r_{200}/5$, when rescaled to the cluster radius $r_{200}$.  
Although this choice does not completely remove 
from $\tilde{\vec u}(\vec x)$  the laminar flow patterns present in the ICM, 
nevertheless it is supposed to filter out most of the contribution to 
${\vec u}(\vec x)$. In a recent paper \cite{va09a} studied the development of 
turbulence in the ICM using the adaptative grid AMR cosmological code ENZO and
argued that for their simulated cluster the ICM velocity field at 
length scales $\simgt 300$~kpc is
dominated by laminar flows. Moreover, the turbulent energy density radial 
profiles constructed from a filtered field using a length scale of $H/2$ do
not show appreciable differences from those derived here using $H$ as 
a filtering scale.
   \begin{figure*}
   \centering
   \includegraphics[width=17.2cm,height=13.2cm]{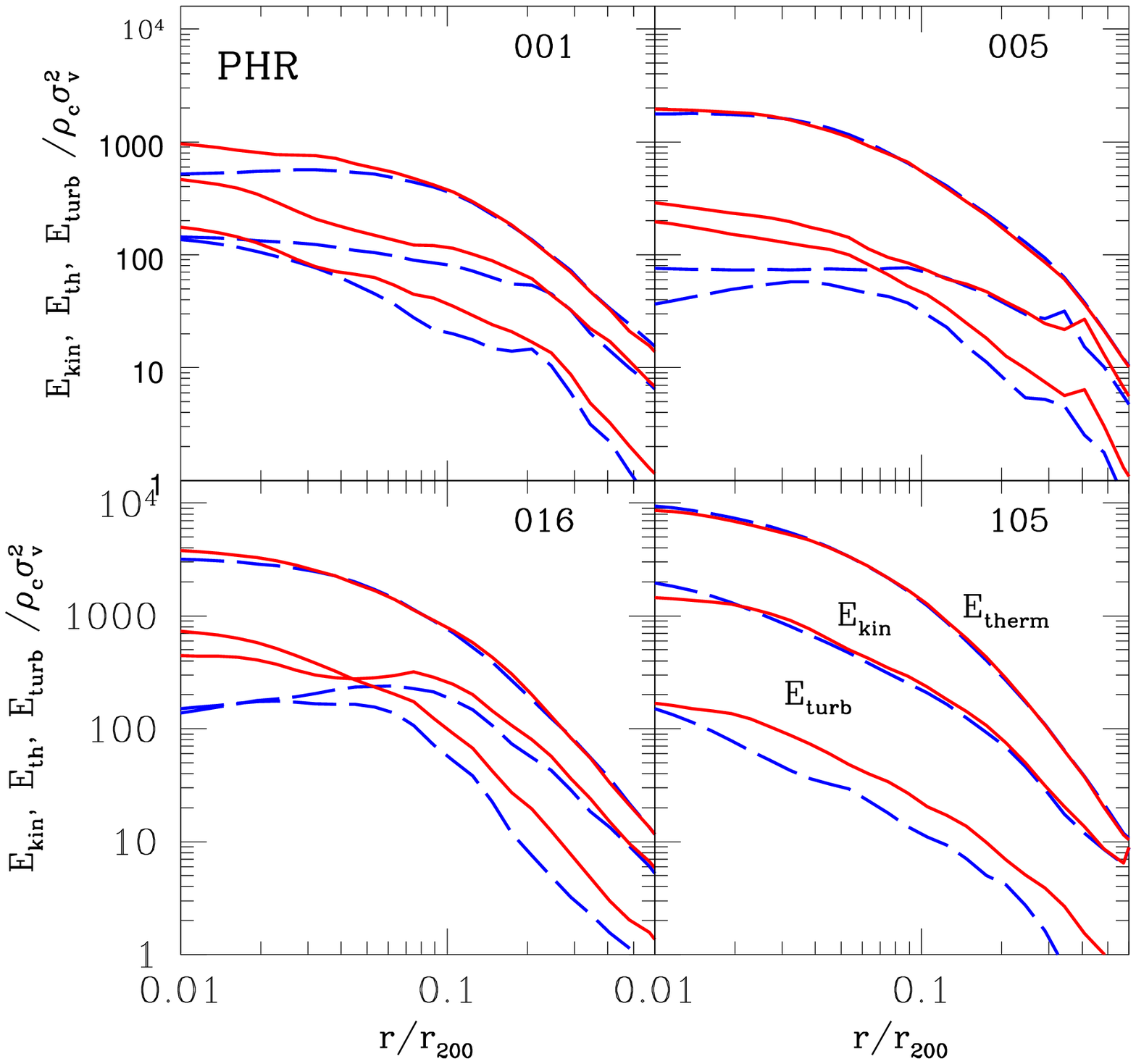}
 \caption{Final energy density radial profiles constructed from the HR
runs of the dynamically perturbed test clusters.
Each panel shows: the profile of the thermal energy density 
 $E_{th}= {3 k_B T(r) \rho_g(r)}/{2 \mu m_p} $, the kinetic energy density
 profile $E_{kin}=\rho_g \vec u^2/2 $ and the turbulent one 
$E_{turb}=\rho_g \tilde{\vec u}^2/2$; the 
 turbulent velocity field $\tilde{\vec u}(\vec x)$ is defined  according to
the procedures discussed in sect. \ref{clusterb.sec}. The profiles are 
distinguished by the corresponding labels indicated in the bottom  right
panel and have been rescaled in units of $\rho_c \sigma_v^2$.  
 For the sake of clarity, in each panel only profiles of simulations with
two different AV parameters (AV$_0$ and AV$_4$) are shown.}
   \label{PHRekin.fig}%
    \end{figure*}

In order to consistently compare our simulation results with those 
of \cite{va09a}, the radial density profiles are constructed using the 
same definitions. The thermal energy density is defined as $E_{th}=
{3 k_B T(r) \rho_g(r)}/{2 \mu m_p} $, where $T(r)$ is the mass-weighted 
 gas temperature, $k_B$ is the Boltzmann constant, $\mu=0.6$ is the molecular weight
 and $m_p$ the proton mass; $E_{kin}=\rho_g \vec u^2/2 $ 
and $E_{turb}=\rho_g \tilde{\vec u}^2/2$ are, respectively, the
 kinetic and turbulent energy density. 
The total energy density is defined as $E_{tot}=E_{th}+E_{kin}$.
All of the quantities with a radial dependence are estimated by averaging 
over a collection of values; these are calculated according to the SPH 
prescription (\ref{sphvar.eq}) at a set of grid points uniformly spaced in 
angular coordinates 
and lying at the surface of a spherical shell, located at 
distance $r$ from the cluster center. 
The energy density profiles constructed in this way are shown in 
Fig. \ref{PHRekin.fig} for the clusters of the perturbed subsample. The profiles
have been rescaled in units of $\rho_c \sigma_v^2$ and for the sake of 
clarity for each cluster only the profiles extracted from the runs of 
two AV test cases (AV$_0$ and AV$_4$) are shown.
 The corresponding ratios $E_{turb}/E_{tot}$ are shown in Fig. 
\ref{PHRekinr.fig}.

A first conclusion to be drawn from the radial behavior of the profiles is
the significant impact on the ratio $E_{turb}/E_{tot}$  of the numerical 
viscosity scheme adopted in the simulation.
As a consequence of the suppression of viscous damping of random gas
motion, for the runs AV$_4$ the ratio $E_{turb}/E_{tot}$  at any radial
distance is higher than that of the corresponding runs AV$_0$ by a factor
ranging between $\sim 30\% $ and $\sim100\%$. 
The ratio is nearly  constant for the cluster $005$ ( $\sim 8\%$)  
and $016$  ( $\sim 10\%$ ) and increases
from $\sim2\%$ to $\sim 5\%$ for cluster $105$. Finally, 
for cluster $001$  it decreases  ($\sim 10\%$ at $r=r_{200}/100$) as $r$ increases
 ($\sim 5\%$ at $r=r_{200}/2$).  
A similar behavior is present for the profiles of the relaxed cluster 
subsample. 
   \begin{figure*}
   \centering
   \includegraphics[width=17.2cm,height=13.2cm]{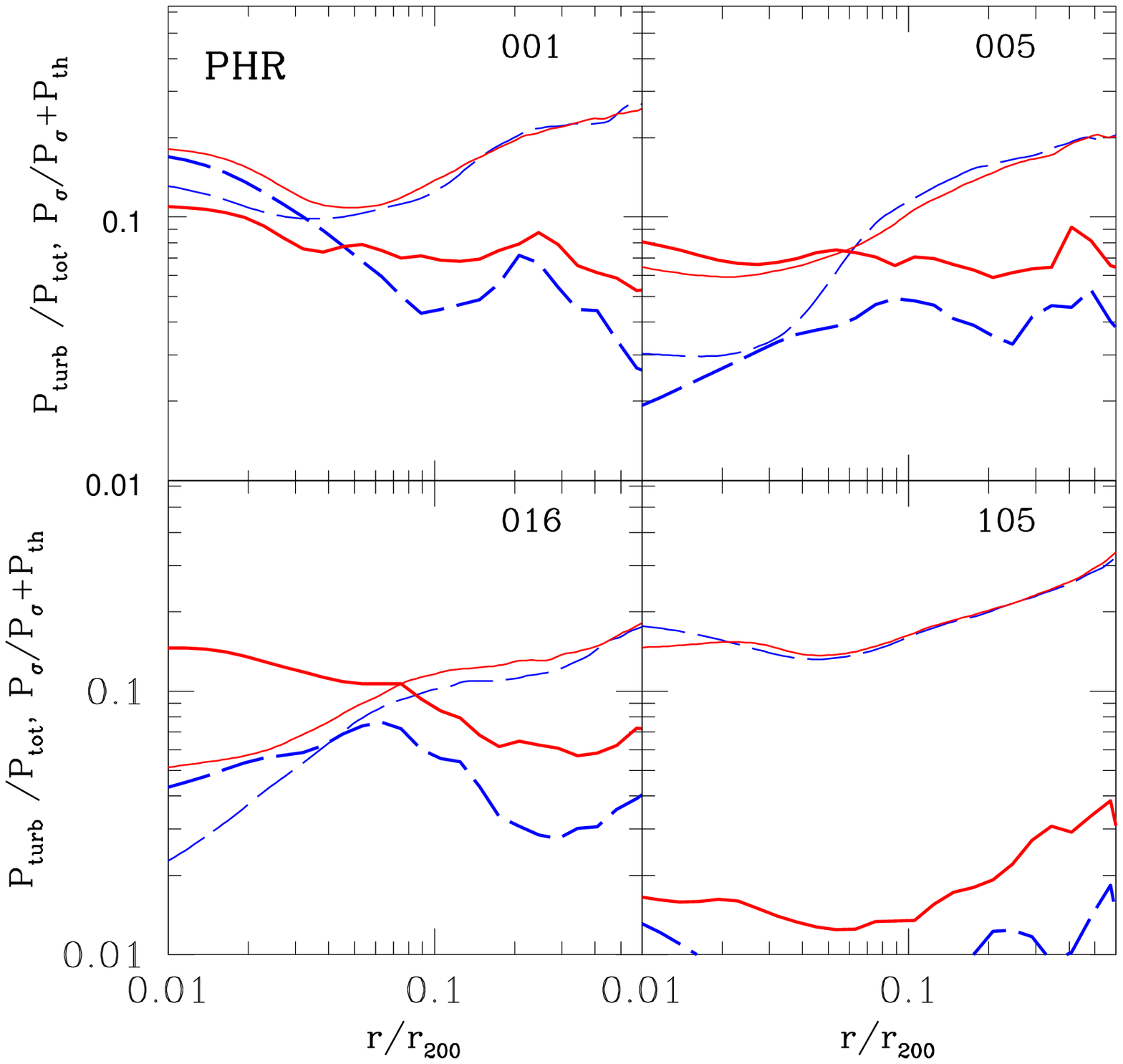}
   \caption{Ratio of the turbulent to the total pressure, 
$P_{tu}/P_{tot}\equiv E_{turb}/E_{tot}$, for the profiles of 
Fig. \ref{PHRekin.fig}. Additionally, for the same test runs, are shown 
(thin lines) the ratios of the `velocity' pressure to the total one : 
$P_{\sigma}/ \left[P_{\sigma}+P_{th}\right]= \frac{1}{3}\sigma^2(r) /\left[
kT(r)/\mu m_p +\sigma^2(r)/3 \right]$.}
   \label{PHRekinr.fig}%
    \end{figure*}

These findings indicate the lack of any systematic dependence of 
the ratio $E_{turb}/E_{tot}$  on the cluster radial distance and illustrate
the sensitivity of the turbulent energy density profile on the 
dynamical history of the individual cluster. 
In order to extract statistically meaningful results 
about the contribution of random gas motion to the ICM energy budget, it is 
then necessary to use a very large simulated sample \citep[ say $\simgt 100$ 
clusters, see, e.g., ][]{pi08}.

It is difficult to assess the consistency of the profiles presented here with
those obtained by \cite{va09a}, since the results indicate large cluster-to-
cluster variations among the individual turbulent energy density profiles and 
in \cite{va09a} the profile is shown for only one cluster.
However, the authors argue that, for their cluster, the turbulent energy budget
is smaller by a factor $\sim5-6$ than previous SPH results. 
The ratio $E_{turb}/E_{tot}$ lies here in the range between a few percent and 
$\sim 5-10\%$, and similar values for this ratio are quoted 
by \cite{va09a}.
Moreover, a visual inspection shows a strong similarity between the profiles 
of cluster $105$ and those derived by \cite{va09a} for their simulated 
cluster ( Fig. 6, bottom right).
As already stressed in Sect. \ref{clustera.sec}, a proper study of turbulence 
in the ICM using an SPH code awaits simulations of much higher resolution than 
those presented here, nevertheless these results are encouraging and support 
the view that SPH simulations can be used to investigate this topic, 
provided that the numerical viscosity effects are properly treated 
and the medium is adequately resolved.

An important issue which needs to be investigated is the impact of the 
 numerical viscosity scheme on the cluster mass bias estimated
from simulations.  Recently, the relative importance of different
mass bias components has been studied in detail \citep{pi08} using a large set
 of N-body/SPH simulations of galaxy clusters. Apart from the bias originating 
from the use of observational quantities, such as spectroscopic temperatures,
and those induced by assuming specific models for the measured profiles, 
it was found that mass underestimates are dominated by the  hydrostatic 
equilibrium assumption and by the absence in the Jeans equation for the 
mass of pressure support terms due to random gas motions.
In particular, for relaxed clusters the mass bias due to these terms
increases with the distance $r$ from the cluster centre and can be as high as 
$\sim5-10\%$ for $r$ ranging between $r_{2500}$ and $r_{200}$.
Given that the importance of galaxy clusters as cosmological probes relies 
on the use of accurate mass estimators, it is therefore important 
to estimate in the Jeans equation  the dependency of these gas motion terms
on numerical viscosity.
 
For doing this, the relative importance of the velocity pressure terms is quantified
by constructing the radial profile of  

   \begin{equation}
\frac{P_{\sigma}}{P_{\sigma}+P_{th}}=\frac{\sigma^2(r)/3}
{\sigma^2(r)/3+kT(r)/\mu m_p} ~,
  \label{pres.eq}
   \end{equation}
 
where $\sigma(r)$ is the rms gas velocity dispersion. The latter is evaluated 
at the same radial shells used to evaluate the 
energy density profiles.
 In analogy with Eq. \ref{pres.eq}, a relative turbulent pressure term can 
be defined, $P_{turb}/P_{tot}$, for which 
$P_{turb}/P_{tot}\equiv E_{turb}/E_{tot}$.
The quantities $P_{\sigma}/(P_{\sigma}+P_{th})$ can then be directly compared 
with the turbulent-to-total energy density ratios and these are shown in Fig. 
\ref{PHRekinr.fig} as thin lines. For the four test clusters, the ratio 
increase with the radial distance $r$ and 
$P_{\sigma}/(P_{\sigma}+P_{th})\sim 10-20\%$  at $r=r_{500}\sim 0.6 r_{200}$,
while at the same distance 
$P_{\sigma}/(P_{\sigma}+P_{th})\sim 10\%$  for the relaxed subsample.

 This range of values for the contribution of random gas motions to the total 
pressure is in accordance with the estimates extracted
from previous simulations \citep{ras06,nag07a,pi08,lau09}. 
Moreover, the profiles of $P_{\sigma}/(P_{\sigma}+P_{th})$ show a weak 
dependency
on the numerical viscosity scheme used in the simulations.
Differences between the profiles are significant in the cluster cores, but at 
$r=r_{500}$ the profiles of runs AV$_0$ differ by a few per cent from those of
runs AV$_4$, indicating therefore that pressure terms due to random gas 
motions are already estimated with good approximation in SPH simulations 
in which a standard AV scheme is implemented. The physical consequences of this result will be discussed in the conclusions, here it is worth discussing 
the origin of the different radial behavior of the ratio 
$P_{\sigma}/(P_{\sigma}+P_{th})$ with respect to that of $E_{turb}/E_{tot}$.

The rms gas velocity $\sigma(r)$ is defined as 
   \begin{equation}
\sigma(r)= \sqrt {<u^2(\vec r)>-<u(\vec r)>^2}~,
  \label{sgma.eq}
   \end{equation}
where the field velocity $\vec u(\vec r)$ is calculated 
according to the SPH prescription at the grid points of the spherical shell and 
brackets denote averages  performed over the set of points.
This definition is in accordance with the procedures used to compute the
energy density profiles; however the gas velocity dispersion calculated 
following the standard definition, i.e. by averaging over the gas particles 
contained within the shell of radial 
coordinate $r$ and thickness $\Delta r$,  does not differ significantly
 from that obtained using Eq. (\ref{sgma.eq}).
The different radial behavior of the rms gas velocity dispersion $\sigma(r)$
from that of the local velocity field $\tilde{\vec u}(r)$  can be 
ascribed to the different procedures adopted for computing the two 
velocities, which in turn define the two fields.
   \begin{figure*}
   \centering
   \includegraphics[width=17.2cm,height=13.2cm]{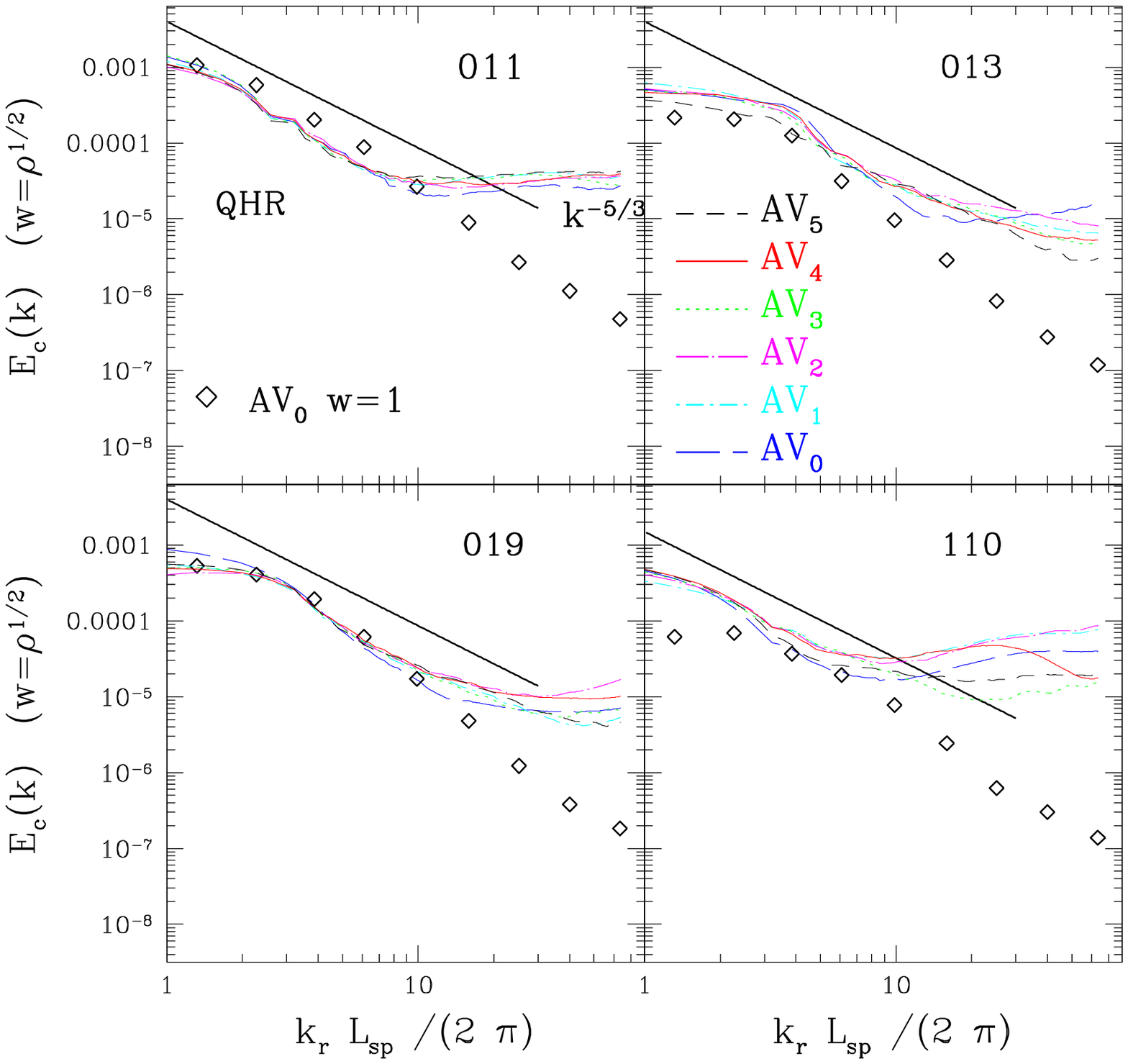}
   \caption{As in Fig. \ref{QHRc.fig},  compressive components of the 
density-weighted velocity power spectra are shown at the present epoch 
for the cooling HR runs of the relaxed cluster subsample.}
   \label{QHRcc.fig}%
    \end{figure*}

Because of the filtering procedures, the latter is dominated by the small-scale 
random gas motion with a coherence length scale $\sim 200-300$~kpc. For the 
rms gas velocity $\sigma(r)$, the contribution of laminar infall motion is
not removed and it is mixed with that of turbulent motion.
Therefore, while the local velocity field
 $\tilde{\vec u}(r)$  can be properly defined as a turbulent velocity field
\citep{va09a}, the dispersion $\sigma(r)$  is just a measure of the gas 
velocity variance at the scale $r$. This is the reason why the ratio 
 $E_{turb}/E_{tot}$ does not show a well-defined dependency  with radius, 
while $\sigma(r)$ increases with $r$ as the streaming motion due to
accretion from the infalling material becomes progressively more important.
A similar viewpoint is discussed by 
\citet[][see also Maier et al. 2009]{zh10}, who studied  
the generation of vorticity fields by means of 
of hydrodynamic simulations  in order to
 measure the level of turbulence in the intergalactic medium.

To summarize, the notion of turbulent pressure support,  
which accounts for the X-ray mass bias when assuming hydrostatic
 equilibrium for the ICM, is somewhat misleading 
at large cluster radii because the velocity 
terms which are missed in the Jeans equation are dominated by the 
streaming motion of the infalling material, with the local turbulent velocity 
field progressively becoming relatively less important in the cluster outskirts.

   \begin{figure*}
   \centering
   \includegraphics[width=17.2cm,height=13.2cm]{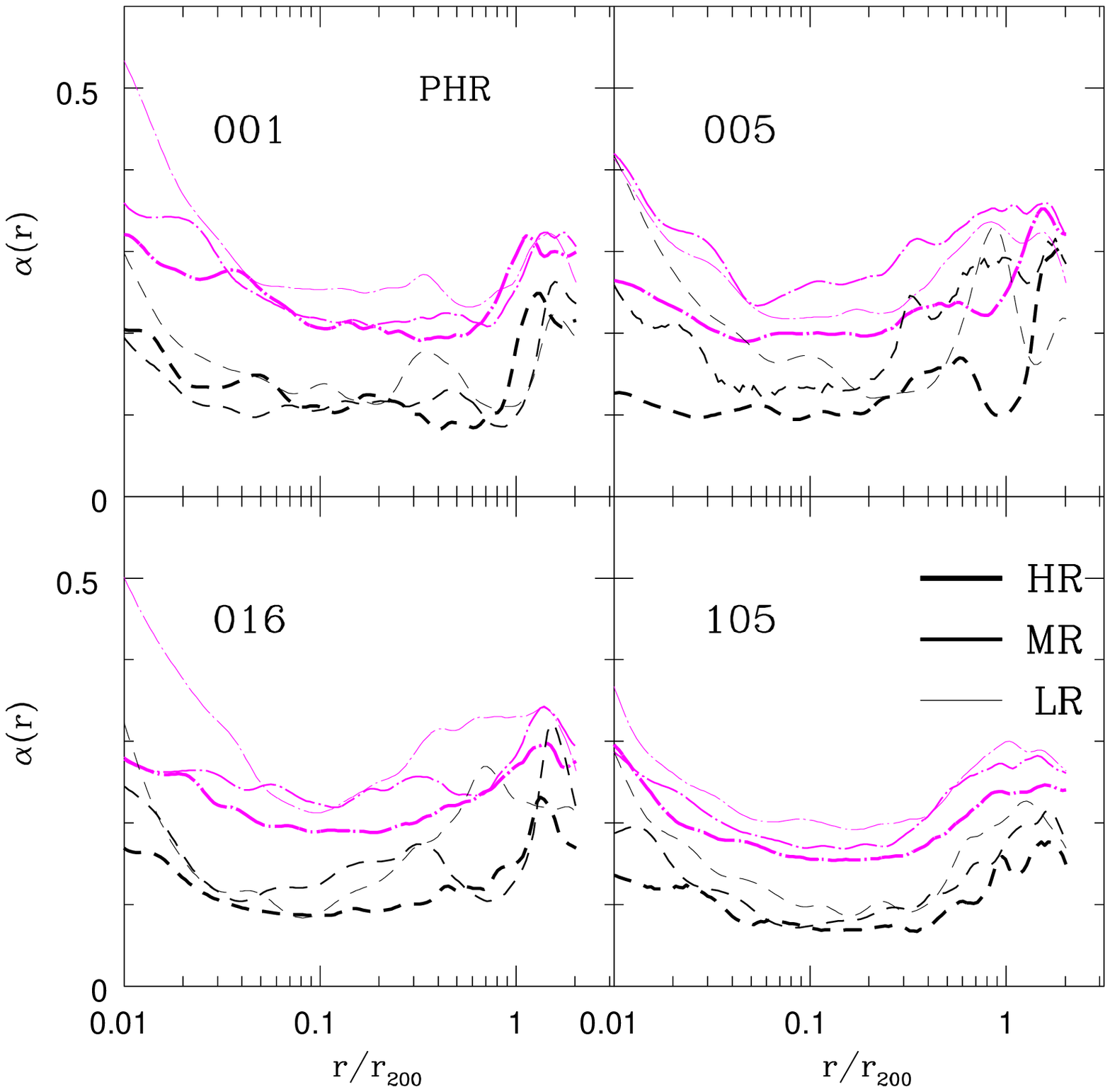}
   \caption{As in Fig. \ref{RESQalfa.fig}, final profiles of the viscosity 
parameter $\alpha$ are shown for the cooling runs of the perturbed test 
clusters with different AV parameters and resolution.}
   \label{RESPcalfa.fig}%
    \end{figure*}

   \begin{figure*}
   \centering
   \includegraphics[width=17.2cm,height=13.2cm]{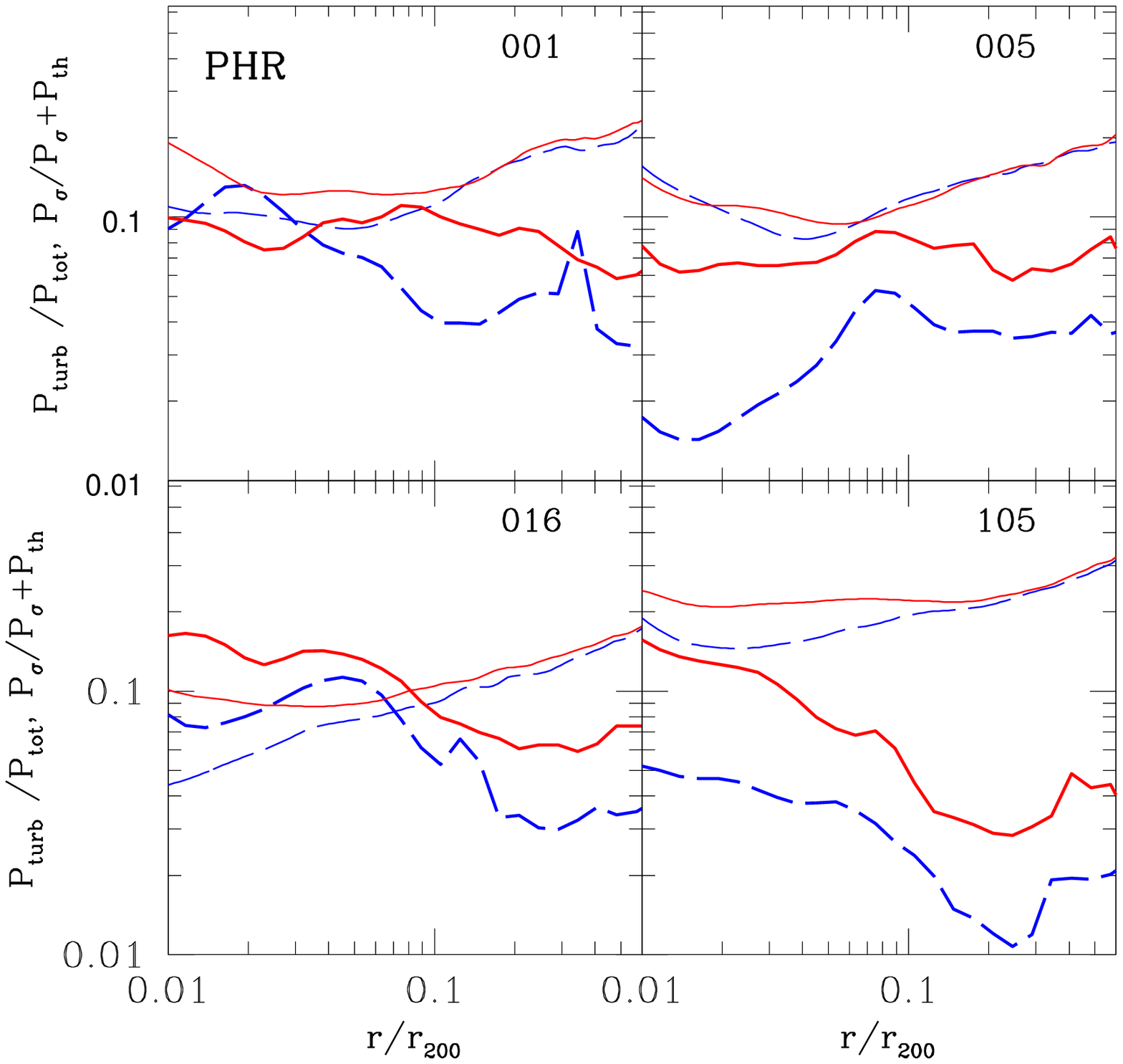}
   \caption{As in Fig. \ref{PHRekinr.fig}, but for the cooling runs.}
   \label{PHRcekinr.fig}%
    \end{figure*}

   \begin{figure*}
   \centering
   \includegraphics[width=17.2cm,height=13.2cm]{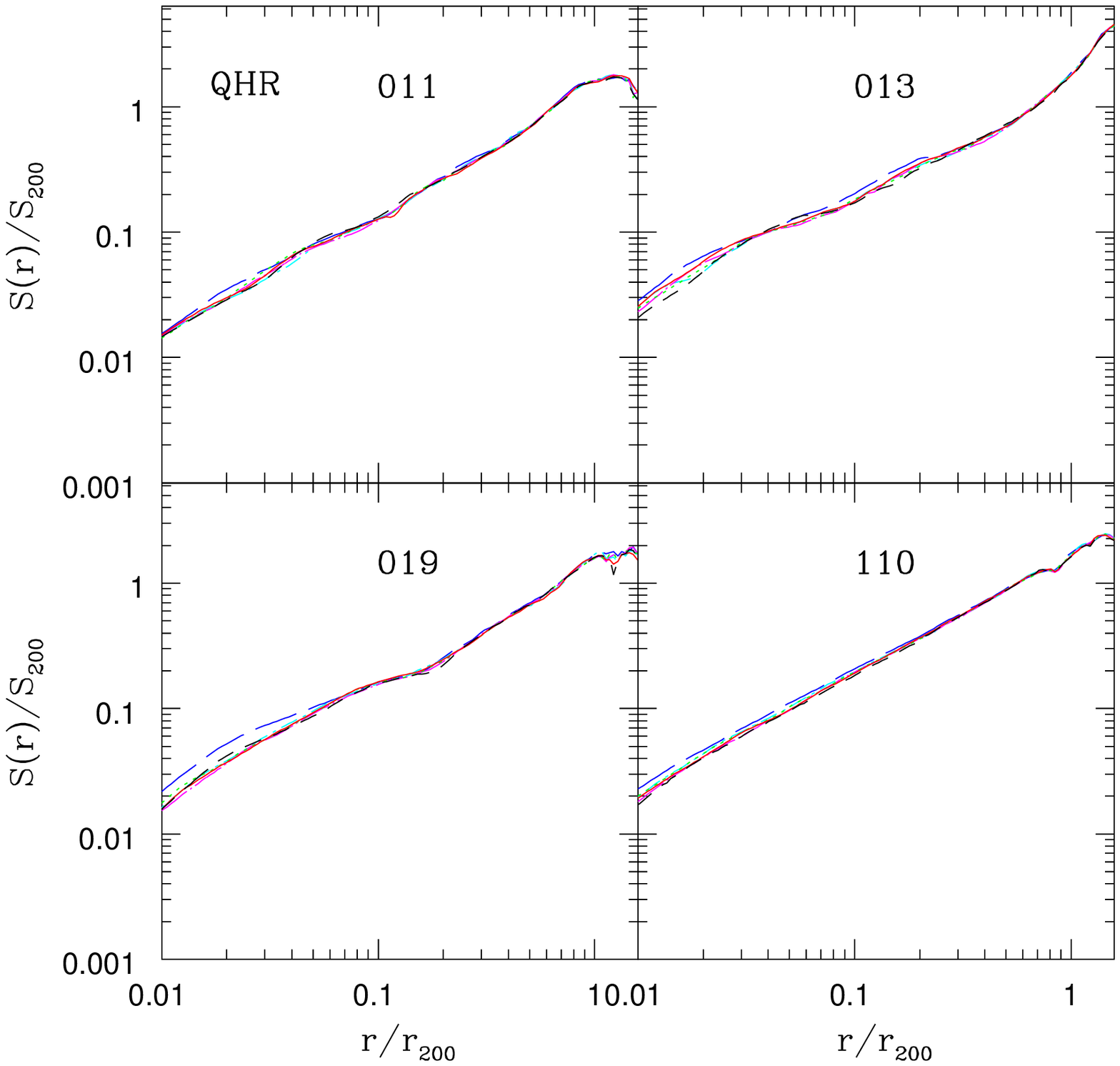}
   \caption{As in Fig. \ref{QHRentr.fig},  but for the cooling runs.}
   \label{QHRcentr.fig}%
    \end{figure*}

   \begin{figure*}
   \centering
   \includegraphics[width=17.2cm,height=13.2cm]{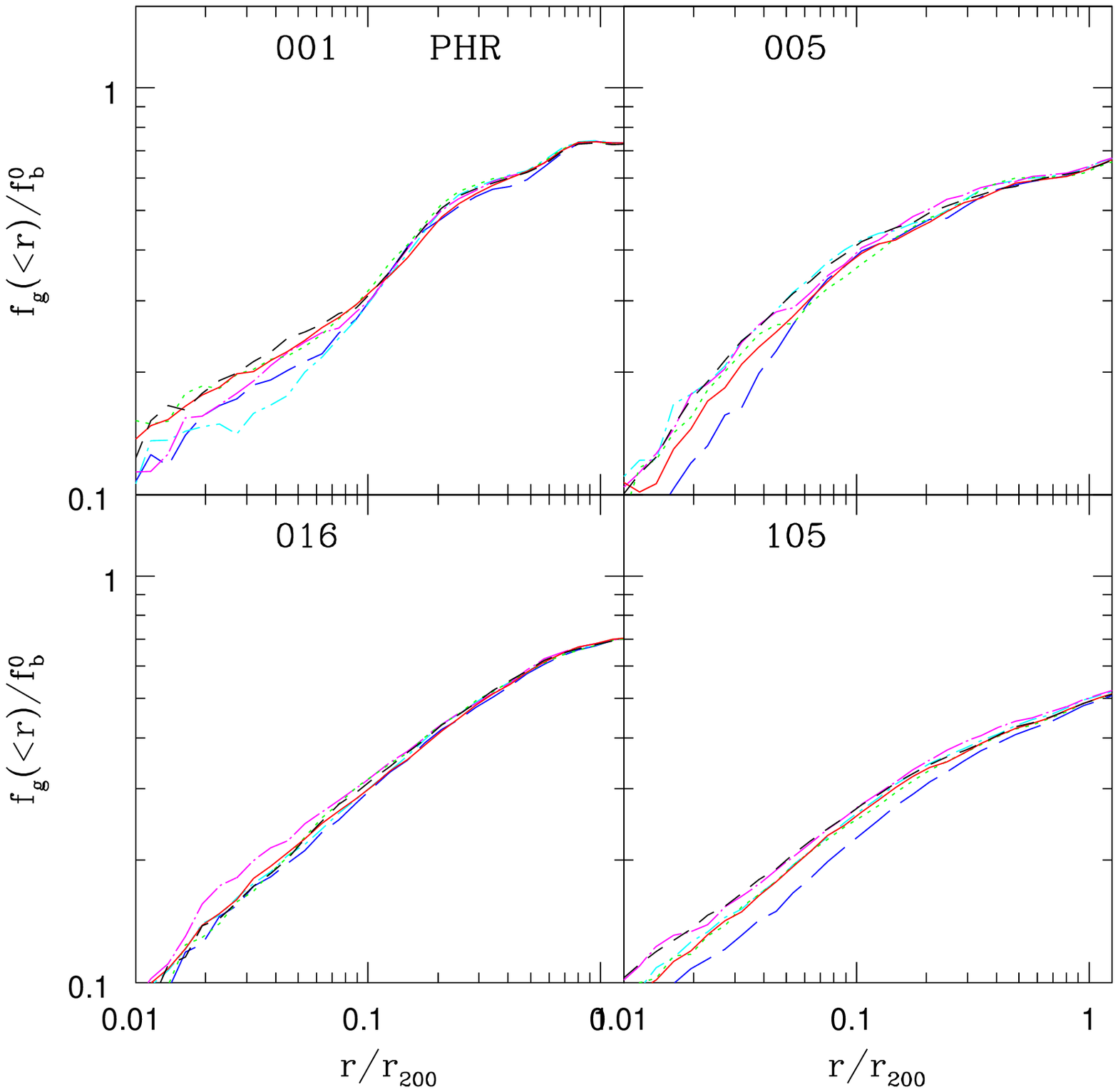}
   \caption{Final radial profiles of the gas 
fraction $f_g(<r)=M_{gas}(<r)/M_{cl}(<r)$ are shown  
 for the HR cooling runs of the perturbed cluster subsample in units of the 
cosmic value $f_b^0=\Omega_b/\Omega_m$.}
   \label{RESPcfgas.fig}%
    \end{figure*}

\subsection{Simulations with cooling and star formation}
\label{clusterc.sec}

The discussion of the previous section was aimed at investigating the effects
of numerical viscosity on ICM properties when the gas evolution of the 
cluster is driven by a relatively simple physics, in which gravity 
is the sole source of structure formation. However, a realistic modeling 
 for the physics of the ICM must incorporate the possibility for the gas to 
cool radiatively, turning cold dense clumps of gas into stars. 
Moreover, energy and metal feedback that follows from supernova explosions 
should be considered as well. 
Hydrodynamical simulations in which these physical processes have been 
taken into account  \citep{mua02,tor03,val03,kay04,nag07b} have shown that
there is substantial agreement between the predicted ICM properties, 
such as the temperature profiles, and the corresponding measurements, 
though significant discrepancies still  remain \citep{bk09}. 
In particular, the model overpredicts the star mass fraction, which is found in simulations to be a factor 
$\sim2-3$ higher than that estimated from observations. This is the 
so-called `overcooling' problem, for which several explanations have 
been proposed \citep{bk09}, although there is not yet a definitive 
solution.

An improvement in the AV scheme used in the SPH code is 
expected to significantly modify the ICM properties of the 
simulated cooling clusters. To investigate these effects, the analyses of 
the previous sections are repeated here using the same set of cluster samples, 
but in the simulations the physical modeling of the gas now includes radiative 
cooling and star formation, as well as energy and metal feedback from 
supernovae \citep{va06}. 
Our analysis of the results of the simulations will not replicate the
thorough discussion of Sections \ref{clustera.sec} and \ref{clusterb.sec}, 
but only the most important results will be presented from the viewpoint of AV
 when 
cooling is incorporated into the simulations.

When applying the spectral analysis of sect. \ref{clustera.sec} to the 
velocity fields produced by radiative simulations, a striking feature 
which emerges from the spectral behavior of the transformed fields 
is the presence at small scales of a driving source which injects 
turbulence into the ICM. The density-weighted velocity power spectra 
for the compressive components of the relaxed cluster subsample are shown
in Fig. \ref{QHRcc.fig} and exhibit the property of having a spectral
behavior which stops decaying at $\tilde k\sim 10$ and thereafter is nearly 
flat at higher wavenumbers, or even grows as in the case of cluster $110$.
These features are in sharp contrast with the corresponding spectra of 
Fig. \ref{QHRc.fig}  for the adiabatic runs and are clearly induced 
by the presence of cooling.

Previous investigations have considered generation of turbulence in
cluster cores to occur either through the interaction of the ambient ICM
with buoyantly rising bubbles created from AGN jet activities \citep{ch01,br02},
 or induced by orbital motion of galaxies \citep{ez04,ki07}.   
The origin of turbulence is interpreted here as being due to the development 
of  dense compact cool gas cores in the inner cluster regions. These are
characterized by the presence of central gas densities for which $\rho/\rho_c
\sim 10^4$, about a factor $\sim 10$ higher than in the corresponding 
adiabatic runs. Following \cite{fu04b} and \cite{fu05}, the interaction of 
these dense
cool cores with the bulk gas motion of the surrounding ICM leads then
to hydrodynamical instabilities and to the production of turbulence.
That the origin of the turbulent injection resides in the interaction of the
cluster core with the ambient medium is confirmed by the radial dependence 
of the viscosity parameters (Fig. \ref{RESPcalfa.fig}), which exhibit 
a rising ramp in the proximity of the cluster centers because of the 
weak shocks which forms near the core.
 Due to resolution issues, it appears difficult to assess in a quantitative
way the small-scale behavior of the derived spectra, although it seems 
unlikely that this effect should not be confirmed in simulations in 
which the spectra are adequately resolved, as similar features are already 
present in the low resolution runs MR and LR. For the same reason no 
attempt is made here to extract any possible dependence of the spectra 
on the numerical viscosity of the simulations, 
as there is a large scatter at small scales between 
spectra of a given cluster but with different AV settings.

These results provide strong support however for the plausibility of models 
in which radiative cooling in the cluster cores is compensated
by turbulent heating of the ICM \citep{fu04b,fu05}. 
This scenario is observationally motivated by the lack of resonant scattering
from X-ray spectra \citep{ch04} as well as from the constraints on diffusion
processes extracted from measured iron abundance profiles \citep{re06,da08},
which indicate how in cluster cores the outward diffusion of iron must
be driven by turbulent gas motion.
Heating of the ICM has been investigated by \cite{de05}, who constructed a 
set of analytical models in which radiative cooling is locally balanced 
by dissipation of turbulent motion, entropy mixing via turbulent diffusion
and heat conduction.
To compensate radiative losses, both the heating rates from turbulent diffusion
and dissipation are in general important for establishing thermal equilibrium. 
The local turbulent heating rate from turbulent dissipation is given by 
   \begin{equation}
\Gamma(r)\sim \frac{\rho {\tilde u}^3}{l}~,
  \label{turbh.eq}
   \end{equation}
where $l$ is the driving length scale and in order to balance radiative 
losses estimates derived from simple models \citep{de05,da08} require
${\tilde u}\sim 100-300$ \ks. Such a range of values is  
 smaller than that found here in the central cluster 
regions for the local velocity ${\tilde u}$ of the simulated clusters by 
roughly a factor $\sim 50\%$,
showing therefore that analytical models might overestimate the amount
of entropy mixing necessary to achieve thermal balance in cluster cores.
Note, however, that in massive clusters a viable mechanism for the dissipation 
of turbulence is collisionless damping \citep{br07},
hence for these clusters the contribution of the heating rate (\ref{turbh.eq})
 to the ICM thermal balance might be overestimated.
Finally, as indicated by Fig. \ref{RESPcalfa.fig}, a certain amount of 
shock heating is present to reduce the effects of cooling in the core. 

High levels of turbulence in cluster cores can also have a significant 
impact on the stability of cold fronts. These structures 
are interpreted as contact discontinuities which originate 
when gas of different entropy is brought into contact \citep{ma07}. 
 In the sloshing scenario \citep[][and references therein]{zu10}
this happens because of the interaction of the cool core gas with substructures 
falling through the main cluster. Accordingly, cold fronts are created as the 
subsonic sloshing of the central gas causes the encounter of gas flows from 
different directions.
However, cold fronts are prone to disruption by hydrodynamical instabilities
\citep{zu10,ro10} and it is unclear if the level of turbulence found in the
cluster cores of the simulations shown here can be reconciled with the 
presence of cold fronts as seen in many clusters. Such a issue is beyond the
scope of this paper and clearly deserves further investigations.  
In particular, hydrodynamic simulations which 
incorporate a proper modeling of the physical viscosity of the ICM can be used 
to investigate the stability of cold fronts in the sloshing scenario 
\citep{zu10}.
The derived constraints on the ICM viscosity should then permit to address
the viability of turbulent heating models to balance radiative losses.

The radial profiles of the energy density ratios, together with the 
relative pressure terms, are shown in Fig. \ref{PHRcekinr.fig}. These 
profiles are the analog for the cooling runs of those of Fig. 
\ref{PHRekinr.fig} for the nonradiative case.
A comparison between the two figures shows that at $r\simlt 0.1 r_{200}$ 
there are no large variations in the $E_{turb}/E_{tot}$  profiles, with 
the exception  of cluster $105$. This follows because at $r\simlt 0.1 r_{200}$
the steepening in $E_{turb}$ is accompanied by a corresponding steepening 
in $E_{tot}$, so that the ratio $E_{turb}/E_{tot}$ retains a profile 
approximately similar to that of the nonradiative case. The same behavior is found 
for the radial profiles of $P_{\sigma}/(P_{\sigma}+P_{th})$, which 
show a very little dependence on the AV 
strength of the simulations at $r\simgt 0.1 r_{200}$. 
These results hold as well for the 
relaxed subsample. Because the physical description of the ICM in these
simulations is more realistic than in the nonradiative case, these results
provide support for the findings of the previous section, indicating that 
numerical viscosity effects have little impact on the bias of hydrostatic mass 
estimates. 
A similar conclusion holds for the entropy profiles, which are  shown
in Fig.  \ref{QHRcentr.fig} and exhibit very little scatter between runs 
with different AV settings. This is interpreted as a consequence of galaxy
formation processes so that the small amount of entropy mixing, which 
was found in the cluster cores for the adiabatic runs, has now been removed.
 
The radial profiles of the gas mass fraction $f_g(<r)=M_{gas}(<r)/M_{cl}(<r)$ 
are shown in Fig. \ref{RESPcfgas.fig} for the perturbed test clusters. 
The quantity  $f_g(<r)$ is an increasing function of the radius $r$, as the 
amount of cooled gas subject to star formation is a decreasing function of 
radius because of the dependence of the cooling function on the gas density. 
At $r=r_{500}\sim 0.6 r_{200}$ the value of $f_{gas}$  is practically 
independent of the AV scheme used in the simulations, whereas at inner radii 
it is difficult to extract meaningful informations about the dependence 
of $f_g(<r)$ on the implemented AV scheme.
For instance, at radii $ r\simlt 0.1 r_{200}$ the highest values of  $f_g(<r)$
are those of runs AV$_5$ for clusters $001,~005$ and $105$. The lowest values
of $f_{gas}$ are found at $ r\simlt 0.1 r_{200}$ for the standard viscosity 
runs AV$_0$ in the case of clusters $005$ and $105$.

According the new AV formulation, turbulent heating of the ICM is higher 
because of the suppression of viscous damping of random gas motion. 
This in turn implies a reduced star formation efficiency because of a higher 
level of turbulent pressure, 
and at a given radius the quantity  $f_{gas}$ would exhibit  
 higher values for those runs where the AV is reduced.  
The scatter between the profiles of Fig. \ref{RESPcfgas.fig}  does not 
allow to reach firm conclusions about this dependency, 
indicating that a very large simulated sample is needed in order 
to clarify this issue.
Finding a statistically significant dependence of $f_{gas}$ on the AV scheme 
used in the simulations would support the viewpoint that turbulence 
might play a significant role in solving the overcooling problem, as suggested 
by  \cite{zh10}. 
 
To summarize, the role of numerical viscosity in simulations which 
incorporate radiative cooling is significant  in cluster cores, where 
the strength of turbulence driven by hydrodynamical instabilities 
 is strongly amplified with respect to the nonradiative case,  but is 
negligible in establishing the shape of the ICM profiles
at cluster radii $ r\simgt 0.1 r_{200}$. 
 Results from the previous section support the view that simulations 
with a much larger number of gas particles ($\simgt 256^3$) are necessary 
in order to consistently investigate the first of these issues, while 
at  radii $ r\simgt 0.1 r_{200}$ the stability of the ICM profiles presented 
here appears quite robust.

\section{Summary and Conclusions}
\label{concl.sec}

In this paper results from hydrodynamical SPH simulations have been
 presented aimed at investigating the role of the SPH numerical viscosity 
scheme in determining ICM properties of simulated galaxy clusters.
For doing this, the standard AV formulation of SPH has been modified according 
to the proposal of \cite{mm97}, where each particle has its own viscosity
parameter $\alpha(t)$ which evolves in time under the local shock conditions.
This time-dependent AV scheme has the main advantage of properly reducing
the amount of AV present in the fluid in regions far from shocks, 
with respect the standard AV scheme where the corresponding 
parameters are held constant throughout the simulation domain.
The new  time-dependent AV scheme has been implemented  in an SPH code
and several numerical tests have been performed to assess 
the validity of the code and its shock resolution capabilities. 
The chosen tests were the Riemann shock tube problem  and 3D collapse of a
cold gas sphere. Both of these problems have been widely studied in the 
literature
and results from previous simulations provide reference profiles against which 
to compare the profiles extracted from the simulations produced here.

For the tests considered the results of simulations performed using the standard
AV scheme are in good agreement with previous findings, showing the 
validity of the code. 
In order to investigate the impact of the  time-dependent AV scheme in reducing
numerical viscosity effects, for a given test case a set of SPH simulations 
which incorporate 
the new AV formulation has been constructed, in which the runs use the same
initial conditions but differ in the settings of the AV parameters 
 which control the time evolution of the individual viscosity coefficients.
The results of these simulations show profiles which are very similar to those 
produced by the fixed AV runs, indicating that the new AV scheme has shock 
resolution capabilities similar to those of the standard one and at the same time 
a reduced AV far from the shocks. 
Moreover, the time evolution of the viscosity parameter $\alpha(t)$ is 
consistent 
with theoretical expectations and at the shock front the coefficients 
reach peak values which are 
nearly independent of the chosen AV settings. Finally, in the cold gas sphere, 
the  entropy profiles of the time-dependent AV runs exhibit  
 reduced pre-shock heating  at the shock front and better agreement with the 
reference 1D PPM numerical solution.

Having established the reliability of the code and the effectiveness
of the new AV formulation, the impact of AV on ICM properties of simulated
clusters has been investigated by extracting results from a large 
ensemble of hydrodynamical cluster simulations.
The ensemble has been constructed by collecting samples of simulated 
clusters, each of the samples being realized by performing simulations 
for the same set of cluster initial conditions but using 
 a different setting of the AV parameters in the SPH code.
The baseline cluster sample with which the simulation ensemble has been constructed 
 consisted of eight clusters chosen from a cosmological simulation
ensemble with cluster virial masses ranging from 
 $M_{200}\sim 5\cdot 10^{13}h^{-1} M_{\odot}$   up to $M_{200}\sim 
6 \cdot 10^{14}h^{-1} M_{\odot}$, the selection criterion  being that of 
constructing a  representative sample of 
cluster dynamical states and masses. 
The stability of the results against the numerical resolution of the 
simulations has been tested by constructing a set of mirror ensembles
in which the only difference in the cluster simulations with respect to the 
corresponding one of the reference ensemble was in the number of simulation 
particles. Moreover, the ensembles of adiabatic cluster simulations 
have been replicated following the same prescriptions but with the
gas being allowed to cool radiatively and being subject to star formation.
The final outcome of these procedures is a set of hydrodynamical cluster 
simulations which allow  the dependence of ICM gas velocities on the AV scheme 
used in the simulations to be studied in a systematic way.  
Here, the main results are summarized.

The velocity Fourier spectra of the simulated clusters have in common 
certain features which allow several conclusions to be drawn about the 
generation of random gas velocities. All of the spectra exhibit a 
maximum at the smallest spectral cube wavenumber, showing that 
 merging and accretion processes driven by gravity at cluster scales are 
the primary sources of energy injection into the  ICM. The spectra begin to 
manifest their dependence on dissipative effects at length scales 
$l_{diss}\sim r_{200}/10\sim100-300$~kpc, below which the 
less dissipative spectra are those extracted from runs in which the AV 
settings correspond to the shortest decay time scale for the viscosity
parameter $\alpha(t)$. 
The range of values for $l_{diss}$ appears to be in accordance with previous 
findings 
\citep{su06, ma09}, suggesting that the detected spectral features are
not an artifact due to resolution effects.
These behaviors are consistent with those derived 
from the structure function radial profiles and are common to all of the
clusters, regardless of the dynamical state of the cluster.

Spectral decomposition into a transverse and a longitudinal part allows 
the resolution dependence of the spectra to be studied and useful 
insights to be obtained into the resolution requirements which must be fulfilled in 
order to adequately resolve turbulence in SPH simulations of galaxy 
clusters. The longitudinal component of the velocity power spectrum 
$E_c(k)$  is found to exhibit an approximate Kolgomorov scaling over nearly 
a decade, whilst the shearing part $E_s(k)$ has a much steeper fall off.
In analogy with previous findings \citep{fe10}, this behavior is interpreted as 
a resolution effect, as the number of computational elements needed to
adequately resolve rotational motion is higher than that necessary
for longitudinal modes, where only one degree of freedom is present.
Using scaling arguments derived from previous simulations  (K09, 
Price \& Federrath 2010) a number of
$N\sim512^3$ gas particles is then needed to adequately  describe
turbulence driven spectra in SPH simulated clusters. 
However, this estimate might be 
too pessimistic, as the SPH resolution  is not constant but increases 
with the particle density toward the cluster center, although it appears 
unlikely that less than $\sim256^3$ particles can be used to describe $E_s(k)$
 over a decade.
 
Entropy is a fundamental thermodynamic variable and the corresponding average 
radial profiles have been chosen to investigate  the dependence of the 
final ICM thermal properties on the AV scheme used in the simulations. As a 
function of the AV simulation parameters for the entropy profiles of a 
given cluster, a very small scatter is found with the largest differences 
in most cases being $\simlt 50\%$ within radial distances 
 $ r\simlt 0.1 r_{200}$.
The weak dependency of the entropy profiles on the AV parameters of the 
simulations indicates that the amount of entropy mixing in cluster cores due
to numerical viscosity effects can be considered negligible.
As a consequence, the discrepancy between SPH and Eulerian codes  in the 
 core entropy produced in adiabatic cluster simulations can be primarily 
related to the treatment of fluid discontinuities in the SPH formulation, 
rather than to AV effects.
  
At variance with the entropy profiles, the radial behavior of the 
turbulent energy density profile $E_{turb}(r)$  exhibits a significant 
dependence on the chosen AV scheme. There is an increase in the turbulent-to-total 
energy density ratio as less dissipative AV runs are considered, with the
largest variations occurring at $ r\simlt 0.1 r_{200}$.
The range of values for the ratio $E_{turb}/E_{tot}$ is in accordance with 
previous findings obtained using the AMR Eulerian code ENZO \citep{va09a} 
and provides support for the view that differences between results extracted 
from SPH and grid based codes can be reconciled using  an adequate number of 
simulation particles in SPH simulations and properly treating the AV.
 
A striking result is the lack of any significant dependence at large 
cluster radii of the rms gas velocity $\sigma(r)$ on the AV parameters 
used in the simulations. As this quantity dominates in the Jeans 
equation the mass correction terms to the hydrostatic equilibrium equation, 
it follows that numerical viscosity effects can be considered negligible 
in determining the corrections to the hydrostatic X-ray 
cluster mass estimates. Hence, previous SPH simulations in which the 
AV is treated according to the standard formulation, already provide 
numerically reliable estimates for the X-ray mass bias \citep{pi08}.
One important consequence of this result is that the agreement between 
the measured hydrostatic X-ray and weak lensing mass ratio 
 \citep{mh08,zh08,zh10b} with the X-ray mass bias predicted by 
simulations \citep{zh10b}, strongly supports
the already outlined scenario \citep{pi08}, in which the level of non-thermal
pressure present in the ICM is not significant and the gas physics outside 
cluster cores is well described by the current SPH simulations.

In simulations which incorporate radiative cooling, a significant feature is
the presence in the inner cluster regions of a much higher level of turbulence,
which is produced by the hydrodynamical instabilities generated by the 
interaction of the compact cool core with the ambient medium.
Moreover, the range of values of the local turbulent velocity is in accordance
with the constraints required by the turbulent heating model in order to balance
radiative losses \citep{de05}.

To address in a quantitative way the viability of the turbulent heating 
model, a proper treatment would be needed of the small-scale 
fluid instabilities which develop near the cluster cores. This 
requires the use of simulations with a much larger number of particles 
 and at the same time incorporating into the SPH equations the 
thermal diffusion terms which have been proposed \citep{pr08,wa08} 
to correctly model fluid instabilities in SPH.
The latter is particularly relevant in order to reliably estimate the 
radial profile of the local velocity ${\tilde u}(r)$, which is an 
important factor in establishing the local balance between the turbulent
heating rate and radiative losses \citep{de05}.
Such simulations will still miss some basic ingredients of the 
physics describing the thermodynamics of the ICM in cluster cores, 
in particular the effects due to thermal conduction, heating from AGN jets, 
physical viscosity  and 
instabilities driven by the presence of magnetic fields \citep{pa10}.
Nonetheless, the information provided by these simulations 
will probably shed light on the role played by turbulence in establishing 
the observed bimodal distribution for cluster core entropies   \citep{pra10}
and the morphological distinction between cool-core and non cool-core 
clusters.

To summarize, the results presented  here indicate that turbulence is likely
to play a significant role in some open issues of ICM physics, such as 
the cooling flow and the overcooling problems. To properly investigate these
issues it would be necessary however to further improve  the hydrodynamic SPH 
equations, so as to correctly treat fluid discontinuities, and to use 
 a much larger number of particles.
For a given test cluster, a comparison between the results produced by 
these simulations and those extracted from simulations obtained 
using an Eulerian AMR code is likely to give interesting insights 
into both the physical problem and also the numerical framework.

%These runs are aimed at studying the effects on the final gas profiles of the 
%new artificial viscosity scheme in contrast with those obtained from the 
%standard artificial viscosity formulation.

%
%                                                One column figure
%----------------------------------------------------------- S_vib
%
%______________________________________________________________

\begin{acknowledgements}
 The author is grateful to R. Brunino for kindly providing the glass-like 
initial conditions with which the numerical tests of Sect. \ref{testa.sec} 
were done and to M. Steinmetz for providing the PPM solution used in 
Sect. \ref{testb.sec}.  
S. Kitsionas and F. Vazza are also gratefully acknowledged 
 for helpful comments which clarified some of the issues discussed in 
Sect. \ref{cluster.sec}.
\end{acknowledgements}

\end{document}